\documentclass[a4paper,12pt,oneside]{article}
\usepackage[utf8]{inputenc}
\usepackage[T1]{fontenc}
\usepackage{graphicx}
\usepackage{mdwlist}
\usepackage{amsthm}
\usepackage{bm}
\usepackage{blindtext}
\usepackage{array}
\usepackage{multirow}
\usepackage{mdwlist}
\usepackage{amsthm}
\usepackage{bm}
\usepackage{blindtext}
\usepackage{enumitem}
\usepackage[numbers]{natbib}
\usepackage{url} 
\usepackage{graphicx}
\usepackage{caption}    
\usepackage{amsmath}
\usepackage{amsfonts}
\usepackage{mathrsfs}
\usepackage{amssymb}
\usepackage{enumitem}
\usepackage{multicol}
\usepackage{fancyhdr}
\usepackage{makeidx}
\usepackage{mdframed}
\usepackage{xcolor}
\usepackage{fourier-orns}
\usepackage{geometry}
\usepackage{parskip}
\usepackage{listings}
\usepackage{setspace}
\usepackage{imakeidx}
\usepackage[
            pdftex
            ,plainpages=false
            ,pdfpagelabels
            ,pagebackref = false
            ,hyperindex
            ,breaklinks=true 
            ,colorlinks=true 
            ,citecolor=black 
            ,linkcolor=black
            ,urlcolor=black
            ,filecolor=black
            ,bookmarksopen=true
            ]{hyperref}
\usepackage[nopostdot,nonumberlist]{glossaries}
\usepackage{datetime}
\usepackage{pdfpages}
\usepackage{dashrule}
\usepackage{glossaries}
\newcommand{\dd}{\mathrm{d}}
\newcommand{\keywords}[1]{\textbf{\textit{keywords:}} #1}
\newcommand{\palavraschave}[1]{\textbf{\textit{palavras-chave:}} #1}
\newcommand{\EE}{\mathbf{e}}
\newcommand{\RR}{\mathrm{R}}

\def\be{\begin{equation}}
\def\ee{\end{equation}}
\def\ni{\noindent}

\addtocontents{toc}{\protect\setcounter{tocdepth}{2}}
\numberwithin{equation}{section}
\newenvironment{resumo}{\begin{abstract}}{\end{abstract}}
\makeatletter
\def\NAT@def@citea{\def\@citea{\NAT@separator}}
\makeatother
\makeindex[name=authors, title=Author Index]
\makeindex[name=topics, title=Subject Index]
\makeglossaries

\newdateformat{customdate}{\dayofweekname{\THEDAY}{\THEMONTH}{\THEYEAR}, \THEDAY\space \monthname[\THEMONTH] \THEYEAR}

\begin{document}
\thispagestyle{empty}

\begin{figure}[!h]
\centering
	\includegraphics[scale=0.5]{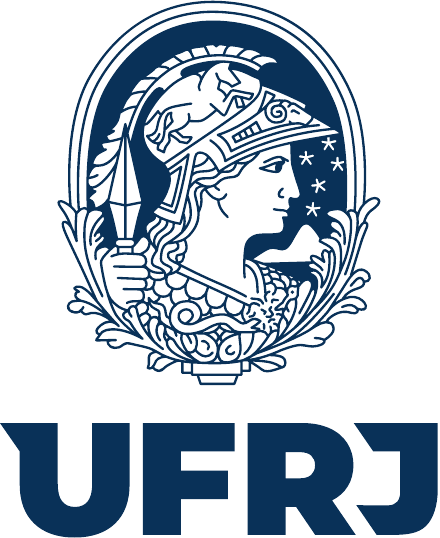}
\end{figure}
\vspace{15pt}

\begin{center}

\textbf{UNIVERSIDADE FEDERAL DO RIO DE JANEIRO}

\textbf{PHYSICS INSTITUTE}

\vspace{30pt}

\setstretch{1.5}
{\LARGE \textbf{The Warp Drive: Superluminal Travel within General Relativity}}

\setstretch{1}
\vspace{25pt}

{\Large {\bf Osvaldo Luiz dos Santos Pereira}}

{\small{\bf ORCID: 0000-0003-2231-517X}}

\vspace{35pt}

\begin{flushright}
\parbox{10.3cm}{Thesis submitted to the Graduate Program 
in Physics of the Physics Institute, Universidade Federal do Rio de Janeiro, in partial fulfillment of the requirements to the degree of Doctor in Physical Sciences.}

\vspace{18pt}

{\large {\bf 1st Adviser: Marcelo Byrro Ribeiro}} \hspace{88pt} 

\small {ORCID: 0000-0002-6919-2624} \hspace{195pt}

\vspace{12pt}

{\large {\bf 2nd Adviser: Everton Murilo Carvalho de Abreu}} 

\small {ORCID: 0000-0002-6638-2588} \hspace{195pt}

\end{flushright}

\vspace{90pt}

\textbf{Rio de Janeiro}

\textbf{\customdate 15 August 2025}

\end{center}
\thispagestyle{empty}

\vspace*{2cm}
\begin{center}
{\Large\textbf{The Warp Drive: Superluminal Travel within General Relativity}}
\\ 
{\large\textbf{Osvaldo Luiz dos Santos Pereira}}\\[1em]
{\normalsize Marcelo Byrro Ribeiro}\\
{\normalsize Everton Murilo Carvalho de Abreu}
\end{center}

\vspace{1em}

\noindent
Tese de Doutorado submetida ao Programa de Pós-Graduação em Física, Instituto de Física, da Universidade Federal do Rio de Janeiro -- UFRJ, como parte dos requisitos necessários à obtenção do título de Doutor em Ciências (Física).

\vspace{1em}
\noindent Aprovada por:
\vspace{1em}
\begin{flushright}
\makebox[10cm]{\hdashrule[0.5ex]{12cm}{0.4pt}{1pt 1.5pt}}\\
\makebox[10cm]{Prof. Dr. Marcelo Byrro Ribeiro, IF-UFRJ}\\
\makebox[10cm]{(Presidente e Orientador)}

\vspace{1em}
\makebox[10cm]{\hdashrule[0.5ex]{12cm}{0.4pt}{1pt 1.5pt}}\\
\makebox[10cm]{Prof. Dr. Everton Murilo Carvalho de Abreu, DEFIS-UFRRJ}\\
\makebox[10cm]{(Coorientador)}

\vspace{1em}

\makebox[10cm]{\hdashrule[0.5ex]{12cm}{0.4pt}{1pt 1.5pt}}\\
\makebox[10cm]{Prof. Dr. Sérgio Eduardo de Carvalo Eyer Jorás, IF/UFRJ}

\vspace{2em}

\makebox[10cm]{\hdashrule[0.5ex]{12cm}{0.4pt}{1pt 1.5pt}}\\
\makebox[10cm]{Prof. Dr. Carlos Augusto Domingues Zarro, IF/UFRJ}

\vspace{2em}

\makebox[10cm]{\hdashrule[0.5ex]{12cm}{0.4pt}{1pt 1.5pt}}\\
\makebox[10cm]{Prof. Dr. Jose Carlos Neves de Araujo, INPE}

\vspace{2em}

\makebox[10cm]{\hdashrule[0.5ex]{12cm}{0.4pt}{1pt 1.5pt}}\\
\makebox[10cm]{Prof. Dr. Carlos Alexandre Wuensche de Souza, INPE}
\end{flushright}

\vfill

\begin{center}
Rio de Janeiro, RJ -- Brasil\\
August de 2025
\end{center}
\newpage
\includepdf[pages=1, fitpaper=true]{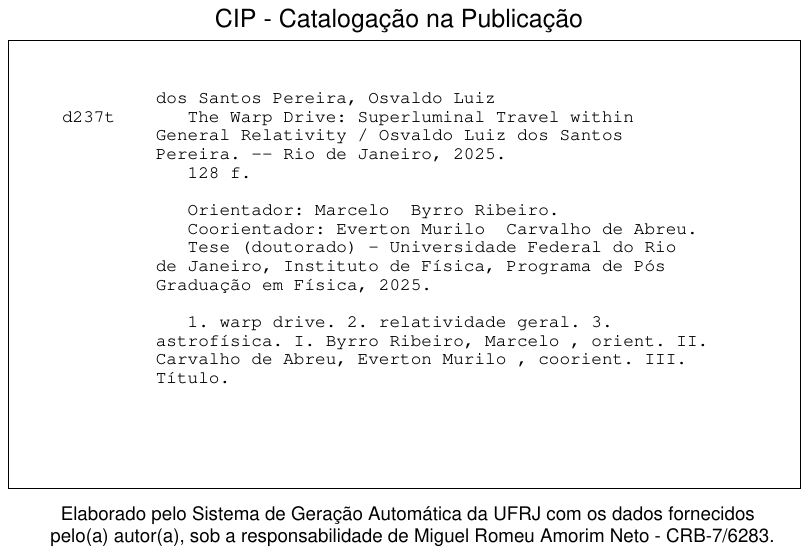}      
\newpage	
\begin{abstract}
\addcontentsline{toc}{section}{Abstract}
In 1994, Miguel Alcubierre proposed that the well-known special relativistic limitation that particles cannot travel with velocities higher than light speed can be bypassed when such trips are considered globally within specific general relativistic frameworks, using a warped region of spacetime in the shape of a bubble that transports particles with mass traveling through spacetime with superluminal speed. Although initial results indicated this scenario as nonphysical, since it would seem to require negative mass-energy density, recent theoretical analyses suggest that such a nonphysical situation may not always be true. This thesis presents newfound solutions for the Einstein field equations, considering the Alcubierre warp drive spacetime metrics. The central premise is to study the fluid matter as the gravity source, rather than the more common vacuum or negative energy sources, to explore the potential for generating superluminal velocities, or \textit{warp speeds}, through a warped region in the spacetime. Such solutions have various matter-energy sources: dust particles, perfect fluid, quasi-perfect fluid with anisotropic pressures, charged dust, and a perfect fluid within a cosmological constant spacetime. 

A connection between some of these solutions featuring shock waves described by a Burgers-type equation with a term on the right-hand side of the equation purely dependent on time is also shown. This could mean warp drives are closely related to vacuum energy and possibly have topological effects such as shock waves.

\end{abstract}
\newpage					
\begin{resumo}
\addcontentsline{toc}{section}{Summary}
Em 1994, Miguel Alcubierre propôs que a conhecida limitação da relatividade especial segundo a qual partículas não podem viajar a velocidades superiores à velocidade da luz pode ser contornada quando tais viagens são consideradas globalmente dentro de estruturas específicas da relatividade geral, utilizando uma região deformada do espaço-tempo na forma de uma bolha que transporta partículas com massa, viajando através do espaço-tempo a velocidades superluminais. Embora os resultados iniciais tenham indicado que este cenário seria não físico, uma vez que pareceria exigir densidade de massa-energia negativa, análises teóricas recentes sugerem que essa situação não física pode nem sempre ser verdadeira. Esta tese apresenta soluções recentemente encontradas para as equações de campo de Einstein considerando as métricas de espaço-tempo da propulsão de dobra de Alcubierre. A premissa central é o estudo da matéria fluida como fonte da gravidade, em vez das fontes mais comuns de vácuo ou energia negativa, para explorar o potencial de gerar velocidades superluminais, ou \textit{warp speeds}, através de uma região deformada no espaço-tempo. Tais soluções são dotadas de várias fontes de matéria-energia, a saber, partículas de poeira, fluido perfeito, fluido quase perfeito com pressões anisotrópicas, poeira carregada e um fluido perfeito dentro de um espaço-tempo com constante cosmológica. 

\palavraschave{dobra espacial, m\'etrica de Alcubierre, poeira, fluido perfeito, fluido anisotrópico, ondas de choque}
\end{resumo}		
\newpage
\begin{center}
{\LARGE \textbf{Declaration of Originality}}
\end{center}
\vspace{2cm}
I certify that I am the sole author of this thesis. All used materials, references to the literature, and the work of others have been cited. This thesis has not been presented for examination anywhere else. The original results presented here were published in Refs.\,\cite{nos1,nos2,nos3,nos4,nos5}
\newpage
\begin{center}
{\LARGE \textbf{Dedication}}
\end{center}
\vspace{2cm}
I dedicate this work to everybody who somehow helped me become who I am today and will be in the future.
\newpage
\begin{center}
{\LARGE \textbf{Acknowledgments}}
\end{center}
\vspace{2cm}
I thank my advisers, Prof. Marcelo Byrro Ribeiro and Prof. Everton Murilo Carvalho de Abreu, for their discussions and guidance throughout my PhD. 
\newpage	
\tableofcontents
\newpage
\addcontentsline{toc}{section}{List of Figures}
\listoffigures
\newpage
\addcontentsline{toc}{section}{List of Tables}
\listoftables
\section*{Introduction}	
\addcontentsline{toc}{section}{Introduction}

\index[authors]{Alcubierre} 
Alcubierre \cite{Alcubierre1994} advanced a model based on a specific general relativistic spacetime geometry in which massive particles can travel at superluminal speeds if they are located inside a specially designed spacetime distortion. A spacetime distortion refers to the warping or curvature of spacetime caused by the presence of mass and energy. This curvature affects the paths of objects, including light. Essentially, in the theory of General Relativity, gravity is not a force, but a manifestation of this curvature.
The physics of this possible propulsion method, named as \textit{\gls{warp drive}} \index[topics]{warp drive} after science fiction literature, consists of a spacetime metric forming a spacetime distortion, called \textit{warp bubble}\index[topics]{warp bubble}, such that it generates an expansion behind the distortion and a contraction in front of it. That makes it possible for a particle with mass to be inside of a local lightcone with subluminal speed, inside the warp bubble, as special relativity requires. In contrast, outside, the particle is propelled at superluminal speeds. 

\index[authors]{Alcubierre, M.}
\index[authors]{Arnowitt, R.}
\index[authors]{Deser, S.}
\index[authors]{Misner, C.\,W.}
As proposed by Alcubierre, the warp drive metric is based on a general spacetime that uses the 3+1 formalism first published by Arnowitt, Deser, and Misner \cite{adm1, adm2}. It consists of a boost in the direction of one of the spatial coordinates, described by the \textit{shift vector}\index[topics]{shift vector}, a function of spacetime coordinates, and the product of two functions: the velocity of the center of the warp bubble and the shape of the part that controls the shape of the bubble, as the \textit{regulating function}\index[topics]{regulating function}. Although Alcubierre did not present a complete and thorough solution of the \gls{Einstein equations} with his proposed spacetime geometry, he gave an example of a regulating function in the form of a hat function \cite{Alcubierre1994}. One of the main caveats of the Alcubierre warp drive spacetime is the requirement for negative mass-energy density and the violation of dominant and weak energy conditions.

\index[authors]{Ford, L.\,H.}
\index[authors]{Roman, T.\,A.}
\index[authors]{Pfenning, M.\,J.}
\index[authors]{Ford, L.\,H.}
Ford and Roman \cite{FordRoman1996} derived quantum inequalities \index[topics]{quantum inequalities}  for free massless quantized scalar fields, electromagnetic fields, and massive scalar fields in 4-dimensional \gls{minkowskispacetime} and concluded that these constraints have the form of an uncertainty principle limitation on the magnitude and duration of negative mass-energy densities and that the exotic solutions of \gls{Einstein equations}, such as wormholes and warp drive, would have significant limitations in their viability. Pfenning and Ford \cite{Pfenning1997} followed this last work and calculated the upper bound limits necessary for the warp drive viability, concluding that the energy required to create a warp bubble is ten orders of magnitude greater than the total mass of the entire visible universe, which is also negative.

\index[authors]{Van de Broeck, C.}
van de Broeck \cite{Broeck1999} showed how a minor modification of the Alcubierre warp drive geometry could reduce the total energy required to create a warp bubble. He presented an adaptation of the original warp drive metric where the total negative mass would be of the order of just a few solar masses. He suggested that further similar modifications could reduce even more, perhaps drastically, the total energy requirement for generating a superluminal warp bubble. 

\index[authors]{Krasnikov, S.\,V.}
Krasnikov \cite{Krasnikov1998} approached the hyperfast interstellar travel problem in general relativity by discussing the possibility of whether or not a mass particle can reach a remote point in spacetime and return sooner than a photon would. Krasnikov argued that a mass particle cannot win this race under reasonable assumptions for globally hyperbolic spacetimes\index[topics]{globally hyperbolic}. He discussed the specific spacetime topologies in detail, but with the constraint that tachyons would be required for superluminal travel. He conjectured a spacetime modification device that could be used to make superluminal travel possible without \gls{tachyon}. Such spacetime was named as \textit{Krasnikov tube} by Everett and Roman \cite{EveretRoman1997}, and they generalized the metric proposed by Krasnikov by hypothesizing a tube along the path of the particle connecting Earth to a distant star.

\index[authors]{Everett, A.} 
\index[authors]{Roman, T.\,A.}
The Krasnikov metric \index[topics]{Krasnikov metric} cannot shorten the time for a one-way trip from Earth to a distant star, but it can make the time for a round trip arbitrarily short for clocks on Earth. However, the Everett and Roman \cite{EveretRoman1997} extension of the Krasnikov metric has the property that the spacetime is flat inside the tube. The lightcones are opened to allow the superluminal to travel in one direction. They also mentioned that even though the Krasnikov tube does not possess closed-timelike curves\index[topics]{closed-timelike curves}, it is possible to construct a time machine with a system of two non-overlapping tubes and demonstrated that the Krasnikov tube also requires thin layers of \gls{negative energy density} and significant total negative energy.

\index[authors]{Lobo, F.\,S.\,N.}
\index[authors]{Crawford, P.}
\index[authors]{Everett, A.} 
\index[authors]{Roman, T.\,A.}
Lobo and Crawford  \cite{Lobo2002} discussed the Krasnikov metric in detail and addressed the violation of the weak energy condition and the viability of superluminal travel. These authors concluded that with the imposition of the weak energy condition, the \textit{Olum theorem} \index[topics]{Olum theorem} \cite{Olum1998} prohibited superluminal travel and pointed out the necessity of further research on spacetimes with closed timelike curves and the need for a precise definition of superluminal travel. They argued that one can construct metrics that allow superluminal travel but are flat Minkowski spacetimes. The quantum inequalities, brought from quantum field theory, were also discussed by Everett and Roman in Ref.\,\cite{EveretRoman1997}. 

\index[authors]{Nat\'ario, J.}
Nat\'ario \cite{Natario2002} proposed a new version of the warp drive theory with zero expansion \index[topics]{zero expansion} and questioned the effects that occur in the warp drive in opposition to his newly proposed metric, namely, the zero expansion \index[topics]{Nat\'ario warp drive metric} \textit{Nat\'ario warp drive metric}. Nat\'ario discussed the nature of the warp drive spacetime symmetries with a series of propositions and corollaries, such as the proof that the warp drive spacetime is flat whenever the tangent vector to the Cauchy surfaces is a Killing vector field for the Euclidean metric despite being time-dependent as a particular case. In addition, it is also flat wherever the tangent vector is spatially constant. He also showed that nonflat warp drive spacetime violates both the \textit{weak energy condition} (WEC)\index[topics]{zero expansion} and the \textit{strong energy condition} (SEC)\index[topics]{strong energy condition}, as well as the Alcubierre warp drive metric, can be obtained from\index[topics]{Nat\'ario metric} a particular choice of coordinates. The spherical coordinates were the choice of charts for the zero expansion of the Nat\'ario warp drive spacetime.

\index[authors]{Lobo, F.\,S.\,N.}
\index[authors]{Visser, M.}
\index[authors]{Nat\'ario, J.}
Lobo and Visser \cite{LoboVisser2004} pointed out that the warp drive theory is an example of reverse engineering of the solutions of the \gls{Einstein equations} where one defines a specific spacetime metric and then one finds the matter distribution responsible for the respective geometry. These authors verified that the class of warp drive spacetimes necessarily violates the classical \gls{energy conditions} even for low warp bubble velocity. Hence, this is the case for a geometric choice, not superluminal properties. They proposed a more realistic warp drive by applying linearized gravity to the weak warp drive with non-relativistic warp bubble velocities. 

\index[authors]{White, H.\,G.}
\index[authors]{Lee, J.}
\index[authors]{Cleaver, G.}
\index[authors]{Mattingly, B.}
\index[authors]{Kar, A.}
\index[authors]{Gorban, M.}
\index[authors]{Julius, W.}
\index[authors]{Watson, C.}
\index[authors]{Ali, M.\,D.}
\index[authors]{Baas, A.}
\index[authors]{Elmore, C.}
\index[authors]{Lee, J.}
\index[authors]{Shakerin, B.}
\index[authors]{Davis, E.}
\index[authors]{Cleaver, G.}
White \cite{White2003, White2011} described how an experiment to detect warp drives could be implemented at the Advanced Propulsion Physics Laboratory. He also pointed out that the expansion behind the warp bubble and contraction in front of it is due to the nature of the warp drive function and that the distortion of spacetime may be interpreted as a kind of Doppler effect \index[topics]{Doppler effect} or stress and strain on spacetime. Lee and Cleaver \cite{LeeCleaver2016} argued that external radiation might affect the warp drive and that the warp field interferometer proposed by White \cite{White2003, White2011} could not detect spacetime distortions. Mattingly et al. \cite{Mattingly2020} discussed curvature invariants in the Nat\'ario warp drive.

\index[authors]{Bobrick, A.}
\index[authors]{Martire, G.}
Bobrick  and Martire \cite{Bobrick2021} proposed a general warp drive spacetime that encloses all warp drive definitions, removing any alleged issues with the original Alcubierre warp drive. They presented a general subluminal model with spherical symmetry and positive energy solutions that satisfy the quantum energy inequalities. This reduces the warp drive requirement for \gls{negative energy density} by two orders of magnitude. They claimed that any type of warp drive, including the original one \cite{Alcubierre1994}, is a place of regular or exotic material moving in inertial form with a certain speed.

\index[authors]{Lentz, E.\,W.} 
\index[authors]{Fell, S.\,D.\,B.}
\index[authors]{Heisenberg, L.}
Lentz \cite{Lentz2021} claimed that the original warp bubble proposed by Alcubierre \cite{Alcubierre1994} can be physically interpreted as hyper-fast gravitational solitons and presented the first solution for superluminal solitons in general relativity satisfying the WEC and the momentum conditions for the conventional sources of stress, like energy and momentum, that do not require large amounts of negative energy. The basis for his work is to assume that the shift vector components obey a kind of wave equation, giving rise to a positive energy geometry. Fell and Heisenberg \cite{Fell2021} also addressed the warp drive as gravitational solitons and shed light on the Eulerian energies and their relation to the WEC, raising the possibility for superluminal spacetimes with viable amounts of energy density. 

\index[authors]{Quarra, C.\,J.}
Quarra \cite{Quarra2021} established that within the scope of general relativity, specific gravitational waveforms can result in \gls{geodesic}s that arrive at distant points earlier than light signals and presented as an example of a waveform that can be used to manifest superluminal behavior.

\index[authors]{Santiago, J.}
\index[authors]{Schuster, S.}
\index[authors]{Visser, M.}
Santiago, Schuster and Visser \cite{Santiago2021a} claimed that generic warp drive metrics violate the \textit{\gls{Null Energy Condition}} (NEC) and discussed how \textit{Eulerian observers} \index[topics]{Eulerian observers} are privileged, meaning that these observers may perceive positive energy densities, causing the impression of viable warp drive. They also argue that for the warp drive to become possible, all timelike observers must observe positive energy densities. They stated that any warp drive spacetime will unavoidably violate the \gls{energy conditions}. In a subsequent work \cite{Santiago2021b}, they claimed that exotic spacetimes, such as the warp drive one, will always violate \gls{energy conditions}. They also provided other examples of such spacetimes, relating them to wormholes, tractor beams, and stress beams.  

\index[authors]{Shoshany, B.}
\index[authors]{White, H.\,G.}
\index[authors]{Vera, J.}
\index[authors]{Bruccoleri, A.\,H.\,A.\,R.}
\index[authors]{MacArthur, J.}
For a discussion on the basics of warp drive theory, see Refs.\,\cite{AlcubierreLobo2017, Shoshany2019}. For an analysis on the possibility of warp drive features in \gls{negative energy density} distribution of an experimental Casimir cavity \index[topics]{Casimir cavity} see Ref.\,\cite{White2021} from White, Vera, Bruccoleri and MacArthur. 

\index[authors]{Santos-Pereira, O.\,L.}
\index[authors]{Abreu, E.\,M.\,C.}
\index[authors]{Ribeiro, M.\,B.}
\index[authors]{Burgers, J.\,M.}
In a series of articles Santos-Pereira, Abreu and Ribeiro \cite{nos1, nos2, nos3, nos4, nos5} proposed to couple the warp drive geometry to simple known sources of matter and energy and to solve the Einstein equations with them. We discovered that the Burgers-type equation \index[topics]{Burgers-type Equation} which describes the dynamics of \textit{shock waves} \index[topics]{shock waves} \cite{Burgers1948}, appears naturally on the geometric side of \gls{Einstein equations} and that this could lead to a \textit{warp bubble} \index[topics]{warp bubble} creation mechanism to be considered as \textit{shock waves} \index[topics]{shock waves} moving in spacetime at superluminal speeds. 

\index[authors]{Santos-Pereira, O.\,L.}
\index[authors]{Abreu, E.\,M.\,C.}
\index[authors]{Ribeiro, M.\,B.}
In our first study \cite{nos1}, we proposed the dust of particles as the source. We solved the \gls{Einstein equations} for the Alcubierre warp drive spacetime geometry since the \gls{Alcubierre metric} was not advanced initially as a solution of the \gls{Einstein equations} but as a spacetime geometry proposed without a source of gravity field. All Einstein equation solutions for this geometry containing dust lead to \gls{vacuum solution}. We also concluded that these solutions connect the Alcubierre metric to the \gls{Burgers equation}, which describes shock waves moving through an inviscid fluid.  

\index[authors]{Santos-Pereira, O.\,L.}
\index[authors]{Abreu, E.\,M.\,C.}
\index[authors]{Ribeiro, M.\,B.}
Ref.\,\cite{nos2} proposed new solutions for the \gls{Einstein equations} using the perfect fluid and a quasi-\gls{perfect fluid} with anisotropic pressure as a source. In Ref.\,\cite{nos1}, we have already shown that the Alcubierre metric has dust as a source that connects this geometry to the \textit{\gls{Burgers equation}}, which describes the shock waves moving through an inviscid fluid but leading the solutions back to vacuum. The same happened to two of four solutions subcases for the \gls{perfect fluid}. Other solutions for the \gls{perfect fluid} indicate the possibility of warp drive with positive matter density, but at the cost of a complex solution for the warp drive regulating function.\index[topics]{regulating function}

Solutions were also obtained regarding an \textit{anisotropic quasi-perfect fluid}\index[topics]{anisotropic quasi-\gls{perfect fluid}}, indicating that warp speeds could be created with positive matter density. Weak, Dominant, Strong, and Null \gls{energy conditions} were calculated for all studied subcases, which satisfied the \gls{perfect fluid} and created constraints in the anisotropic quasi-\gls{perfect fluid} quantities, allowing for a positive matter density and the formation of a warp bubble. Summing up the previous results, energy-momentum tensors describing more complex forms of matter or field distributions generate solutions for the \gls{Einstein equations} with the warp drive metric, where a negative matter density might not be a strict precondition for attaining warp speeds.

\index[authors]{Santos-Pereira, O.\,L.}
\index[authors]{Abreu, E.\,M.\,C.}
\index[authors]{Ribeiro, M.\,B.}
In Ref.\,\cite{nos3}, we analyzed the solutions of the \gls{Einstein equations} with a charged dust energy-momentum tensor as a source for warp velocities. The \gls{Einstein equations} with the \gls{Cosmological Constant}\index[topics]{cosmological constant} are written, and all solutions with energy-momentum tensor components for electromagnetic fields generated by charged dust are presented as well as the respective \gls{energy conditions}. The results show an interplay between the \gls{energy conditions} and the electromagnetic field such that, in some cases, the former can be satisfied by both positive and negative matter density. The dominant and null \gls{energy conditions} are violated. A result connects the electric energy density with the cosmological constant to the effects of the electromagnetic field on the bubble dynamics.

\index[authors]{Abell\'an, G.}
\index[authors]{Bolivar, N.}
\index[authors]{Vasilev, I.}
Following our prescription to investigate the warp drive spacetime with fluid sources as matter-energy content, Abell\'an, Bolivar, and Vasilev published two articles \cite{Abellan20232, Abellan20231} where they reproduced the same results found by us \cite{nos1, nos2, nos3} in different coordinate charts to verify if the change of geometry in the warp drive metric causes implications on the need for negative energy densities to create the warp bubble. They shared the same conclusions that we had \cite{nos1, nos2, nos3, nos4} that the warp drive has everywhere \gls{negative energy density} should be analyzed carefully since the violation of all the \gls{energy conditions} can be avoided. The warp bubble could even be produced in the laboratory \cite{White2003, White2011, Bobrick2021, Lentz2021}.

This thesis is organized in the following manner. The first chapter presents basic concepts and fundamentals regarding general relativity, such as \gls{Einstein equations}, notation, conventions, \gls{energy conditions}, \gls{ADM formalism}, and tetrad formalism\index[topics]{tetrad formalism}, used to discuss the warp drive theory. The second chapter discusses the warp drive basics, how it was first proposed by Alcubierre, the construction of this spacetime, the choice of ADM parameters made by Alcubierre for his original warp drive, fundamentals to the theory such as discussion of \gls{geodesic} equations\index[topics]{geodesic equations}, hyperbolicity of the spacetime, the Eulerian observers and the induced \gls{geodesic}. It also discusses some caveats found in the warp drive theory since the first publication by Alcubierre in 1994 \cite{Alcubierre1994}. Chapter three presents a methodological approach to analyze the warp drive spacetime. Chapter four presents the dust solution to the warp drive, discussing the results found in Ref.\,\cite{nos1}. Chapter five presents the result for the \gls{perfect fluid} from Ref.\,\cite{nos2}. Chapter six presents the result for the \gls{anisotropic fluid} from Ref.\,\cite{nos2}. Chapter seven presents the results for the charged dust from Ref.\,\cite{nos3}. Chapter eight presents the result of adding a \gls{Cosmological Constant} and altering the warp drive geometry coupled to the \gls{perfect fluid} from Ref.\,\cite{nos4}. At the end of the thesis, a chapter with conclusions and final remarks for this thesis. Several appendices contain the SAGE Manifold codes used in this work. All of the calculated components of Einstein, Ricci, and Riemann tensors are presented in these appendices. A small introduction to the thought experiment proposed by Alcubierre \cite{Alcubierre1994} as to how the warp bubble can be propelled through spacetime is presented at the final section of the appendix.

\section{Fundamentals}

This chapter presents the essential tools used in this thesis. The first section presents a notation summary, explaining, for example, how to represent covariant and contravariant tensors, how labels with Latin and Greek letters range, and which simplifications of physical units are used throughout this thesis. However, this thesis avoids the extensive use of abbreviations, and they are explicitly mentioned in the text whenever employed. 

The second section presents equations from General Relativity used in this thesis to discuss warp drive theory and the results shown in this work. It also examines the construction of essential tensors such as the \gls{riemanntensor}, Ricci tensor\index[topics]{Ricci tensor}, and Einstein tensor using the \gls{Christoffel Symbols}\index[topics]{Christoffel Symbols} and the \gls{metrictensor}. It also presents the conventions of signs in the Einstein equations \index[topics]{Einstein equations} and the use of the cosmological constant.\index[topics]{cosmological constant} 

\index[authors]{Alcubierre, M.}
The third section presents a small summary of energy condition inequalities and how they can be interpreted as limiting physical conditions for several types of observers regarding their 4-velocities in an attempt to define energy densities as positive everywhere. Section four summarizes how the electromagnetic tensor can be incorporated into Einstein's equations. Section five summarizes basic geometrical fundamentals on the ADM formalism and how the main parameters that describe kinematics and dynamics are related and contained in the warp drive theory proposed by Alcubierre \cite{Alcubierre1994}. Section six presents the fundamentals of tetrad formalism with simple examples within the warp drive context.

\index[authors]{Hartle, J.\,B.}
\index[authors]{D'Inverno, R.}
\index[authors]{Vickers, J.}
Essential reference books on General Relativity used in this thesis are Hartle's \cite{Hartle2003}, D'Inverno's and Vickers \cite{dInverno2022}. This work uses the metric sign convention 
\be
(-+++) \,.
\ee
Partial derivatives are denoted by colon $(,)$ or interchangeably,  
\be
\partial_\mu = \frac{\partial}{\partial x^\mu} \,.
\ee
The operator $\nabla_\nu$ and the semicolon $(;)$ are used interchangeably to denote the covariant derivative. As usual, a lower index means covariant components, \textit{i.e.}, vectors $A_\mu$, tensors $A_{\mu\nu}$, and an upper index means contravariant components of vectors $A^\mu$ or tensors $A^{\mu\nu}$. Mixed tensors are denoted by $A^\mu{}_\nu$ or $A_\mu{}^\nu$. Latin letters taken as index range from 1 to 3, and Greek letters, when taken as index, range from 0 to 3. Index in Latin capital letters will represent tetrad index ranging from 0 to 3, for example $e^A{}_\mu$. Lating bold letters represent vectors and operators, for example $\mathbf{A}$. 

\subsection{General Relativity}

General Relativity is a fundamental theory of physics that describes the behavior of gravity and the structure of space and time. Albert Einstein developed it in the early 20th century, and it is considered one of the pillars of modern physics. It provides a framework for understanding the behavior of massive objects, such as planets, stars, and black holes, as well as the overall structure and evolution of the universe. Unlike Newton's law of gravity, which assumes that gravity is a force acting at a distance, General Relativity describes gravity as the curvature of spacetime caused by the presence of matter and energy. 

The  General Relativity theory has had profound implications for our understanding of the universe, including the prediction of black holes, the existence of \gls{gravitational wave}, and the discovery of the expanding universe. Numerous experiments and observations have confirmed General Relativity, and it remains one of the most well-tested theories in human history. The following tensor gives the Einstein equations
\be
G_{\mu\nu} + \Lambda g_{\mu\nu} \equiv R_{\mu \nu}  - \frac{1}{2}R \,g_{\mu \nu}  + \Lambda g_{\mu\nu}
= \kappa T_{\mu\nu}\,,
\label{eeq}
\ee
where $\Lambda$ is a cosmological constant, and Einstein's gravitational constant is defined by
\be
\kappa = \frac{8\pi G}{c^4}\,,
\ee 
where $G$ is Newton's gravitational constant and $c$ is the speed of light. This work uses the natural units $G = c = 1$, hence, $\kappa = 8\pi$. $G_{\mu\nu}$ is the Einstein tensor, which is symmetric and divergence-free, $R_{\mu\nu}$ is the Ricci tensor, $g_{\mu\nu}$ is the metric tensor, and $R$ is the Ricci scalar curvature given by the following expression
\be
R = g^{\mu\nu} R_{\mu\nu} \,.
\ee
The Riemann curvature tensor is four-dimensional and describes the curvature of spacetime in terms of the distribution of matter and energy as
\be
R^\alpha{}_{\beta \mu \nu} = 
\frac{\partial}{\partial x^\mu} \Gamma^\alpha{}_{\nu \beta} 
- \frac{\partial}{\partial x^\nu} \Gamma^\alpha{}_{\mu \beta} 
+ \Gamma^\alpha{}_{\mu e}
\Gamma^\sigma{}_{\nu \beta} - \Gamma^\alpha{}_{\nu \sigma} 
\Gamma^\sigma{}_{\mu \beta} \,,
\ee
where the Christoffel symbols of the second kind $\Gamma^\mu {}_{\alpha \beta}$ are expressed as metric derivatives 
\be
\Gamma^\mu{}_{\alpha \beta} = g^{\mu \nu} \Gamma_{\nu \alpha \beta}  
= \frac{1}{2} \, g^{\mu \nu} 
\left(\frac{\partial g_{\nu \alpha}}{\partial x^\beta}  + 
\frac{\partial g_{\nu \beta}}{\partial x^\alpha}  - 
\frac{\partial g_{\alpha \beta}}{\partial x^\nu} \right)\,,
\ee
where $\Gamma_{\nu \alpha \beta}$ is the Christoffel symbol of the first kind, and its definition can also be readily seen from the above equation. The Christoffel symbols, also known as the Levi-Civita connection coefficients, are a set of coefficients used to define the covariant derivative of a vector or tensor in curved spacetime. They describe how a vector field changes as it is parallel transported along a curved path and is defined in terms of the metric tensor and its derivatives. Specifically, the Christoffel symbols of the first kind are given by
\be
\Gamma_{\sigma\alpha\beta} = g_{\sigma\nu}{\Gamma^\nu}_{\alpha\beta}\,.
\ee
The Riemann tensor has 256 dependent components, reflecting the complexity of spacetime's curvature. However, many of these components are related by symmetry properties. The Ricci curvature tensor $R_{\mu\nu}$ is a contracted version of the Riemann tensor that describes the curvature of spacetime. The scalar curvature is the trace of the Ricci tensor. Ricci scalar and tensor curvature determine whether solutions of the Einstein equations are vacuum ones when $R = 0$ and $R_{\mu\nu} = 0$. The Ricci curvature tensor $\RR_{\alpha \beta}$ is the contracting of Riemann tensor
\be
R_{\alpha \beta} = R^\mu{}_{\alpha \mu \beta} = 
\frac{\partial}{\partial x^\mu} \Gamma^\mu{}_{\beta \alpha} 
- \frac{\partial}{\partial x^\beta} \Gamma^\mu{}_{\mu \alpha} 
+ \Gamma^\mu{}_{\mu \sigma} \Gamma^\sigma{}_{\beta \alpha} 
- \Gamma^\mu{}_{\beta \sigma} \Gamma^\sigma{}_{\mu \alpha},
\ee
where the Ricci tensor is symmetric $R_{\alpha\beta} = R_{\beta\alpha }$.

The Christoffel connection coefficients define the covariant derivative, which takes a vector or tensor derivative in curved spacetime. The covariant derivative is compatible with parallel displacement, meaning that if a vector or tensor is parallel displaced along a curve, its covariant derivative taken along that curve is zero. The process of parallel displacement is used extensively in General Relativity to define the concept of a geodesic, which is the path that a free particle follows in curved spacetime under the influence of gravity. The curvature of spacetime determines the geodesic deviation from a straight line in flat spacetime, which we observe as the effect of gravity. The covariant derivative of a vector or tensor field in a curved space is
\be
\nabla_\mu A^\nu = \partial_\mu A^\nu + 
\Gamma^{\nu}{}_{\mu\sigma}A^{\sigma}\,,
\ee
where $\nabla_{\mu}$ is the covariant derivative, $A^{\nu}$ is a contravariant tensor field, $\partial_{\mu} \equiv \partial/\partial x^\mu$ is the partial derivative along the coordinate $x^{\mu}$, and $\Gamma^{\nu}_{\mu\lambda}$ are the Christoffel symbols of the second type, as mentioned above. The covariant derivative of a covariant tensor field $(A_{\alpha \beta})$ is
\be
\nabla_\mu A_{\alpha \beta} = A_{\alpha \beta;\mu} 
\equiv \partial_\mu A_{\alpha \beta}
- \Gamma^\nu{}_{\mu \alpha} A_{\nu \beta}
- \Gamma^\nu{}_{\mu \beta} A_{\alpha \nu} \,,
\ee
and the covariant derivative of a contravariant tensor $A_{\alpha \beta}$ is given by 
\be
\nabla_\nu A^{\alpha \beta} = A^{\alpha \beta}{}{}_{;\nu} 
\equiv \partial_\nu A^{\alpha \beta}
+ \Gamma^\alpha{}_{\nu \mu} A^{\mu \beta}
+ \Gamma^\beta{}_{\nu \mu} A^{\alpha \mu}
\ee
where $A^{\alpha\beta}$ is a contravariant tensor field. 

\subsection{Energy conditions}

Energy conditions in General Relativity are criteria designed to encapsulate the notion of positive energy in the theory of gravity. These conditions play a crucial role in the theoretical foundations of General Relativity and its predictions about the universe. General Relativity, proposed by Einstein in 1915, is a theory of gravitation that describes gravity not as a force between masses, as in Newton's law of gravitation, but as a result of the curvature of spacetime caused by mass-energy. 

\index[authors]{Hawking, S.\,W.}
\index[authors]{Ellis, G.\,F.\,R.}
The energy conditions are mathematical inequalities that relate the energy density and pressure within a region of spacetime. They are used to ensure that the energy-momentum tensor, which describes the distribution of matter and energy in spacetime, behaves in a way that aligns with our physical intuitions about energy. These conditions are not derived from the theory of General Relativity but are imposed on it to exclude nonphysical solutions and make supposedly meaningful physical predictions. For a more detailed exposition of the energy conditions, see Ref.\,\cite{HawkingEllis1973}. The energy conditions are named weak energy conditions (WEC), dominant energy conditions (DEC), strong energy conditions (SEC), and null energy conditions (NEC).

\subsubsection{Weak Energy Conditions}

The \gls{Weak Energy Condition} is equivalent to the assumption that the energy density measured by any observer is non-negative. The energy-momentum tensor at each point of the spacetime obeys the inequality
\be
T_{\alpha \sigma} \, u^\alpha u^\sigma \geq 0,    
\ee
for any timelike vector $(u_\alpha u^\alpha < 0)$. This is also true 
for any null vector $(k_\alpha k^\alpha = 0)$. 

\subsubsection{Dominant Energy Conditions}

The \gls{Dominant Energy Condition} stipulates that the weak energy condition must be upheld, and the energy flow vector is non-spacelike. For every timelike vector $u_a$, the following inequalities must be satisfied
\be
T^{\alpha \beta} \, u_\alpha u_\beta \geq 0, \quad \text{and} 
\quad F^\alpha  F_\alpha  \leq 0, 
\ee
where 
\be
F^\alpha = T^{\alpha \beta} u_\beta
\ee
is defined as a non-spacelike vector, that is, a future-pointing causal vector, so the physical interpretation of this inequality is that the flow of mass and energy may never exceed the speed of light. On any orthonormal basis, the energy dominates the other components of the energy-momentum tensor, 
\be
T^{00} \geq |T^{ab}|, \ \text{for each} \ a, b.
\ee
\index[authors]{Hawking, S.\,W.}
\index[authors]{Ellis, G.\,F.\,R.}
Hawking and Ellis \cite{HawkingEllis1973} suggested that this condition must hold for all known forms of matter and that it should be the case in all situations. This condition ensures that the energy density dominates over the pressure, provided that the energy flow does not exceed the speed of light. This is crucial for the causality principle in relativity. So there will be no \gls{causality violation}.

\subsubsection{Strong Energy Conditions}

The \gls{Strong Energy Condition} is stricter than the weak one. It requires a timelike convergence condition given by the following inequality
\be
R_{\mu\nu} v^\mu v^\nu \geq 0\,,
\label{tccsec}
\ee
\index[authors]{Landau, L.\,D.}
\index[authors]{Raychaundhuri, A.}
where $v^\nu$ is any \gls{timelike vector} and $R_{\mu\nu}$ is the Ricci tensor. Equation \eqref{tccsec} was first noted as a congruence condition by Landau \cite{Landau1975} and independently by Raychaudhuri \cite{Raychaudhuri1955} in what is now known as the Raychaudhuri equation \cite{Raychaudhuri1955}. The term in the equation \eqref{tccsec} is responsible for the monotonic decrease of vorticity along a geodesic; physically, this means that gravity is always attractive. From the theory of General Relativity, we know that the trace-reversed form of Einstein's equations is given by
\be
R_{\mu\nu} = \kappa\left(T_{\mu\nu} - \frac{1}{2}T g_{\mu\nu}\right)\,.
\label{tracerev}
\ee 
To derive the above formula, take the trace of the energy-momentum tensor by contracting it with $g^{\mu\nu}$ on both sides of the Einstein equations, resulting in
\be
g^{\mu \nu} R_{\mu \nu} - \frac{1}{2} R g^{\mu \nu} g_{\mu \nu} - \Lambda g^{\mu\nu} g_{\mu\nu} = \kappa g^{\mu \nu} T_{\mu \nu}\,,
\ee
and since $g^{\mu\nu} g_{\mu\nu} = 4$ in Minkowski spacetime, it implies that the expression relating the trace of the Ricci tensor and the trace of the energy-momentum tensor, when $\Lambda = 0$, is given by
\be
R = - \kappa T\,,
\ee
and substituting this expression into the Einstein equations results in Eq.\,\eqref{tracerev}. So, the strong energy condition, considering $\Lambda$ a geometrical factor, can be put in the form where the energy-momentum tensor must obey the following inequality
\be
\left(T_{\alpha \beta} - \frac{1}{2}T \, g_{\alpha \beta} \right) 
u^\alpha u^\beta \geq 0\,,
\ee
for any timelike vector $u^\alpha$. This requirement is stronger than the weak energy condition. Beyond ensuring positive energy density, the strong energy condition also incorporates the effects of pressure. It requires that the energy density and pressure sum are non-negative for all observers. This condition was important in proving the singularity theorems, including those predicting the existence of black holes.

\subsubsection{Null Energy Conditions}

The \gls{Null Energy Condition} is the limit where the strong and the weak energy conditions are satisfied for observers with null 4-velocity if the following inequality is satisfied
\be
T_{\alpha \sigma} \, k^\alpha k^\sigma \geq 0\,,
\ee
where $k^\alpha$ is any \gls{null vector}. To calculate the null energy condition, suppose the following null vector $k^\alpha$. Also, the timelike convergence condition seen in equation \eqref{tccsec} reaches its limit for observers with null 4-velocity, which means that the following inequality is known as the null convergence condition
\be
R_{\mu\nu} k^\mu k^\nu \geq 0\,,
\label{nccnec}
\ee
$k^\mu$ is a null vector. This energy condition is a lighter version of the weak energy condition. It applies to null (light-like) vectors and ensures that the energy density along any null vector is non-negative. It is useful in theorems related to the structure of spacetime, such as the formation of singularities and the properties of black hole horizons.

These energy conditions are essential for studying cosmology, black holes\index[topics]{black holes}, and wormholes\index[topics]{wormholes}. They help understand the universe's behavior at large scales, the formation of structures in the cosmos, and the theoretical possibilities of phenomena like time travel, wormholes, and warp drives. However, it is important to note that these conditions are classical, and their applicability in the quantum realm, where the energy density can be negative due to quantum effects, remains a subject of ongoing research.

\subsection{General Relativity and electromagnetism}

It is well known that Maxwell electromagnetism\index[topics]{Maxwell electromagnetism} is consistent with special relativity. The Lorentz force law states that the Maxwell equations are valid for any inertial reference system, and they can be put in what is known as the covariant formulation, which means to describe electromagnetism in special relativity language in a manifestly invariant form under Lorentz transformations. This formalism is constructed in flat spacetime with the Minkowski metric in Cartesian coordinates given by
\be
\dd s^2 = - \dd t^2 + \dd x^2 + \dd y^2 + \dd z^2\,\,.
\label{Mink}
\ee

Due to the manifest covariance of the Maxwell equations in spacetime notation, if covariant derivatives replace partial derivatives, the extra terms cancel out. The equations remain the same, making it possible to substitute the Minkowski metric $\eta_{\mu\nu}$ for a curved spacetime metric $g_{\mu\nu}$ in a general curvilinear coordinate system. The following expression represents the 4-gradient, which can be applied to tensors and vectors
\be
\partial^\mu = \frac{\partial}{\partial x^\mu} = 
\left(\frac{\partial}{\partial t}, - \nabla \right) \,\,,
\ee
and we have that $g_{\alpha \beta}$, and the 4-gradient in covariant form is given by 
\be
\partial_\alpha = g_{\alpha\beta} \partial^\beta\,.
\ee
The electric field $\mathbf{E}$ and the magnetic field $\mathbf{B}$ are described by the electromagnetic 4-potential $(A^\mu)$, which is a 4-vector with the electric scalar potential $\phi$ as its first component and the magnetic vector potential $\mathbf{A}$ as the other three components
\be
A^{\alpha }=\left(\phi,{\mathbf{A}}\right)\,\,.
\ee
The electric and magnetic fields above can be written in tensor notation, such as
\be
B^a = \epsilon^{abc}\Big(\partial_b A_c - \partial_c A_b \Big) \,,
\label{magvec}
\ee
where $\epsilon^{abc}$ is the Levi-Civita symbol\index[topics]{Levi-Civita symbol} which is either equal to $1$ if the triple of symbols $(abc)$ are even permuted, or it is equal to $-1$ if the permutation is odd or it is equal to $0$ if there are any repeated indexes. Eq.\,\eqref{magvec} can also be put in familiar vector form
\be
\mathbf{B} = \nabla \times \mathbf{A} \,,
\ee
where $\nabla$ is the nabla\index[topics]{nabla operator}, or del, operator used in mathematics, in subjects such as vector calculus as a vector differential operator.

The electric field components are written in terms of the vector potential as
\be
E_a = \partial_{{}_0} A_a - \partial_a A_{{}_0}\,\,.
\ee
The above equations for $B^a$ and $E^a$ implies that the field strength tensor $F^{\alpha \beta}$ is 
\be
F_{\alpha \beta } = \partial_{\alpha }A_\beta - \partial_{\beta}A_\alpha \,.
\ee 

In matrix form, the field strength $F^{\alpha \beta}$ can be written as,
\be
F^{\alpha \beta} = 
\begin{pmatrix}
0   & -E_1 & -E_2  & -E_3 \\
E_1 & 0    & -B_3  & B_2 \\
E_2 & B_3  & 0     & -B_1 \\
E_3 & -B_2 & B_1   & 0
\end{pmatrix} \,\,,
\ee
where the electric field $\mathbf{E}$ and the magnetic field  $\mathbf{B}$ can be written in vector form 
\begin{align}
\mathbf{E} &= (E_1, E_2, E_3)\,,
\\
\mathbf{B} &=	(B_1, B_2, B_3) \,.
\end{align}
The field strength $F^{\alpha \beta}$ is traceless, an important feature of the description of Maxwell equations in curved spacetime. The following expression gives the covariant form of the electromagnetic tensor in a general curved spacetime,
\be
F_{\alpha \beta} = g_{\alpha \mu} \,g_{\beta \nu}\, F^{\mu \nu} \,.
\ee
The energy-momentum tensor for the electriomagnetic field in a curved spacetime can be written as 
\be
T_{\alpha\beta}^{(\text{elec})} = \frac{1}{4\pi}\left(
\frac{1}{4} g_{\alpha\beta}F_{\gamma\nu}F^{\gamma\nu}
- g^{\gamma\nu}F_{\alpha\gamma}
F_{\beta \nu}\right).
\ee
The above equation is the electromagnetic contribution to the total energy-momentum tensor. If an extra term containing matter density $\mu$ 
\be
T_{\mu\nu}^{(\text{dust})} = \mu\, u_{{}_\alpha} u_{{}_\beta}
\ee 
\index[authors]{D'Inverno, R.}
\index[authors]{Vickers, J.}
If added to the electromagnetic energy-momentum tensor, it will describe a charged matter with density $\mu$ and proper velocity $u_\alpha$ \cite{dInverno2022}. The energy-momentum tensor is
\begin{align}
\nonumber T^{\alpha\beta} &= T^{\alpha\beta}_{\text{(dust)}} + T^{\alpha\beta}_{\text{(elec)}} \\
&= \frac{\mu}{4\pi} u^\alpha u^\beta  + 
\frac{1}{4\pi}\left(\frac{1}{4}g^{\alpha\beta}F_{\nu\sigma}F^{\nu\sigma}
- F^{\alpha\nu}F^\beta{}_\nu\right),
\label{eemt}
\end{align}
which is the \textit{Electromagnetic Energy-Momentum Tensor} (EEMT) for dust embedded in an electromagnetic field in a curved spacetime. Since the field strength $F_{\alpha\beta}$ is traceless, it implies that the electromagnetic part $T_{\alpha\beta}^{\text{(elec)}}$ of the total energy-momentum tensor is also traceless
\be
T_{\alpha\beta}^{\text{(elec)}} g^{\alpha\beta} = 0 \,.
\ee
If we substitute the metric $g_{\mu\nu}$ for the Minkowski metric $\eta_{\mu\nu}$, then the following expression gives the matrix form of this energy-momentum tensor
\be
T_{\alpha \nu} =
\begin{pmatrix}
\mu + \frac{1}{2}(E^2 + B^2) & -S_1 & -S_2 & -S_3 \\
-S_1 & -\sigma_{11} & -\sigma_{12} & -\sigma_{13} \\
-S_2 & -\sigma_{21} & -\sigma_{22} & -\sigma_{23} \\
-S_3 & -\sigma_{31} & -\sigma_{32} & -\sigma_{33}
\end{pmatrix}.
\label{emtmatrix}
\ee
where $E$ is the norm of the electric field $\mathbf{E}$, $B$ is the norm of the magnetic field $\mathbf{B}$, $\mu$ is the matter density, and $S_1$, $S_2$ and $S_3$ are the components of the Poynting vector 
\be
\mathbf{S} =  \mathbf{E} \times \mathbf{B} \,.
\ee 
The above equation is written in natural units where the vacuum permeability $\mu_0$ equals one. The $\sigma_{ij}$ are the nine components of the Maxwell stress tensor\index[topics]{Maxwell stress tensor}
\be
\sigma_{ij} = E_i E_j + B_i B_j - \frac{1}{2}\left(E^2 + B^2\right)
\delta_{ij}\,,
\label{maxst2}
\ee
Eq.\,\eqref{maxst2} represents the Maxwell stress tensor $\sigma_{ij}$, which encapsulates the distribution of electromagnetic stresses, the momentum fluxes, in space due to electric and magnetic fields. Specifically, it describes the flux of the $i$-th momentum component across a surface of constant $x^j$. The structure of $\sigma_{ij}$ highlights two competing physical contributions: the first term, $E_i E_j + B_i B_j$, corresponds to the directional components of the electromagnetic field tensions, or pressures, whereas the second term, $\frac{1}{2}(E^2 + B^2)\delta_{ij}$, acts as an isotropic pressure that tends to compress the field. This form ensures the tracelessness of the electromagnetic stress-energy tensor in vacuum, $\sigma_{ii} = 0$, in line with the fact that the electromagnetic field is conformally invariant in four-dimensional spacetime. The balance between anisotropic field stresses and isotropic energy density plays a central role in phenomena such as radiation pressure, electromagnetic wave propagation, and the self-consistent coupling of fields to the curvature in Einstein-Maxwell systems.

\subsection{Arnowitt-Deser-Misner formalism}

\index[authors]{Darmois, G.}
\index[authors]{Lichnerowicz, A.}
\index[authors]{Choquet-Bruhat, Y.}
\index[authors]{Dirac, P.\,A.\,M.}
Historically, the 3+1 approach has been put forward by Darmois \cite{Darmois1927}, independently by Lichnerowicz \cite{Lichnerowicz1944}, and also by Yvone Choquet-Bruhat \cite{Bruhat1952, Bruhat1956}. In 1958, 3+1 formalism started to be used to construct the Hamiltonian form of General Relativity by P.~A.~M. Dirac and later by Arnowitt, Deser, and Misner. 

The 3+1 formalism became popular in the numerical relativity community in 1970. It is used to rewrite the Einstein equation as an initial value problem and construct the Hamiltonian form of General Relativity. This method is based on the concept of the hypersurface, which is independent of whether the given spacetime is a solution of the Einstein equations or not. 

\index[authors]{Arnowitt, R.}
\index[authors]{Deser, S.}
\index[authors]{Misner, C.\,W.}
\index[authors]{Ellis, G.\,F.\,R.}
\index[authors]{Thorne, K.\,S.}
\index[authors]{Macdonald, D.}
This formalism aims to create a Hamiltonian formulation of General Relativity and achieve the canonical quantization of the gravitational field. As part of this program, Misner and Wheeler \cite{MW1957, ADM1960} wrote down the Maxwell equations in 3+1 formalism, Arnowitt, Deser, and Misner (ADM) wrote down the 3+1 Maxwell equations for point charges. In 1973, George Ellis formulated Maxwell's equations in 3+1 formalism \cite{Ellis1973}, and following his effort, significant progress was made in promoting the use of 3+1 formalism in astrophysics research, including studies of black holes, pulsars, jets, and many other fields. The reader may refer to Thorne and MacDonald \cite{KipMac} for a thorough exposition. 

ADM formalism is a way to describe spacetime in General Relativity using a $3+1$ decomposition of spacetime in time and space, hence the name $3+1$ formalism. It considers time as a separate parameter that evolves, forming a foliation of spacetime and defining hypersurfaces for each specific measured time $t$. This formalism is described as successive spacelike hypersurfaces 
\be
\Sigma_{t_1} \ , \ \Sigma_{t_2} \ , \ \Sigma_{t_3} \ , \ ... 
\ee 
Each surface $\Sigma_{t}$ represents the state of space at a particular time $t$, and the collection of these surfaces illustrates the evolution of the spacetime geometry. 

Mathematically, this \gls{spacetime foliation} is expressed by decomposing the four-dimensional spacetime metric $g_{\mu\nu}$ into a spatial three-metric $\gamma_{ij}$ on each hypersurface slice $\Sigma_t$ and the time parameter as a timelike 4-vector that flows through the slices. This ADM formalism uses special functions that govern the kinematics and the system's dynamic evolution. One of those functions is the \textit{lapse function} \index[topics]{lapse function} $\alpha$, which determines the time progression rate between different hypersurface slices. Another function is the \textit{\gls{shiftvector}} \index[topics]{shift vector} $\beta^i$, which relates the spatial coordinates of one slice to the next, describing the direction of moving particles onto the hypersurface and how fast those particles move on the surfaces. Time flows perpendicular to the hypersurfaces, determined for a fixed, constant, and unique time. Every vector embedded in the hypersurfaces is spatial by default.

\begin{figure}[ht]
 \centering
 \includegraphics[scale=1.6]{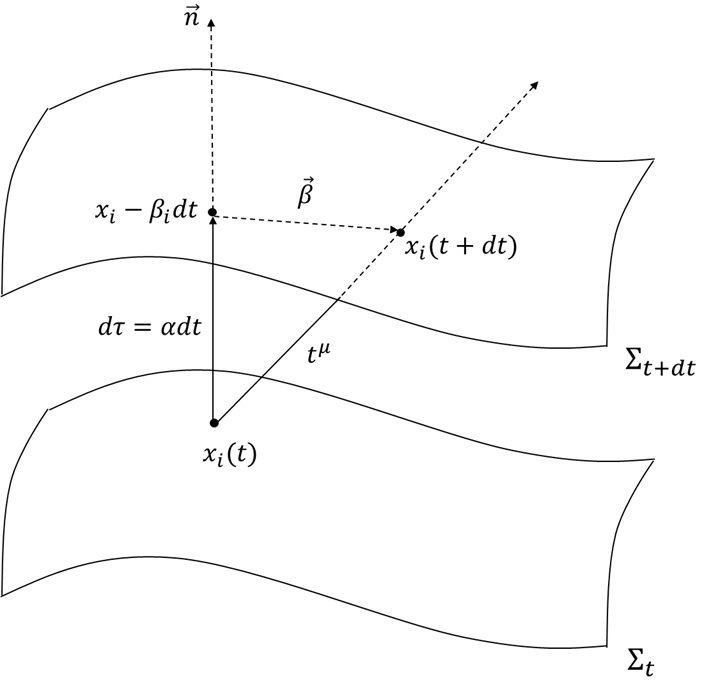}
 \caption[Schematic representation of the 3+1 decomposition of spacetime.]{Schematic representation of the 3+1 decomposition of spacetime with hypersurfaces of constant time $t$ coordinate $\Sigma_t$ and $\Sigma_{t+\dd t}$ foliating the spacetime. For each time-sliced hypersurface, two parameters explain the dynamics and kinematics of the ADM formalism, the surface metric $\gamma_{ij}$ and the \gls{extrinsic curvature} tensor $K_{ij}$.}
 \label{foliation}
\end{figure}

The time evolution of the geometry is governed by the Einstein equations, which in this formalism breaks down into the Hamiltonian \index[topics]{Hamiltonian} and momentum constraints, and the evolution equations for the spatial $\gamma_{ij}$ and $K_{ij}$, the \gls{extrinsic curvature}\index[topics]{extrinsic curvature} of the slices. This curvature quantifies how a spatial slice is embedded into the four-dimensional spacetime and the degree to which spacetime is curved. Fig.\,\ref{foliation} illustrates the main parameters of the ADM formalism, giving a geometrical intuition to those parameters, and how the spacetime is divided into foliations.

The vector $\vec{n}$, with coordinates $n^\mu$, is normalized and normal to the hypersurfaces: it projects points in time $t$ from one hypersurface onto the next. The 4-vector $t^\mu$ connects points in time $t$ on neighbor slices $\Sigma_t$ and $\Sigma_{t+\dd t}$ at times $t+\dd t$. The shift vector $\vec{\beta}$ connects these points in times $t$ and $t + \dd t$ in the same hypersurface $\Sigma_{t+\dd t}$ carrying only spatial information. This means that the shift vector is purely geometrical. However, it is common in this ADM formalism\index[topics]{ADM formalism} to define a 4-vector for the shift vector $\beta^\mu$ that contains spacetime time coordinate information. These three vectors $n^\mu, t^\mu, \beta^\mu$ form an equilateral triangle with right angles between the normal vector $\vec{n}$ and the shift vector $\vec{\beta}$. 

\index[authors]{Alcubierre, M.}
Alcubierre \cite{Alcubierre1994} used this ADM formalism to construct his original Warp Drive. One of the primary purposes of using this formalism is that it can make spacetime globally \gls{hyperbolic} with the correct choice of parameters, ensuring no closed spacetime curves and no going back in time. 

The ADM equations consist of constraint equations satisfied on each slice. Evolution equations describe how $\gamma_{ij}$ and $K_{ij}$ change over time. It describes how the shape of the space changes from one slice to the next and is related to the time derivative of the spatial metric. The spacetime interval $\dd s^2$ in the ADM formalism is
\be
\dd s^2 = -\alpha^2 \dd t^2 
+ \gamma_{ij} (\dd x^i + \beta^i \dd t)(\dd x^j + \beta^j \dd t)\,.
\ee
The \gls{lapsefunction}\index[topics]{lapse function} relates the proper time $\dd\tau$ experienced by an observer moving normally between the hypersurfaces to the coordinate time $\dd t$
\be
\alpha = \frac{1}{\sqrt{-g^{00}}} = \frac{\dd\tau}{\dd t} \,.
\ee
The shift vector $\beta^i$ is defined from the off-diagonal components $g_{0i}$ of the spacetime metric
\be
\beta^i = \gamma^{ij} g_{0j} \,,
\ee
where $\gamma^{ij}$ is the inverse of the spatial metric, defined as the metric induced on each spatial hypersurface, and $h_{\mu\nu}$ is the projection of the spacetime metric onto the hypersurface
\be
h_{\mu\nu} = g_{\mu\nu} + n_\mu n_\nu \,.
\label{admhype}
\ee
For the purely spatial components, this simplifies to:
\be
h_{ij} = g_{ij}\,.
\ee
Since $g_{ij}$ components are not $\dd t$ dependent, it makes them strictly spatial. The ADM formalism parameters are the lapse function $\alpha$, the shift vector $\beta^i$, and the spatial induced metric $\gamma_{ij}$. Those parameters are connected to the spacetime metric components $g_{\mu\nu}$ by 
\begin{align}
\alpha &= \sqrt{-g_{00}} \,, 
\\ \beta^i &= \frac{g_{0i}}{g_{00}}\,, 
\\ \gamma_{ij} &= g_{ij} \,.
\end{align}
These relations allow the decomposition of the spacetime metric into its spatial and temporal parts, which is particularly useful in numerical relativity, where the Einstein equations are solved as an initial value problem on each spatial slice. 

The \gls{extrinsic curvature} $K_{ij}$ captures the curvature of a hypersurface embedded in a higher-dimensional spacetime. Defined as the Lie derivative of the spatial metric $\gamma_{ij}$ along the unit normal vector $n^\mu$ to the hypersurface $\Sigma$, see Eq.\,(2.3.6) in Ref.\,\cite{Alcubierre2012}
\be
K_{ij} = -\frac{1}{2} \mathcal{L}_{\vec{n}} \gamma_{ij} \,,
\ee
if can be expressed in terms of the covariant derivative of $g_{\mu\nu}$, see Eq.\,(2.3.2) in Ref.\,\cite{Alcubierre2012}
\be
K_{ij} = - h_{i}^{\phantom{\ }b} \nabla_b n_j \,,
\ee
where $h_{a}^{\phantom{\ }b}$ projects tensors from the four-dimensional spacetime onto the three-dimensional hypersurface, $n^\mu$ is the unit normal vector to the hypersurface, and satisfies the condition, and the negative norm defines timelike vectors
\be
n^\mu n_\mu = -1 \,.
\label{normcondunivec}
\ee 
Eq.\,\eqref{normcondunivec} defines $n^\mu$ as a unit timelike normal vector to a spacelike hypersurface in the ADM decomposition. It represents the four-velocity of Eulerian observers\index[topics]{Eulerian observers}, ensuring proper time evolution along the foliation. This normalization anchors the split between space and time in a Lorentzian manifold\index[topics]{Lorentzian manifold}.

\subsubsection{The Lie Derivative}

The Lie derivative represents the change of a tensor field along the flow of a vector field. In the ADM formalism, the Lie derivative measures the change in the spatial metric $\gamma_{ij}$ as one moves along a direction orthogonal to the spatial hypersurfaces if one chooses to. This is crucial for defining the extrinsic curvature\index[topics]{extrinsic curvature} $K_{ij}$, quantifying how the hypersurface is embedded in the whole spacetime. Physically, $K_{ij}$ encodes the rate of change of the spatial geometry as seen by Eulerian observers and enters directly into the Hamiltonian formulation of General Relativity, governing the evolution of the geometry via Einstein’s equations.

For a tensor field $T$ and a vector field $V$, the Lie derivative of $T$ along $V$ is written as
\be 
\mathcal{L}_V T \,,
\ee 
is defined as coordinate-free. For the case of the \gls{extrinsic curvature} $K_{ij}$, the vector field $V$ is the unit normal $n^\mu$ to the hypersurface. The Lie derivative of a rank-2 tensor like the spatial metric $\gamma_{ij}$ along the unit normal $n^\mu$ is
\be
\mathcal{L}_{n} \gamma_{ij} = n^\mu \nabla_\mu \gamma_{ij} + \gamma_{\mu j} \nabla_i n^\mu + \gamma_{i\mu} \nabla_j n^\mu\,.
\ee
Accounting for the direct change of $\gamma_{ij}$ in the direction of $n^\mu$, and the change due to the twisting of spatial coordinates as one moves along $n^\mu$. The \gls{extrinsic curvature} is 
\be
K_{ij} = \frac{1}{2\alpha} 
\left( \dot{\gamma}_{ij} - \nabla_i \beta_j - \nabla_j \beta_i \right)
\ee
where $\dot{\gamma}_{ij}$ denotes the time derivative of the spatial metric. The \gls{extrinsic curvature} plays a crucial role in the dynamics of General Relativity, entering the Hamiltonian and momentum constraints of the ADM formalism, governing the evolution of spacetime geometry. It also measures how much matter and energy are present within the hypersurface through the Einstein equations. 

\begin{figure}[ht]
 \centering
 \includegraphics[scale=1.2]{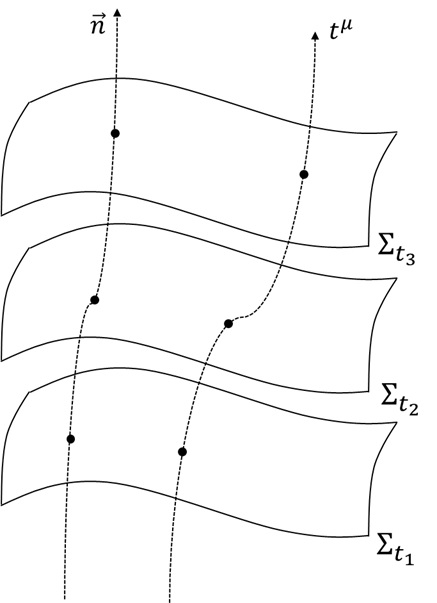}
 \caption[Eulerian observers in ADM formalism.]{Schematic representation of the 3+1 decomposition of spacetime with the world lines of Eulerian observers $\vec{n}$ and coordinate observers $t^\mu$.}
 \label{eulerianobs}
\end{figure}

Fig.\,\eqref{eulerianobs} illustrates how the world lines of two different observers flow through the hypersurfaces of constant time. The lines following the 4-vector $\vec{n}$ are commonly known as the \textit{Eulerian observers}, being associated with timeline geodesics, and following the $t^\mu$ world-line are observers with proper time measured by the spacetime coordinate $t$. This geometric picture underpins the initial value formulation in numerical relativity, where spatial slices evolve over coordinate time.
 
\begin{figure}[ht]
 \centering
 \includegraphics[scale=1.4]{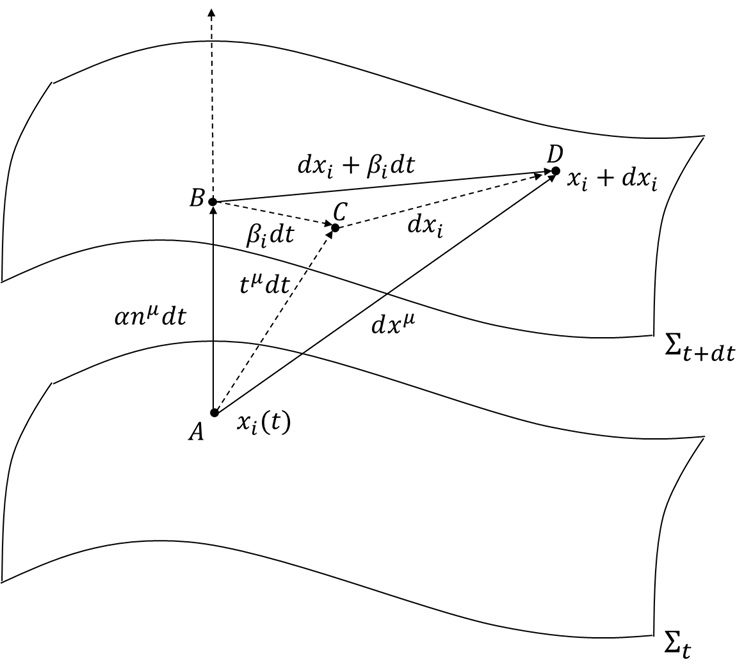}
 \caption{Schematic of 4-vector decomposition on the ADM formalism.}
 \label{adm6}
\end{figure}

Fig.\,\eqref{adm6} illustrates the foliation of spacetime using two neighboring hypersurfaces $\Sigma_{t}$ and $\Sigma_{t + \dd t}$, separated by an elapsed time $\dd t$. This figure has two important triangles $\Delta ABC$ and $\Delta ABD$ relating the geometrical aspects of ADM formalism. Another way to derive the 3+1 formalism general metric is noticing that the triangle $\Delta ABC$ has right angles because of the normal unit vector $n^\mu$. By the Pythagoras theorem in Lorentzian signature, one can find the following expression relating line segments
$\overline{AD}$, $\overline{AB}$, and $\overline{BD}$
\be
\lVert \overline{AD} \rVert^2 = - \lVert \overline{AB} \rVert^2 + \lVert \overline{BD} \rVert^2
\label{admdist}
\ee

The line segment $\lVert\overline{AD}\rVert$ is just the metric distance $\dd s$, the spacetime interval between points $A$ and $D$. The line segment $\lVert\overline{AB}\rVert$ is the timelike distance between the neighboring hypersurfaces $\Sigma_t$ and $\Sigma_{t + \dd t}$, analogous to the distance between two planes in Euclidean geometry. The line segment $\lVert\overline{BD}\rVert$ is the spacelike distance between points on the sliced hypersurface $\Sigma_{t + \dd t}$. Inserting these expressions in Eq.\,\eqref{admdist} leads to the following result
\be
\dd s^2 = - \lVert \alpha n^\mu \dd t \rVert^2 +
\lVert \dd x^i + \beta^i \dd t\rVert^2\,,
\label{normadm}
\ee
expanding the terms in the above Eq.\,\eqref{normadm} in terms of its spacetime coordinates for each 4-vector leads to the following result
\be
\dd s^2 = - \alpha^2 \dd t^2 + \gamma_{ij}\left(\dd x^i + \beta^i \dd t\right) \left(\dd x^j + \beta^j \dd t\right) \,,
\ee
which is the ADM formalism general metric $g_{\mu\nu}$. This geometric derivation of the metric gives great intuition into how the ADM formalism parameters behave. The minus sign in front of the term $\lVert \alpha n^\mu \dd t \rVert^2$ appears because we are not using the usual Euclidean norm to calculate distances; we are using the Minkowski norm induced by the Minkowski metric space, which carries a minus sign in its signature. For a deeper discussion on Minkowski spacetime see Ref.\,\cite{dInverno2022}. As an example, for a general vector in Minkowski space given by $\{t,x,y,z\}$, the norm for this 4-vector is given by
\be
\lVert (t,x,y,z) \rVert^2 = -t^2 + x^2 + y^2 +z^2\,,
\ee
so the distance between two points $P1 = (t_1,x_1,y_1,z_1)$ and $P2 = (t_2,x_2,y_2,z_2)$ in Minkowski space, induced by its norm, is given by
\be
\dd(P1,P2) = \lVert P_2 - P1 \rVert^2 = -(t_2-t_1)^2 + (x_2-x_1)^2 + (y_2-y_1)^2 + (z_2-z_1)^2\,.
\ee
The spatial metric $\gamma_{ij}$ makes its presence because it carries information about how spatial coordinates behave on the sliced hypersurface $\Sigma_t$.

\subsection{Tetrad formalism}

In General Relativity, the tetrad formalism provides a powerful and flexible framework for describing spacetime by introducing a locally flat reference frame in a curved manifold \index[topics]{curved manifold} at each point. Instead of using the coordinate-based metric tensor $g_{\mu\nu}$ directly, the tetrad formalism expresses the metric in terms of a set of four linearly independent basis vectors, known as the tetrads, or \textit{vielbeins}\index[topics]{vielbeins}, in four dimensions. These vectors define a local orthonormal frame\index[topics]{local orthonormal frame}, allowing for a more explicit geometric interpretation and simplifying the incorporation of spinor fields and gauge-theoretic approaches to gravity.

The fundamental idea behind the tetrad formalism is to relate the general curved spacetime, described in a coordinate basis, to a locally flat Minkowski space at each point. This is achieved by introducing a set of tetrad (or \textit{vielbein}) fields $e^\mu_a$, which map between the coordinate basis indices (Greek letters) and the locally Minkowski basis indices (Latin letters). The metric tensor in this formulation is given by
\be
g_{\mu\nu} = \eta_{ab} e^a_{\mu} e^b_{\nu}\,,
\ee
where $\eta_{ab}$ is the Minkowski metric of special relativity. The tetrads thus serve as a bridge between General Relativity and Special Relativity at the local level.

One of the main advantages of tetrad formalism \index[topics]{tetrad formalism} is its utility in studying spacetimes with strong gravitational fields, particularly in connection with spinor fields\index[topics]{spinor fields}, the Einstein-Cartan theory\index[topics]{Einstein-Cartan theory}, and supergravity\index[topics]{supergravity}. Moreover, it provides a convenient way to express the equations of motion in a form reminiscent of gauge theories, making it useful in modern approaches to quantum gravity\index[topics]{quantum gravity}.

\index[authors]{Cartan, E.}
\index[authors]{Weyl, H.}
\index[authors]{Einstein, A.}
The idea of tetrads was developed independently by \'Elie Cartan \cite{Cartan1922}, Einstein \cite{Einstein1928}, and Hermann Weyl \cite{Weyl1929}. Cartan tried to create another general theory of gravitation that could account for torsion and spin, while Einstein and Weyl introduced the concept of frame fields. In General Relativity, a frame field is called a tetrad (or vierbein from the German language). The traditional formalism of General Relativity uses coordinate basis $\{\EE_\mu\} = \{\partial_\mu\}$, but it is always possible to define an orthogonal and locally inertial reference frame, defining a basis at a point in spacetime using tangent vectors\index[topics]{tangent vectors}
\be
\EE_{a} = e^\mu{}_a \, \partial_{\mu} \,.
\label{tetradtan}
\ee
The Latin index is used for the non-coordinate basis, while the Greek index is used for the coordinate basis. In Eq.\,\eqref{tetradtan}, $e^\mu{}_A$ are the components of the tetrad $e_{a}$. Defining a set of dual-vectors that form a dual-basis \index[topics]{dual basis} (normal or co-tangent) described as 1-forms\index[topics]{1-form}
\be
\EE^a = e^a{}_\mu \dd x^\mu \,,
\ee
so the components of the tetrad and the co-tetrad \index[topics]{co-tetrad} are mutually orthonormal
\be
e^a{}_\mu e^\mu{}_b = \eta^a{}_{b}\,,
\label{tetradortho}
\ee
and $\eta^a{}_{b}$ is the Minkowski metric on a non-coordinate basis. When dealing with Riemannian manifolds, the Kronecker delta \index[topics]{Kronecker delta} $\delta^a{}_b$ can be used instead. The metric tensor operator, taken in its matrix representation $\mathbf{g}$ can be written in terms of a coordinate basis as \index[topics]{coordinate basis}
\be
\mathbf{g} = g_{\mu\nu} \dd x^\mu \dd x^\nu  
\label{tetcoord}
\ee
where $g_{\mu\nu}$ is given, the metric components are evaluated with the coordinate basis.
\be
g_{\mu\nu} = \mathbf{g}\left(\partial_\mu,\partial_\nu\right) \,,
\ee
where $\mathbf{g}(,)$ is a bilinear operator that takes two input parameters in their slots. Likewise, we can express the metric tensor in terms of a tetrad basis
\be
\mathbf{g} = g_{ab} \, \EE^a \EE^b\,,
\ee
and remembering that we can write the tetrad in terms of 1-forms \index[topics]{1-form} $\EE^a = e^a{}_\mu \dd x^\mu$, inserting this definition in the above equation results in the following expression
\be
\mathbf{g} = g_{ab}(e^a{}_\mu \dd x^\mu) \, (e^b{}_\nu \dd x^\nu) \,,
\ee
reorganizing the above equation for a clear interpretation leads to
\be
\mathbf{g} = (g_{ab} \, e^a{}_\mu e^b{}_\nu ) \, \dd x^\mu \, \dd x^\nu 
\ee
and remembering that the matrix representation of the metric tensor $\mathbf{g}$ can be represented in coordinate basis as Eq.\,\eqref{tetcoord}, we arrive at the following identity
\be
\mathbf{g} = g_{\mu\nu} \, \dd x^\mu \dd x^\nu = (g_{ab} \, e^a{}_\mu e^b{}_\nu ) \, \dd x^\mu \, \dd x^\nu\,,
\ee
we then derive an important relation between the metric tensor components $g_{\mu\nu}$ and the tetrad vectors $e^A{}_\mu$ given by the formula
\be
g_{\mu\nu} = g_{ab} \, e^a{}_\mu e^b{}_\nu \,,
\label{tetnoncoord}
\ee
the metric components are given on a tetrad non-coordinate basis as 
\be
g_{ab} = \mathbf{g}\left(\EE_a,\EE_b\right) = \eta_{ab}
\ee
and from the orthogonality condition from  Eq.\,\eqref{tetradortho} it is easy to deduce that Eq.\,\eqref{tetnoncoord} can be linear transformed into the following representation  
\be
g_{ab} = g_{\mu\nu} \, e^\mu{}_a e^\nu{}_b\,.
\ee
The above expression is easily derived by remembering the duality condition of tetrads
\be
e^a{}_{\mu} e^\nu{}_{a} = \eta_{\mu}{}^\nu \,,
\ee
\be
e^\mu{}_{a} e^b{}_{\mu} = \eta_{a}{}^b.
\ee
As a worked example, one may find tetrads for the Alcubierre warp drive metric \cite{Alcubierre1994}
\be
\dd s^2 = - (1 - \beta^2) \dd t^2 + 2 \beta \, \dd t \dd x 
+ \dd x^2 + \dd y^2 + \dd z^2
\ee
where $\beta = \beta(t,x,y,z)$ is the spacetime coordinates shift vector in the context of the 3+1 formalism. This function carries information about the velocity and shape of the warp bubble, which is responsible for transporting mass particles through spacetime in a superluminal fashion, so defining two natural local tetrads 
\be
\EE_a = e^\mu{}_a \, \partial_\mu\,,
\ee
the local reference frames of either a co-moving or a static observer. An example of a co-moving tetrad \index[topics]{co-moving tetrad} for the warp drive can be found in Ref.\,\cite[Chap. 2.7]{MuellerGrave2010}
\begin{align}
\EE_0 &= \partial_t + \beta \, \partial_x \,,\\
\EE_1 &= \partial_x \,,\\
\EE_2 &= \partial_y \,,\\
\EE_3 &= \partial_z \,,
\end{align}
Here $\EE_0$ represents the observer's time direction, which includes the velocity shift $\beta$, and $\EE_1, \EE_2, \EE_3$ are purely spatial and correspond to the standard Cartesian basis. Since this observer moves with the warp bubble, they do not experience relative motion. This means they would measure themselves as being at rest within the bubble. 

An example of a static local tetrad for the Alcubierre drive can be represented by
\begin{align}
\EE_0 &= \frac{1}{\sqrt{1 - \beta^2}} \, \partial_t \,,
\label{e0tetrad} \\
\EE_1 &= \frac{\beta}{\sqrt{1-\beta^2}} \,  \partial_t 
+ \sqrt{1-\beta^2} \,  \partial_x \,, \label{e1tetrad} \\
\EE_2 &= \partial_y \,,\\
\EE_3 &= \partial_z \,.
\end{align}
Here $\EE_0$ is the rescaled time component, accounting for the time dilation effect due to the movement of the warp bubble, and $\EE_1$ mixes time and space components, reflecting that for a static observer, motion in time is intertwined with space due to the presence of the shift vector $\beta$. These tetrads show that an observer outside the warp bubble perceives time and space differently.

\index[authors]{Alcubierre, M.}
The tetrads help us understand the different perspectives of observers in and outside the warp bubble. The co-moving observer is attached to the observer inside the warp bubble and cannot move with velocities faster than the speed of light, respecting the laws of Special Relativity \cite{Alcubierre1994}, remaining well-behaved at all times. In contrast, the static frame experiences significant relativistic distortions for large $\beta$, implying that the warp bubble allows for apparent faster-than-light travel without locally breaking special relativity. While the co-moving tetrad is valid everywhere in the warp drive spacetime, the static tetrad can only be defined where $\beta < 1$, as Eq.\,\eqref{e0tetrad} and Eq.\,\eqref{e1tetrad} show. Also, notice that $\beta > 1$ is not allowed as shown in Eqs.\,\eqref{e0tetrad} and \eqref{e1tetrad}, since this would mean that the tetrad would be a complex-valued function. This is an interesting result since the function $\beta$ can be interpreted as the velocity of the warp bubble, also carrying the information of the form of this bubble. 
\section{Warp Drive basics}

\index[authors]{Alcubierre, M.}
This section presents the basic concepts of warp drive Theory. The first section introduces the Alcubierre warp drive spacetime, relating the ADM formalism to the choice of parameters to construct the original warp drive metric. Geometrical aspects of the warp bubble and asymptotic behavior are introduced, as presented by Alcubierre with the help of a step function, to make the warp drive metric asymptotically flat\index[topics]{asymptotically flat}. The volume expansion formula in terms of the extrinsic curvature is presented. Section two discusses the geometrical aspects of geodesics and lightcones in warp drive spacetime, demonstrating that the geometry described by Alcubierre is Lorentzian and globally hyperbolic with free-falling observers along the geodesics. The tilting of lightcones near the walls of the warp bubble is discussed qualitatively. The third section discusses the caveats of the warp drive theory discovered since its first publication in 1994. There is a need for negative energy densities, \gls{exoticmatter}\index[topics]{exotic matter} and event horizons\index[topics]{event horizon} within the warp bubble and other aspects. 

\subsection{Introducing the Alcubierre Warp Drive metric}

Alcubierre \cite{Alcubierre2012} made use of the 3+1 formalism metric to define the warp drive metric
\begin{align}
\nonumber {ds}^2 &= - \dd\tau^2 = g_{\mu \nu} \dd x^\mu \dd x^\nu, \\ 
&= - \left(\alpha^2 -\beta_i\beta^{i}\right) \dd t^2 + 2 \beta_i \dd x^i \dd t + \gamma_{ij} \dd x^i \dd x^j,
\label{metric1}
\end{align}
where $\alpha$ is the lapse function, $\beta^i$ is the shift vector and $\gamma_{ij}$ is the spatial metric for the foliated hypersurfaces. $\alpha$ and $\beta^i$ are functions to be determined, $\gamma_{ij}$ is a positive-definite induced metric on each spacelike hypersurface. This spacetime is globally hyperbolic \cite{Alcubierre1994, Alcubierre2012, Gourgoulhon2012}, maintaining the causal structure. The lapse function $\alpha$ is defined as
\be
\alpha(t,x^i) = \frac{\dd t}{\dd\tau},
\label{lapsefunc}
\ee
where the proper time $\dd\tau$ is measured by \textit{Eulerian observers} moving along the normal direction to the hypersurfaces. In General Relativity, Eulerian observers are stationary observers at rest for a chosen spatial coordinate system. They provide a reference frame for describing spacetime and its properties. These observers are fixed at specific spatial coordinates. Their world lines are used to define a coordinate system for the spacetime. Alcubierre defined the Eulerian observers by defining their 4-velocity
\be
n^\alpha = \frac{1}{\alpha}(1,\beta^1,\beta^2,\beta^3).
\ee
Alcubierre assumed the following choice for those parameters
\begin{align}
\alpha &= 1, 
\label{alphawarp drivemetric}
\\ \beta^1 & = - v_s(t)f\left[r_s(t)\right]
\label{betawarp drivemetric}
, \\ \beta^2 &= \beta^3 = 0,
\label{betayzwarp drivemetric}
\\ h_{ij} &= \delta_{ij}.
\label{hwarp drivemetric}
\end{align}
So, the warp drive metric assumes the following form
\be
\dd s^2 = - \left[1 - v_s(t)^2 f(r_s)^2\right] \dd t^2 
- 2 v_s(t) f(r_s) \dd x \, \dd t + \dd x^2 + \dd y^2 
+ \dd z^2. 
\label{warp drivemetric1}
\ee
The function $v_s(t)$ is interpreted as the center of the bubble velocity within the curve $x_s(t)$, given by
\be
v_s(t) = \frac{\dd x_s(t)}{\dd t}.
\ee
$f(r_s)$ is the \gls{regulating form function} of the warp bubble, and $r_s(t)$ is the Euclidean distance to the curve $x_s(t)$ connecting the bubble's center to a generic spacetime point. 
\be
r_s(t) = \sqrt{\left[x - x_s(t)\right]^2 + y^2 + z^2}.
\label{radiusfunc}
\ee
An example of regulating function $f(r_s)$, Alcubierre gave 
\be
f(r_s) = \frac{\tanh\left[\sigma(r_s + R)\right] - \tanh\left[\sigma(r_s - R)\right]} {2 \tanh(\sigma R)},
\label{regfunction}
\ee
where $\sigma$ is inversely related to the warp bubble thickness, and $R$ is a parameter proportional to the radius of the warp bubble. The regulating function \index[topics]{regulating function} approaches a step function in the limit of a really thin bubble wall, $\sigma \to \infty$:
\be
\lim_{\sigma \to \infty} f(r_s) =
\begin{cases}
1, & \text{for} \ \ r_s  \in [-R,R] \\
0, & \text{otherwise}. 
\end{cases}
\label{stepfunc}
\ee

\begin{figure}[ht]
 \centering
 \includegraphics[scale=0.7]{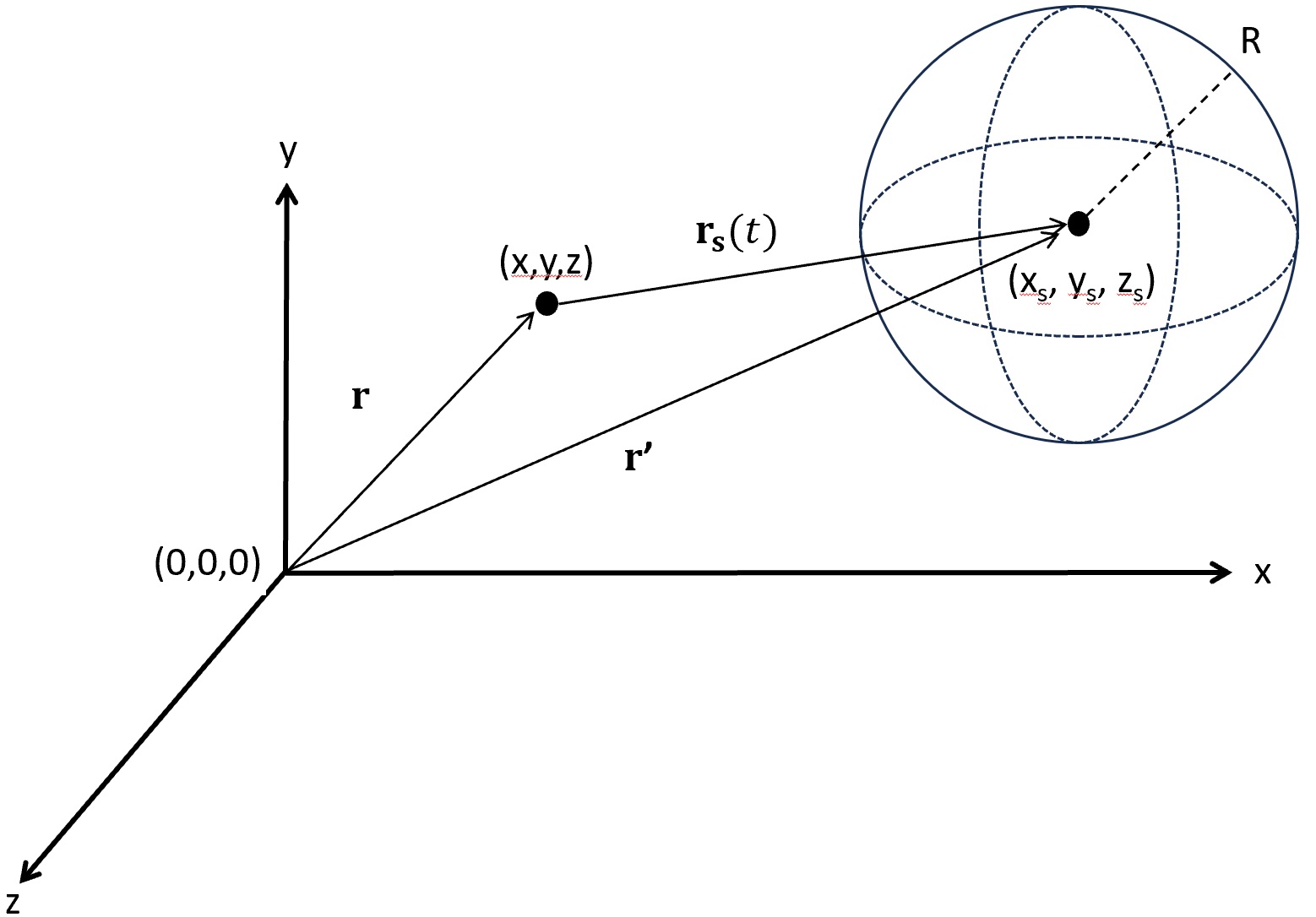}
 \caption[Schematics for the bubble center distance to a point in spacetime.]{Diagram to demonstrate the distance $r_s(t)$ to the point $x_s(t)$ at the center of the warp bubble. The bubble's center is $(x_s, y_s,z_s)$ in general, but Alcubierre chose the spatial axis with the x-axis in the direction of the bubble movement.}
 \label{warp_bubble_dist}
\end{figure}

Fig.\,\ref{warp_bubble_dist} shows a schematic representation of the warp bubble. 

Fig.\,\ref{warp_bubble_dist} does not to explicitly depict the curve $x_s(t)$ described by the center of the bubble. But you can imagine that its center describes a regular path in space with respect to inertial observers. To describe his warp drive, Alcubierre made the coordinate choice of $y_s = z_s = 0$, so the warp bubble's center changes coordinates only considering the direction of $x_s(t)$ coordinate. The vector sum $|\mathrm{r} - \mathrm{r'}|$ results in Eq.\,\eqref{radiusfunc}.

Alcubierre proposed a mechanism to propel a particle with mass remaining inside a local lightcone inside the warp bubble described in Fig.\,\ref{warp_bubble_dist}, in a flat spacetime, respecting the laws of special relativity and the light speed limits. Outside, observers far away from the bubble and from other gravitational interactions would also be in flat spacetime. 

The only curved region would be inside the walls of the warp bubble, which has a limited thickness. Eq.\,\eqref{stepfunc} describes that the limit of a thin bubble results in the step function. The form of the function $f[r_s(t)]$ in Eq.\,\eqref{regfunction} has a limiting behavior being approximated asymptotically to a step function: it varies continuously inside the warp bubble in the limit defined by the region $[-R,R]$ as shown in Eq.\,\eqref{stepfunc}, reaching the maximum value of 1 inside that region, and being equal to 0 outside.

Since the lapse function defined in Eq.\,\eqref{lapsefunc} is defined as $\alpha = 1$ by Alcubierre, it implies that there is no time dilation between proper time $\tau$ and coordinate time measurements $t$. This choice for the lapse function also guarantees that the metric determinant always has a Lorentzian signature, which makes the spacetime proposed by Alcubierre well-behaved. 

Alcubierre chose the Eulerian (free-fall) observers to analyze the physical properties of his warp drive. These Eulerian observers are always in free fall following timelike geodesics induced by the metric, preventing closed timelike curves that can break causal structure. The shift vector $\beta$ can boost the warp drive in the $x$-coordinate direction, also carrying information about the expansion and shape of the created warp bubble. For faraway observers, the spacetime is flat everywhere except on the walls of the warp bubble. This fact can be seen from the step function behavior approaching zero outside the bubble, making the warp drive metric asymptotically flat. 

The curvature of warp drive spacetime is contained in the extrinsic curvature tensor $K_{\mu \nu}$ for a hypersurface of constant time, as discussed above, see Ref.\,\cite{Alcubierre1994}
\be
K_{\mu \nu} = \frac{1}{2\alpha}\left(\beta_{\nu;\mu} 
+ \beta_{\mu;\nu} - \gamma_{\mu\nu,t}\right),
\label{extcurve1}
\ee
The volume expansion is calculated in terms of the trace of the extrinsic curvature
\be
\theta = - \text{Tr}\,K = K^\alpha_{\phantom{a} \alpha},
\ee
where $\text{Tr}$ is the trace operator of the curvature tensor $K$. The above formula is written in the coordinate-free language, which is general for any coordinate chart. For the warp drive, Alcubierre defined the shift vector as the product of the warp bubble center velocity and the shape function, $\beta = - v_s(t) f(r_s)$, so inputting this expression in the formula for the expansion volume leads to the following expression
\be
\theta = - v_s \frac{(x - x_s)}{r_s}\left(\frac{\partial f}{\partial r_s}\right). 
\ee
The above expression shows that the volume expansion of the bubble is directly proportional to the total partial derivative of the shape function for the Euclidean distance $r_s$. It is worth remembering that the shape function defined by Alcubierre in Eq.\,\eqref{regfunction} is just an \textit{ad-hoc} example, not a result of Einstein's equation solutions for the warp drive.

Eq.\,\eqref{extcurve1} shows that the extrinsic curvature for surface levels of constant time is inversely proportional to the lapse function $\alpha$. This means that the spacetime can be approximately flat depending on large values of $\alpha$. Still, since Alcubierre made a choice $\alpha = 1$, this is not a feature one should worry about for his metric, but it is clear that the hypersurfaces are Euclidean spaces because of $\gamma_{ij} = \delta^i_j$. The determinant of the Alcubierre warp drive is equal to $-1$, i.e., it has a Lorentzian signature, being globally hyperbolic, prohibiting closed timelike curves. For the 3+1 metric, the determinant equals $-\alpha^2$, so trying to construct different classes of warp drives with a non-homogeneous lapse function implies time dilation between measurements of the proper time for observers outside and far away from the bubble following the warp drive metric geodesics, and it may even lead to causality violation.

\subsection{Geodesics and lightcones}

Substitution of $x = x_s(t)$ in the warp drive metric in Eq.\,\eqref{warp drivemetric1} shows that a particle inside the bubble moves on a timelike curve regardless of the $v_s(t)$ value. Considering a test particle moving on the $x$ direction with a shift vector $\beta$, and regarding $\dd y = \dd z = 0$
\be
\dd s^2 = - \dd t^2 + (\dd x - \beta\dd t)^2,
\ee
considering $\dd s^2 = - \dd \tau^2$, where $\tau$ is an affine parameter, and remembering that the shift vector is $\beta = - v_s f(r_s)$, substituting $\dd x = \dd x_s = v_s \dd t$ in the warp drive metric  leads to
\be
- \dd \tau^2 = - \dd t^2 + (1 - f) v_s^2 \dd t
\ee
in the limiting condition of a very thin warp bubble wall, inside the warp bubble, the regulating function approaches a step function, and $f = 1$ inside the warp bubble, then 
\be
\dd s^2 = - \dd\tau^2 = - \dd t^2.
\ee
This demonstrates that mass particles inside the bubble are always in free fall on a timelike geodesic because $\alpha = 1$. Even though the coordinate acceleration can be arbitrary, the proper acceleration along the particle's path will always be zero. Using proper time $\tau$ as an affine parameter, calculating the Euler-Lagrange equations for the warp drive metric
\be
\mathcal{L} = \left(\frac{\dd s}{\dd \tau}\right)^2 = 
- \dot{t}^2 + \left(\dot{x} - \beta \dot{t}\right)^2,
\ee
where dots mean differentiation concerning the proper time $\tau$, then the Euler-Lagrange equations namelly
\be
\frac{\dd}{\dd\tau}
\left(\frac{\partial\mathcal{L}}{\partial \dot{x}^\mu}\right)
- \frac{\partial\mathcal{L}}{\partial x^\mu} = 0.
\ee
Become:
\begin{align}
&x: \ \frac{\dd}{\dd\tau}(\dot{x} - \beta\dot{t}) + \beta_x \dot{t} (\dot{x} - \beta \dot{t}) = 0,
\label{elgeod1}
\\
&t: \ \ddot{t} + \frac{\dd}{\dd\tau}\left[\beta(\dot{x}-\beta\dot{t})\right] + \beta_t \dot{t} (\dot{x}-\beta\dot{t}) = 0,
\label{elgeod2}
\end{align}
where the lower index on $\beta_x$, and $\beta_t$ indicates partial derivatives. It follows that
\be
\frac{\dd x}{\dd t} = \beta
\ee
is a solution to the geodesic equations. Since $\dot{t} = 1$, then one possible 4-velocity vector is 
\be
\dot{x}^\mu = (1, \beta, 0, 0).
\label{euleobsvel}
\ee
Eq.\,\eqref{euleobsvel} is the 4-velocity for the Eulerian observers, thus demonstrating that the proper acceleration of the mass particle inside the bubble is always zero, independently of $x_s(t)$, and $\beta$. This is readily seen from
\be
\dot{x}_\mu = g_{\mu\nu} \dot{x}^\nu = (-1,0,0,0),
\ee
which makes those observers timelike, since
\be
\dot{x}_\mu \dot{x}^\mu = - 1,
\ee
which immediately implies $\ddot{x} = 0$, demonstrating that proper acceleration is zero.

\begin{figure}[ht]
 \centering
 \includegraphics[scale=0.45]{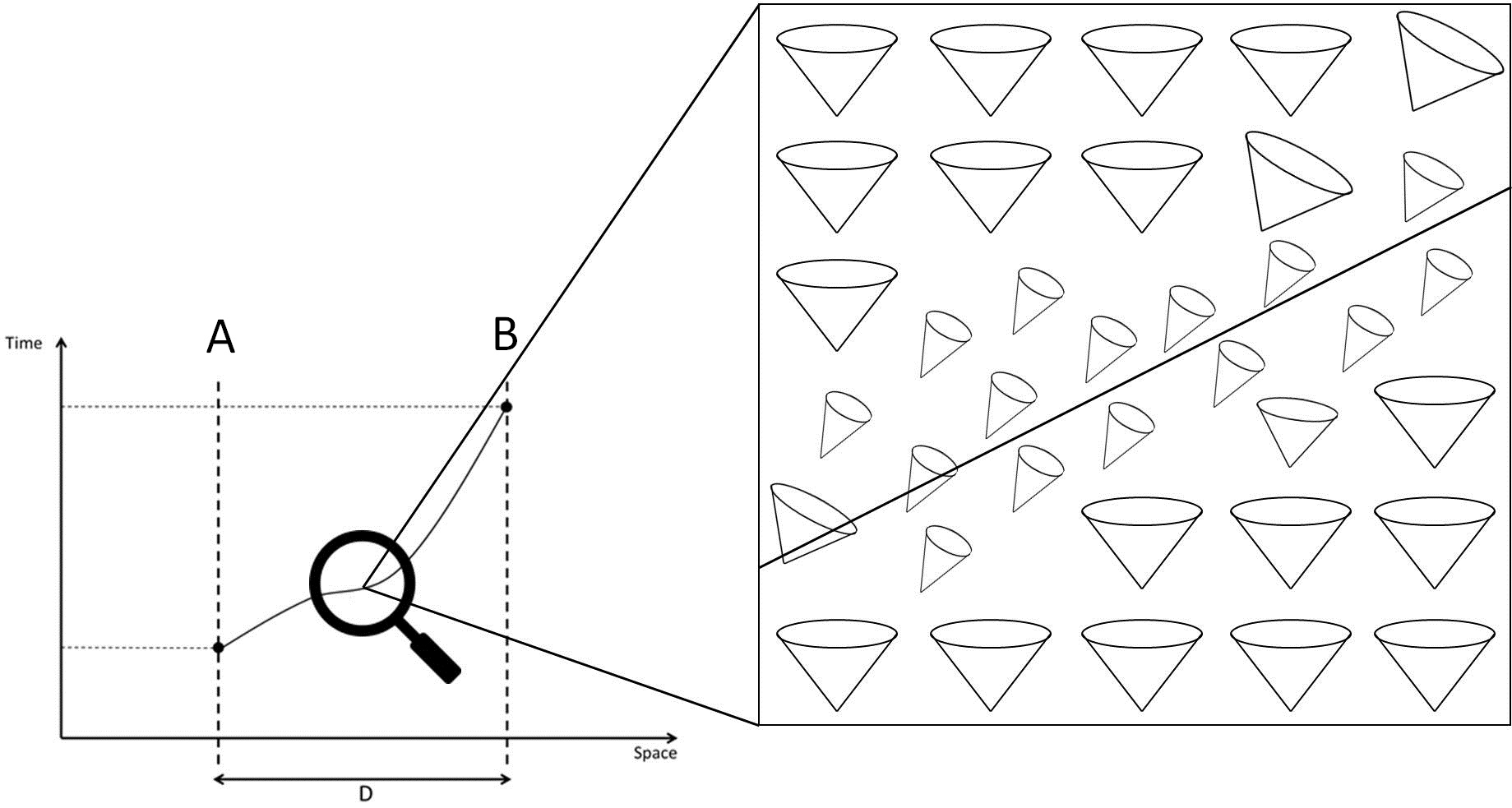}
 \caption[Light cones in warp drive spacetime.]{Light cones in warp drive spacetime being tilted by the gravitational interaction near the warp bubble exterior walls. The figure was based on Figure 7.2, on page 145, from Ref.\,\cite{Hartle2003}.}
 \label{lightcone_1}
\end{figure}

Another property of warp drive spacetime is that the warp bubble distorts spacetime around it, tilting the lightcones of observers that get near its walls. The null trajectories of a particle with zero mass are $45$ degree lines, i.e., \textit{null lightcones}, with
\be
\dd x = \pm \dd t,
\ee
where $\dd\tau$ is the proper time measured by the local observers and $\dd t$ is coordinate time increment. Considering the Alcubierre warp drive metric
\be
\dd s^2 = - \dd t^2 + (\dd x - \beta\dd t)^2 + \dd y^2 +  \dd z^2,
\ee
without loss of generality $\dd y = \dd z = 0$, and setting $\dd s^2 = 0$ for null trajectories, leads to
\be
\frac{\dd x}{\dd t} =\beta \pm 1,
\label{fastlight}
\ee
when $\beta = 0$, the $\pm 1$ corresponds to the usual null rays with $45^\circ$ angles forming the lightcone. When $\beta \neq 0$, it demonstrates that the lightcones are tilted near the warp bubble wall. On the interior of the bubble $f(r_s) = 1$, so the lightcones inside the warp bubble are tilted with angles less than $45^\circ$, but outside $f(r_s) = 0$ and the light rays are the usual $45^\circ$ lines in flat spacetime. The trajectory of the mass particle inside the warp bubble is always a timelike curve for any value of the center speed $v_s(t)$. Fig.\,\eqref{lightcone_1} shows how lightcones behave near the warp bubble.

This tilting of lightcones near the warp bubble walls, either outside or inside the bubble, creates strange behaviors, such as horizon events. It is possible to find points inside the warp bubble where photons never reach the bubble walls, making it outside the lightcones of the observer inside the bubble. The following section will discuss this in detail as warp drive caveats.

\subsection{Warp drive caveats}

Alcubierre constructed his warp drive metric \cite{Alcubierre1994} very carefully, choosing the parameters for the ADM general metric with care, trying to build a globally hyperbolic spacetime with Lorentzian signature. He wanted to avoid closed, timelike curves and the breaking of causal structures. He proposed a thought experiment describing how a spaceship would be carried away inside a warped bubble in a superluminal fashion. He demonstrated that the warp bubble acceleration is limitless. He also created a mechanism using his metric so that the spaceship inside the bubble would remain inside its local lightcone, respecting the light speed limit and global causality. The only issue explicitly discussed by Alcubierre was the apparent violation of the weak energy condition by his warp drive metric. He concluded by inspecting the Einstein tensor components when contracted with the 4-velocity from a class of observers in free fall. Considering Einstein's equations, Alcubierre concluded that any energy-momentum tensor would always violate the weak energy condition because the expression found was always non-positive. This is true when considering the shift vector a real-valued function or disregarding vacuum solutions. However, this thesis demonstrates that finding Einstein equation solutions with a source and respecting the energy conditions is possible.

Saying again, with the choice of lapse function $\alpha = 1$ and using ADM formalism with foliated spacetime and Cauchy surfaces, Alcubierre guaranteed that the warp drive spacetime is globally hyperbolic, not allowing closed timelike curves and ensuring that causal structure is respected. The observers following the warp drive geodesics, named Eulerian observers, are always in free fall; thus, in flat spacetime, they face no gravitational interaction from the warp bubble. Using a shift vector with a non-zero component in only one spatial direction simulated the one-direction motion of a spaceship commonly displayed in sci-fi movies. This choice for the shift vector also simplified the physics in the proposed warp drive since all the information regarding the kinematics and dynamics of the warp bubble could be encapsulated in only one parameter. The form of this shift vector component $\beta[f(r_s(t))]$ in Eq.\,\eqref{betawarp drivemetric}, with a center velocity of the bubble $v_s(t)$ multiplied by a regulating form function $f(r_s)$ with an asymptotic behavior to a step function in the limit of very thin warp bubble walls and a Euclidean distance $r_s(t)$ connecting the center of the bubble to generic points in the spacetime made sure that there is no time dilation between observers being carried away inside the bubble and outside observers in free fall, in flat spacetime. This mechanism ensures that this outside observer would perceive the warp bubble reaching superluminal speed without the interference of having to deal with the Lorentz factor and time dilation issues when analyzing events. 

Since the warp drive theory was published in 1994, several caveats have been found for general warp drive classes over the past three decades. These caveats are mentioned in this section: the \gls{eventhorizon} behavior inside the warp bubble out of its center. \gls{closedtimelikecurve} with slight modifications in the Alcubierre warp drive metric.

\subsubsection{Negative energy density}

This section addresses the possible necessity of negative energy densities for the feasibility of the warp drive. Alcubierre \cite{Alcubierre1994} defined a specific metric for his warp drive via the metric
\be
\dd s^2 = - \dd t^2 + (\dd x + \beta \dd t)^2 + \dd y^2 + \dd z^2.
\ee
He then proceeded to analyze the energy density for his warp drive spacetime as it is measured by a specific class of observers who are in free fall. For this analysis, Alcubierre used the Eulerian observers with 4-velocities given by the following expression
\be
u_\mu = (-1,0,0,0) \ \ , \ \ u^\mu = (1,-\beta,0,0).
\ee
It must be noted that the 4-velocity defined by the above equation is timelike $u_\nu u^\nu = - 1$. The energy density, as measured by those Eulerian observers, is given by 
\be
G^{\mu\nu} u_\mu u_\nu = G^{00} =  -\frac{1}{4} \left[\left( \frac{\partial \beta}{\partial y} \right)^2 + \left( \frac{\partial \beta}{\partial z} \right)^2 \right] \leq 0,
\label{negeng}
\ee
and the above expression equals the component $G^{00}$ of Einstein tensor, so Alcubierre concluded considering Einstein equations $G^{\mu\nu} = 8 \pi T^{\mu\nu}$ that the term $T^{00}$ regarded as the energy density $\rho$,
\be
\rho = T^{\mu\nu}u_\mu u_\nu = T^{00}.
\label{alcenergdens}
\ee
It must be noted that the energy density $\rho$ shown in Eq.\,\eqref{alcenergdens} is a definition made by Alcubierre in his 1994 article. The fact that $\rho = T^{00}$ is not a general result, since the actual definition that Alcubierre made was the contraction of $G^{\mu\nu}/8\pi$ by the covariant velocities $u_\mu$ of the Eulerian observers. If one wants to use the 4-velocity in its contravariant form $u^\mu$, it is necessary to contract $G_{\mu\nu} u^\mu u^\nu$, and since this is a scalar, the results must be the same. However, notice that $T_{00}$ is not the energy density.

The energy-momentum tensor is always negative, and because of that, the warp drive violates the weak and dominant energy conditions. Another way to derive Eq.\,\eqref{negeng} is to consider the Einstein equations in covariant form
\be
T_{\alpha\beta} u^\alpha u^\beta = G_{\alpha\beta} u^\alpha u^\beta
= G_{00} - 2\beta G_{01} + \beta^2 G_{11},
\label{negeng2}
\ee
leading to the same result in Eq.\,\eqref{negeng}, but the expression found in Eq.\,\eqref{negeng2} is always equal to zero for the warp drive metric found by Alcubierre, and the equation below is of great importance in finding solutions to the Einstein equations for the warp drive
\be
G_{00} - 2\beta G_{01} + \beta^2 G_{11} = 0.
\ee
When Einstein's equations are considered, this equation leads to state equations connecting the geometric aspects to matter-energy. The conclusion that warp drive always violates energy conditions must be analyzed within boundary conditions since vacuum solutions imply that Eq.\,\eqref{negeng} is zero, satisfying the energy conditions immediately. 

\subsubsection{Event horizon inside the warp bubble}

Discussing the warp drive spacetime, it is shown that lightcones, when close to the warp bubble, are tilted in angles less than $45^\circ$. This tilting of lightcones generates a horizon problem with a coordinate singularity. This was first pointed out in Ref.\,\cite{AlcubierreLobo2017} by Alcubierre and Lobo. Considering the warp drive metric 
\be
\dd s^2 = - \dd t^2 + (\dd x - \beta \dd t)^2 + \dd y^2 + \dd z^2,
\ee
which is not diagonal since it has cross terms $\dd t \, \dd x$, but can be diagonalized using the following change of coordinates, which is also a tetrad
\be
\dd\tau = \dd t - A \dd x, 
\label{coordtransf}
\ee
where $A$ is some function to be determined by imposing the metric diagonalization\index[topics]{metric diagonalization}. Inserting this coordinate change from Eq.\,\eqref{coordtransf} in the warp drive metric leads to 
\be
\dd s^2 = - (1 - \beta^2)\dd\tau^2 + 2(A - \beta -  A\beta^2)\dd x \dd\tau + (1 - A^2 + A^2\beta^2 + 2 \beta A) \dd x^2 + \dd y^2 + \dd z^2,
\ee
Imposing the metric to be diagonal, where the cross term $g_{x\tau}$ component is zero  
\be
A - \beta -  A\beta^2 = 0,
\ee
results in the following expression for the function $A$ in terms of $\beta$
\be
A = \frac{\beta}{1 - \beta^2}.
\ee
With this function, and the proposed change of coordinates from Eq.\,\eqref{coordtransf}, the warp drive metric is then diagonalized, taking the following form:
\be
\dd s^2 = - (1 - \beta^2) \dd\tau^2 + \left(\frac{1}{1 - \beta^2}\right) \dd x^2 + \dd y^2 + \dd z^2 .
\label{diagwarp drive}
\ee
 The determinant of the diagonalized warp drive metric in the above equation is given by
 \be
 \det(g_{\mu\nu}) = - 1,
 \ee
maintaining the Lorentzian signature of the original warp drive metric, but now the induced spatial metric $\gamma_{ij}$ is no longer extrinsically flat, and its determinant $\gamma$ is given by,
\be
\gamma = \text{det}(\gamma_{ij}) = \frac{1}{1 - \beta^2},
\ee
so if values of the shift vector are given by 
\be
\beta = \pm 1,
\ee 
then the induced spatial metric $\gamma_{ij}$ singular at such points. Noting that those are the exact values of light speed in different directions. Those points are also coordinate singularities for the diagonalized warp drive metric. Imposing the conditions 
\be
\dd y = \dd z = 0,
\ee
disregarding irrelevant coordinates from the warp drive metric and considering
\be
\dd s = 0
\ee
for the line element in Eq.\eqref{diagwarp drive} to calculate null trajectories of a light particle emitted by an observer inside the warp bubble, this leads to the following expression for the $x$-coordinate with respect to the new time coordinate $\tau$
\be
\frac{\dd x}{\dd \tau} = \pm (1 - \beta^2).
\label{evehor1}
\ee
Eq.\,\eqref{evehor1} tells us that for points $x_0$, inside the warp bubble, where the shift vector assumes values $\beta = \pm 1$, if an observer in this inertial rest frame emits photons, then it is possible for those photons to be hovering in those interior points near the warp bubble walls. Any photons emitted by observers inside the warp bubble will not reach the bubble wall, lying outside the forward and backward lightcone of the observer inside the warp bubble. This strange behavior is similar to a black hole's event horizon. See Ref.\,\cite{dInverno2022} for an introduction to black holes \index[topics]{black hole}. To make this argument clearer, remember the definition given by Alcubierre for the shift vector as the product of the bubble speed and the regulating function \index[topics]{regulating function}
\be
\beta = - v_s(t) \, f[r_s(t)].
\ee
The regulating function is a smooth, monotonically increasing function on the bubble's interior, ranging from values in the interval $[0,1]$, equal to 0 in the bubble's center, and reaching values closer to 1 as calculated near the warp bubble walls. Substituting the expression for the shift vector in Eq.\,\eqref{evehor1} leads to the following expression
\be
(1 - v_s f)(1 + v_s f) = 0,
\ee
so in the points $x_0 = r_s(t = t_0)$, the form function $f$ assume  values where
\be
f(x_0) = \pm \frac{1}{v_s},
\ee
presenting event horizon behavior \index[topics]{event horizon behavior}. This important result needs further investigation, and Alcubierre and Lobo first pointed out this horizon behavior by the warp bubble Ref.\,\cite{AlcubierreLobo2017}. It is well known that event horizons can cause inversion of light cones \cite{dInverno2022}. If the lightcones are somewhat reversed, this could imply the possibility of breaking the causal structure as observers transverse the warp bubble walls. Demonstrating this hover frame for photons is an easy task. Considering the line element as an affine parameter defined by
\be
\dd s^2 = - \dd\sigma^2, 
\ee
where the affine parameter $\sigma$ can be taken as the proper time of a co-moving observer, and neglecting irrelevant coordinates $\dd y = \dd z = 0$, from Eq.\,\eqref{diagwarp drive} it follows
\be
\frac{\dd x}{\dd \sigma} = \frac{1 - \beta^2}{\beta} \left(\frac{\dd t }{\dd \sigma} - 1\right) \,,
\ee
since the lapse of proper time is homogeneous, that is 
\be
\alpha = \frac{\dd \sigma}{\dd t} = 1, 
\ee
leads to the result where, at the bubble surface it is true that
\be
\frac{\dd x}{\dd \sigma} = \frac{1 - \beta^2}{\beta} = 0.
\label{evehor2}
\ee

Eq.\,\eqref{evehor1} showed a hover frame \index[topics]{hover frame} for photons inside the warp bubble. Eq.\,\eqref{evehor2} shows that mass particles inside the warp bubble can also hover upon event horizons inside their lightcones $(\beta = 1)$. Interestingly, the shift vector can not be zero, so near the event horizon inside the warp bubble, mass particles perceive a warped region of spacetime. The imposition of a homogeneous lapse function by Alcubierre \cite{Alcubierre1994}, even though it makes the warp drive spacetime Lorentzian and well-behaved regarding the existence of no closed timeline curves, can give rise to event horizons' behavior inside the warp bubble.

\section{Approaching the warp drive with different matter-energy sources}

This chapter presents the methodology of this work, with the fundamental equations found inside Einstein's equations when coupling the warp drive spacetime with different energy-matter sources: dust, perfect fluid, anisotropic fluid, charged dust, and the aid of the cosmological constant \cite{nos1,nos2,nos3,nos4,nos5} \index[topics]{cosmological constant}. The warp drive metric is given by
\be
\dd s^2 = - \left(1 - \beta^2\right)\dd t^2 - 2 \beta \dd x \dd t + \dd x^2 + \dd y^2 + \dd z^2.
\label{alcwarp drivemetric}
\ee
The first section discusses the Einstein tensor components and how the algebraic manipulation leads to important relations that connect the warp drive theory to shock waves via a Burgers-type equation. Relevant equations used to solve Einstein's equations are presented, including a significant relation that connects the geometrical aspects of the warp drive metric, the lapse functions, and the shift vector to matter and energy content, resulting in state equations. Section two discusses the curvature of warp drive geometry, demonstrating that it is not Ricci flat. Still, it is conditionally flat in the region where the Burgers-type equation is solved regarding the shift vector as a dependent variable, and when the shift vector assumes the value of light speed. The third section presents fundamentals about fluids in curved spacetime with different physical aspects such as heat, energy, and momentum flow. The fourth section presents relevant equations connecting the charged dust source to the warp drive geometry, describing all the components of the energy-momentum tensor. 

\subsection{Solving Einstein equations}

The Einstein tensor with the cosmological constant \index[topics]{cosmological constant} $\Lambda$ for the metric in Eq.\,\eqref{alcwarp drivemetric} gives
\be
G_{00} =  \Lambda(1-\beta^2) - \frac{1}{4}(1 + 3\beta^2)\left[\left(\frac{\partial \beta}{\partial y} \right)^2 + \left(\frac{\partial \beta}{\partial z} \right)^2 \right] - \beta \left(\frac{\partial^2 \beta}{\partial y^2} + \frac{\partial^2 \beta}{\partial z^2}\right),
\label{et00}
\ee
\be
G_{01} =  \Lambda \beta + \frac{3}{4} \beta \left[\left(\frac{\partial \beta}{\partial y}\right)^2 + \left(\frac{\partial \beta}{\partial }\right)^2 \right] + \frac{1}{2}\left(\frac{\partial^2 \beta}{\partial y^2} + \frac{\partial^2 \beta}{\partial z^2}\right),
\label{et01}
\ee
\be
G_{02} = - \frac{1}{2} \frac{\partial^2 \beta}{\partial x \partial y} - \frac{\beta}{2} \left(2\frac{\partial \beta}{\partial y} \, \frac{\partial \beta}{\partial x} + \beta \frac{\partial^2 \beta}{\partial x \partial y} + \frac{\partial^2 \beta}{\partial t \partial y}\right),
\label{et02}
\ee
\be
G_{03} = - \frac{1}{2} \frac{\partial^2 \beta}{\partial x \partial z} 
- \frac{\beta}{2} \left(2\frac{\partial \beta}{\partial z} \, \frac{\partial \beta}{\partial x} + \beta \frac{\partial^2 \beta}{\partial x \partial z} + \frac{\partial^2 \beta}{\partial t \partial z}\right),
\label{et03}
\ee
\be
G_{11} = \Lambda - \frac{3}{4} \left[\left(\frac{\partial \beta}{\partial y}\right)^2 + \left(\frac{\partial \beta}{\partial z}\right)^2 \right], 
\label{et11}
\ee
\be
G_{12} = \frac{1}{2}\left(2 \frac{\partial \beta}{\partial y} \, \frac{\partial \beta}{\partial x} + \beta \frac{\partial^2 \beta}{\partial x \partial y} + \frac{\partial^2 \beta}{\partial t \partial y}\right),
\label{et12}
\ee
\be
G_{13} = \frac{1}{2}\left(2 \frac{\partial \beta}{\partial z} \, 
\frac{\partial \beta}{\partial x} + \beta \frac{\partial^2 \beta}{\partial x \partial z} + \frac{\partial^2 \beta}{\partial t \partial z}\right),
\label{et13}
\ee
\be
G_{23} = \frac{1}{2} \frac{\partial \beta}{\partial z} \, \frac{\partial \beta}{\partial y},
\label{et23}
\ee
\be
G_{22} = - \Lambda - \frac{\partial}{\partial x}\left[\frac{\partial \beta}{\partial t} + \frac{1}{2} \frac{\partial}{\partial x} (\beta^2)
\right] - \frac{1}{4}\left[\left(\frac{\partial \beta}{\partial y}\right)^2 - \left(\frac{\partial \beta}{\partial z}\right)^2 \right],
\label{et22}
\ee
\be
G_{33} = - \Lambda - \frac{\partial}{\partial x}\left[\frac{\partial \beta}{\partial t}+ \frac{1}{2} \frac{\partial}{\partial x} (\beta^2)\right]+ \frac{1}{4}\left[\left(\frac{\partial \beta}{\partial y}\right)^2 - \left(\frac{\partial \beta}{\partial z}\right)^2\right].
\label{et33}
\ee
After algebraic manipulation, the following set of relevant equations is obtained
\be
G_{33} + G_{22} = 2 \Lambda + \frac{\partial}{\partial x}\left[\frac{\partial \beta}{\partial t} + \frac{1}{2} \frac{\partial}{\partial x} (\beta^2) \right].
\label{burgerseq}    
\ee
\be
G_{00} + 2 \beta G_{01} + \frac{1}{3} \left(3\beta^2 - 1\right) G_{11} = 0,
\label{expet1}
\ee
\be
G_{33} - G_{22} = \frac{1}{2}\left[\left(\frac{\partial \beta}{\partial y}\right)^2- \left(\frac{\partial \beta}{\partial z}\right)^2\right],
\label{gradpress}
\ee 
Eq.\,\eqref{burgerseq} implies an intrinsic relation of warp drive spacetime and shock waves\index[topics]{shock waves} via the Burgers-type equation. This interesting equation shows that the sum of two static pressures\index[topics]{static pressure} from the energy-momentum tensor can result in shock-wave solutions. Stable solutions to warp bubble formation may depend on how the off-diagonal terms from the energy-momentum tensor contribute to the geometric part of the warp drive. Considering Einstein's equation $G_{\mu\nu} = 8 \pi T_{\mu\nu}$ in Eq.\,\eqref{expet1} helps to find important relations between matter density, static pressures\index[topics]{static pressure}, flows of momentum\index[topics]{flow of momentum}, and the shift vector. From expression 
\be
T_{00} + 2 \beta T_{01} + \frac{1}{3} \left(3\beta^2 - 1\right) T_{11} = 0,
\label{eqstaterel} 
\ee
it is possible to find equations of state \index[topics]{equation of state} relating pressure $p$ and matter density $\mu$
\be
p = \omega \, \mu,
\ee
where $\omega$ may depend on the geometrical aspects of spacetime, such as the shift vector or a non-homogeneous lapse function, which are also functions of spacetime coordinates.

\subsection{Warp Drive curvature}

Considering the ansatz $\beta = \beta(t,x)$, then non-zero Riemann tensor components are
\be
\mathrm{R}_{\phantom{t}t t x}^{t\phantom{t}\phantom{t}\phantom{x}} =  \beta \, \mathrm{R}_{\phantom{t}x t }^{t\phantom{t}\phantom{t}\phantom{x}} = \beta \, \frac{\partial}{\partial x}\left[
\frac{\partial \beta}{\partial t}
+ \frac{1}{2} \frac{\partial(\beta^2)}{\partial x}\right],
\label{curv1}
\ee
and the only non-zero Ricci tensor components are 
\be
R_{00} = \left(\beta^2 - 1\right)\frac{\partial}{\partial x}\left[\frac{\partial\beta}{\partial t} + \frac{1}{2}\frac{\partial}{\partial x}\left(\beta^2\right)\right],
\label{curv2}
\ee
\be
R_{11} = \frac{\partial}{\partial x}\left[\frac{\partial\beta}{\partial t} + \frac{1}{2}\frac{\partial}{\partial x}\left(\beta^2\right)\right].
\label{curv3}
\ee
Both Ricci components can be summarized as
\be
R_{00} = (\beta^2 - 1) R_{11} \,.
\label{curv4}
\ee
So, this spacetime is flat when $\beta = \pm 1$ from Eq.\,\eqref{curv4}, but it is in a luminal regime where the shift vector assumes lightspeed. Nevertheless, the metric is not diagonal when $\beta = 0$, the identification with Minkowski is direct from the warp drive metric in Eq.\,\eqref{alcwarp drivemetric}, which means there is no warp bubble. However, when $\beta = \beta(t,x)$  and when this shift vector satisfies the Burgers-type equation from equations \eqref{curv1}, \eqref{curv2}, and \eqref{curv3}, part of this spacetime becomes flat. This result was the one we found in Refs.\,\cite{nos1,nos2,nos4,nos5} connecting a possible vacuum solution for the Einstein equations and shock waves. This result means the warp drive spacetime is connected to the Minkowski through the warp bubble walls. It is possible to demonstrate that this condition for local flatness is also the same for continuous jump conditions between a warp drive spacetime and Minkowski.

\subsection{Fluid sources}

Considering a general energy-momentum tensor, one can recover familiar sources: dust, perfect fluid, and a special type of anisotropic fluid. In Chap.\,IX, section 19 of Ref.\,\cite{choquet2008general}, the energy-momentum tensor of a viscous dissipative fluid is defined as
\be
 T^{\alpha \beta} = \mu u^\alpha u^\beta + p h^{\alpha \beta} + u^\alpha q^\beta + u^\beta q^\alpha  + \pi^{\alpha \beta},
\label{anisoemt}
\ee
where $h_{\alpha \beta}$ projects tensors onto hypersurfaces orthogonal to $u^\alpha$, being defined as E.\,\eqref{admhype}
\be
h_{\alpha \beta} = g_{\alpha \beta} + u_a\,u_b,
\ee
$\mu$ is the matter density, $p$ is the fluid static pressure, $q^\alpha$ is the heat flux vector and $\pi^{\alpha \beta}$ is the viscous shear tensor. The world lines of the fluid elements are the integral curves of the four-velocity vector $u^\alpha$. The heat flux vector and the viscous shear tensor are transverse to the world lines, so $q_a\,u^a = 0$, and $\pi_{ab}\,u^b = 0$. 

To define the energy-momentum tensor of the anisotropic fluid, I used an ansatz so that it would reduce to the perfect fluid. I defined the folowing tensor in matrix form:
\be
T_{\alpha \sigma} = 
\begin{pmatrix} 
\mu + \beta^2 p  & \quad \!\!\! - \beta D & \quad 0  & \quad 0  \\ 
- \beta D        & \:\: A         & \quad 0  & \quad 0 \\ 
0                & \:\: 0         & \quad B  & \quad 0 \\ 
0                & \:\: 0         & \quad 0  & \quad C 
\end{pmatrix},
\label{emtanisofluid}
\ee
here $p$, $A$, $B$, and $C$ are pressures, $D$ is the momentum density parameter representing this fluid's heat flux, and $\mu$ is the matter density. Making all pressures $(A, B, C)$ and momentum density $(D)$ equal to the pressure $p$ recovers the perfect fluid for the warp drive metric, and the energy-momentum tensor for a perfect fluid is given by
\be
T_{\alpha \beta} = \left(\mu + p\right) \, u_\alpha u_\beta
+ p \, g_{\alpha \beta}.
\ee
In matrix form
\be
T_{\alpha\beta} = 
\begin{pmatrix} 
\mu + \beta^2 p & - \beta p   & 0 & 0  \\ 
- \beta p         & p         & 0 & 0  \\ 
0                 & 0         & p & 0  \\ 
0                 & 0         & 0 & p 
\end{pmatrix}.
\ee
The perfect fluid solutions have equations of state given by Ref.\,\cite{EllisElst}, 
\be
p = p(\mu) = (\gamma - 1)\mu\,,
\ee
\be
\dot{\gamma} = \frac{\dd\gamma}{\dd t} = 0. 
\ee
Ordinary fluids can be approximated with 
\be
1 \leq \gamma \leq 2, 
\ee
where incoherent matter corresponds to $\gamma = 1$, and radiation to $\gamma = \frac{4}{3}$.

The same argument can be made to recover the energy-momentum tensor for dust, assuming the static fluid pressure $p$ to be zero, then 
\be
T_{\alpha \beta} = \mu \, u_{\alpha} u_{\beta},
\ee
In matrix form
\be
T_{\alpha\beta} = 
\begin{pmatrix} 
\mu  & 0 & 0 & 0  \\ 
0    & 0 & 0 & 0  \\ 
0    & 0 & 0 & 0  \\ 
0    & 0 & 0 & 0 
\end{pmatrix}.
\ee
where $\mu$ is the matter density represented by a scalar 
function of the spacetime coordinates, and $u^\alpha$ are 
the observer's 4-velocity components.

\subsection{Charged dust}

In Ref.\,\cite{nos3}, a charged dust was coupled as a source to the warp drive geometry. Due to the complexity of obtaining an analytical and general solution to the Einstein equations, a straightforward configuration was proposed, where the magnetic and electric fields are
\be\mathbf {B} = (0, B_2, B_3),
\label{magfield}
\ee
\be
\mathbf {E} = (E_1,0,0). 
\label{elecfield}
\ee
The energy conditions and the null imposition of the divergence of the energy-momentum were analyzed. A sketch of a possible solution was laid out, and several conditions related to the components of the electromagnetic fields, the shift vector, and the matter density were found. The main result was that, depending on the strength of the electromagnetic field, it was possible to obtain apparent negative energy density with positive matter density. Considering the 4-velocity of the Eulerian observers with 4-velocity 
\be
u^\nu = (1,\beta,0,0),
\ee
where $g^{\alpha\beta}$ is the warp drive metric. Then, the relevant components for the electromagnetic energy-momentum tensor given by Eq.\,\eqref{eemt} are given by the following expressions
\begin{align}
\nonumber T_{00} &= \mu + \frac{1}{2}(B^2 + E^2) + \left(B_3 E_2 - B_2 E_3\right)\beta + \frac{\beta^2}{2}\left(B^2 - 2B_1^2 - E^2 - E_3^2\right) \\
&+ \left(B_2 E_3 - B_3 E_2\right) \beta^3 + \frac{\beta^4}{2}\left(E_2^2 + E_3^2\right),
\label{eemt00}
\end{align}
\be
T_{01} = B_2 E_3 - B_3 E_2 + \frac{\beta}{2}\left(2B_1^2 - B^2 + E^2\right) + \left(B_3 E_2 - B_2 E_3\right) \beta^2 -\frac{\beta^3}{2}\left(E_2^2 + E_3^2\right),
\label{eemt01}
\ee
\be
T_{02} = B_1 E_3 \beta^2 + B_1 B_2 \beta + B_3 E_1 - B_1 E_3,
\label{eemt02}
\ee
\be
T_{03} = - B_1 E_2 \beta^2 + B_1 B_3 \beta - B_2 E_1 + B_1 E_2, 
\label{eemt03}
\ee
\be
T_{11} = \frac{\beta^2}{2} \left(E_2^2 + E_3^2\right) + \frac{1}{2}\left(B^2 - 2B_1^2 + E^2 - 2E_1^2\right) - \left(B_3 E_2 - B_2 E_3\right) \beta, 
\label{eemt11}
\ee
\be
T_{12} = -B_1 E_3 \beta - B_1 B_2 - E_1 E_2, 
\label{eemt12}
\ee
\be
T_{13} = B_1 E_2 \beta - B_1 B_3 - E_1 E_3, 
\label{eemt13}
\ee
\be
T_{22} = \frac{\beta^2}{2}\left(E_2^2 - E_3^2\right) - \left(B_3 E_2 + B_2 E_3\right) \beta + \frac{1}{2} \left(B^2 - 2B_2^2 + E^2 - 2E_2^2\right), 
\label{eemt22} 
\ee
\be
T_{23} = E_2 E_3 \beta^2 + \left(B_2 E_2 - B_3 E_3\right) \beta - (B_2 B_3 + E_2 E_3), 
\label{eemt23} 
\ee
\be
T_{33} = -\frac{1}{2}\left(E_2^2 - E_3^2\right) \beta^2 + \left(B_3 E_2 + B_2 E_3\right) \beta + \frac{1}{2}\left(B^2 - 2B_3^2 + E^2- 2E_3^2\right),
\label{eemt33}
\ee
where $E_1$ is the only component of the electric field given by Eq.\,\eqref{elecfield}, $B_2$ and $B_3$ are the components of the magnetic field given by Eq.\,\eqref{magfield}. When considering the asymptotical behavior of the shift vector $\beta \to 0$, the warp drive metric becomes asymptotically flat, and the energy-momentum tensor components become the usual components when considering Minkowski spacetime
\be
T_{\alpha \nu} =
\begin{pmatrix}
\mu + \frac{1}{2}(E^2 + B^2) & -S_1 & -S_2 & -S_3 \\
-S_1 & -\sigma_{11} & -\sigma_{12} & -\sigma_{13} \\
-S_2 & -\sigma_{21} & -\sigma_{22} & -\sigma_{23} \\
-S_3 & -\sigma_{31} & -\sigma_{32} & -\sigma_{33}
\end{pmatrix}.
\ee
where $S_1$, $S_2$ and $S_3$ are the vector components of the Poynting vector 
\be
\mathbf{S} =  \mathbf{E} \times \mathbf{B},
\ee 
and $\sigma_{ij}$ are the nine components of the Maxwell stress tensor
\be
\sigma_{ij} = E_i E_j + B_i B_j - \frac{1}{2}\left(E^2 + B^2\right)
\delta_{ij}.
\ee
It must be noted that we are using natural units, where the speed of light is equal to unity. Therefore, vacuum permittivity and vacuum permeability are both equal to one.

\section{Dust} 

This chapter demonstrates that coupling dust with the warp drive generates vacuum solutions to Einstein's equations, connecting the warp drive leads to shock waves via a Burgers-type equation. This means that dust is not a sufficient source of energy-momentum to create a stable warp bubble. The warp drive has a local isomorphism with Minkowski since it is asymptotically flat as constructed originally by Alcubierre \cite{Alcubierre1994}, the form function behaves as a step function in the limit when the walls of the warp bubble are really thin, which means that outside the bubble the spacetime is flat. The original warp drive spacetime is not globally flat since the Ricci and Riemann tensors only vanish for certain conditions where the vacuum solutions are satisfied. 

This chapter demonstrates that a particular vacuum solution results in zero curvature in a specific region of spacetime. The vacuum solution suggests that it is an interior solution to a junction of warp drive geometry and possible exterior global spacetime. The energy conditions are trivially satisfied for this solution since it is a vacuum. The main results of this chapter are the fundamental relation between warp drive and shock waves and a possible vacuum solution in a warped region. 

The first section describes how the vacuum solutions are found by coupling the dust with the warp drive. Dust is not a valid source for creating stable solutions with matter content. A Burgers-type equation is presented as a solution to Einstein equations. Section two discusses the curvature of the warp drive spacetime. Regarding the vacuum solution found as a connecting region between the interior of the warp bubble, a warped region with a limited thickness defined by the bubble's wall, it is shown that this spacetime is not Ricci flat but conditionally flat. The third section discusses the results of this chapter. This chapter presents two tables that summarize objectively the results from Einstein equations solutions and the trivial conditions where energy conditions are satisfied.

\subsection{Vacuum solutions for the Einstein equations}

This section presents the vacuum solutions found for the warp drive geometry. Departing from the following linear combination of Einstein tensor components, Eq.\eqref{eqstaterel} gives an important expression that helps to discover equations of state, connecting the warp drive parameters, and the shift vector to matter-energy source parameters 
\be
G_{00} + 2\beta G_{01} + \frac{1}{3}\left(3\beta^2 - 1\right) G_{11} = 0.
\ee
Considering the Einstein equations $G_{\mu\nu} = 8\pi T_{\mu\nu}$ in the above equation, and using the results from Eq.\eqref{eqstaterel} implies that the matter density for the dust is zero, 
\be
\mu = 0.
\ee 
This is a remarkable result since it states that the warp drive needs a more complex source of matter, energy, and momentum to create a stable warp bubble. Regardless, it is still possible to find a warp drive due to geometry only. The results from Einstein's equations are shown below
\be
G_{23} = 8\pi T_{23} = 0,
\ee
lead to the following conditions regarding shift vector partial derivatives
\be 
\frac{\partial \beta}{\partial z} = 0 = \frac{\partial \beta}{\partial y},
\label{dustsol}
\ee
which implies that the shift vector is found to be only a function of $(t,x)$ spacetime coordinates,
\be
\beta = \beta(t,x), 
\ee
and two sets of equivalent solutions for each of the conditions in Eq.\,\eqref{dustsol} are found for the Einstein Equations solutions. Both solutions are separated into two cases.

\ni \textbf{Case 1} begins with the following condition
\be
\frac{\partial \beta}{\partial z} = 0,
\ee 
and as a consequence, it has the following expression result 
\be
\frac{\partial \beta}{\partial y} = 0.
\ee  

\ni \textbf{Case 2} begins with the following condition 
\be
\frac{\partial \beta}{\partial y} = 0,
\ee 
and as a consequence, it has the following expression result
\be
\frac{\partial \beta}{\partial z} = 0.
\ee 
The expressions that result in the vacuum Einstein equations solution for the warp drive geometry are the following
\be
\mu = 0,
\label{mu0}
\ee
\be
\frac{\partial}{\partial x}\left[\frac{\partial \beta}{\partial t} + \frac{1}{2} \frac{\partial}{\partial x} (\beta^2) \right] = 0,  
\label{advec3}
\ee
Eq.\eqref{mu0} implies vacuum since the matter density is null, and Eq.\eqref{advec3} integration is straightforward, leading to a Burgers-type equation below
\be
\frac{\partial \beta}{\partial t}
+ \frac{1}{2}\frac{\partial}{\partial x} (\beta^2) = h(t),   
\label{advec4}
\ee
where $h = h(t)$ is an arbitrary function determined by boundary and initial conditions. In its homogeneous form where $h(t) = 0$, Eq.\,\eqref{advec4} becomes the \textit{inviscid Burgers equation}, a well-known equation occurring in fluid models, as gas dynamics, traffic flows, and conservation laws. It is a quasilinear hyperbolic equation, and its current density is the kinetic energy density. 

A quasilinear hyperbolic equation is a type of partial differential equation where the highest order derivatives appear linearly, but the coefficients of these derivatives can depend on the unknown function and its lower-order derivatives. These equations are often found in the study of wave propagation and conservation laws. Defining the flow density by 
\be
J_f = J_f(\beta),
\ee 
the current density can be a general function of $\beta$ so in this particular case
\be
J_f(\beta) = \beta^2,
\ee
but in general, this equation can be put in the following form
\be
\frac{\partial\beta}{\partial t}+\frac{1}{2}\frac{\partial J_f}{\partial x}  = \nu \frac{\partial^2 \beta}{\partial x^2}.  
\label{eq511}
\ee
where $\nu$ is a diffusion term and Eq.\,\eqref{eq511} is known as the viscous Burgers equation\index[topics]{viscous Burgers equation}. When no diffusion term exists, the equation becomes the inviscid Burgers equation\index[topics]{inviscid Burgers equation}, an equation known for its conservation properties.

It must be noted that the Burgers-type equation found for the vacuum solution of Einstein equations is a particular case of the viscous Burgers equation, where
\be
h(t) = \nu \frac{\partial^2 \beta}{\partial x^2}.
\ee
The phenomena arising from the Burgers equation are the conservation laws and the formation of shock waves, discontinuities that appear after a finite time and then propagate regularly. The one-dimension conservation law implicit in Eq.\,\eqref{eq511} is discussed in Ref.\,\cite{Evans2010}, as it is depicted in
\be
\frac{\partial \beta}{\partial t} +  \frac{\partial}{\partial x}F(\beta) = 0,
\label{bur1}
\ee
where the function $\beta = \beta(t,x)$ is to be determined from initial conditions $\beta(t=0,x) = \beta_0(x)$. The Burgers equation can describe rarefaction and expansion waves. Hence, in the present context, Eq.\,\eqref{eq511} may represent an analogy of a warp bubble being a shock wave formed in vacuum. 

It is important to note that in the warp drive scenario, it is possible to describe different warp drive metrics using different signs of the shift vector as follows
\be 
\beta = \pm v_s(t) f(r_s),
\ee 
Sign $\pm$ has physical significance and may lead to different warp drive geometries. Term
\be
\frac{\partial \beta}{\partial t}
\ee
may be interpreted as a force per unit mass density function, \textit{i.e.}, the time derivative of momentum, and the term 
\be
\frac{1}{2}\frac{\partial (\beta^2)}{\partial x}
\ee
has the physical interpretation of a potential function, \textit{i.e.}, the divergence of the total energy, which in this case is entirely kinetic. The function $h(t)$ determines if the system is dissipative regarding energy flow and momentum. 

\ni Table \ref{tab1dust} summarizes the results obtained for the warp drive coupled with the dust source.

\begin{table}[ht]
\centering
\begin{tabular}{| m{3cm} | m{3cm} | m{5cm} |}
\hline 
Case & Consequence & Results \\ 
\hline 
$1) \ \displaystyle{\frac{\partial \beta}{\partial z} = 0} $ 
& 
$\displaystyle{\frac{\partial \beta}{\partial y} = 0}$
&
$\begin{array} {ll} 
\mu = 0 \\ [6pt]
\beta = \beta(t,x) \\ [6pt]
\displaystyle{\frac{\partial \beta}{\partial t} 
+ \frac{1}{2} \frac{\partial}{\partial x}(\beta^2)
= h(t)}
\\[10pt]
\end{array}$ \\ 
\hline 
$2) \ \displaystyle{\frac{\partial \beta}{\partial y} = 0}$
&
$\displaystyle{\frac{\partial \beta}{\partial z} = 0}$ 
& 
$\begin{array} {ll} 
\mu = 0 \\ [6pt]
\beta = \beta(t,x)\\ [6pt]
\displaystyle{\frac{\partial \beta}{\partial t} 
+ \frac{1}{2} \frac{\partial}{\partial x}(\beta^2)
= h(t)}
\\[10pt]
\end{array}$ \\
\hline 
\end{tabular}
\caption{Summary of results for the warp drive spacetime having dust
matter content.}
\label{tab1dust}
\end{table}

Table\,\ref{tab1dust} shows two equivalent solutions. The same set of equations was reached from different cases, which generated different consequences unfolding in the same solution. This was done for clarity, to explore possible non-unique solutions. Despite this, the solutions are equal. The same approach was used with other results, considering the other energy-momentum sources in the following chapters.

Since the solutions found for Einstein's equations are the vacuum ones, the energy conditions are trivially satisfied since $\mu = 0$.

\begin{table}[!h]
\centering 
\begin{tabular}[c]{l @{\hspace{50pt}} l} 
\hline\hline 
Energy condition & Results \\ [0.5ex] 
\hline 
Weak         &  trivially satisfied \\  
Strong       &  trivially satisfied \\
Dominant     &  trivially satisfied \\
Null         &  trivially satisfied \\ 
[1ex] 
\hline 
\end{tabular}
\caption{Summary for the energy conditions for dust as a source for the warp drive metric.}
\label{dustengcond} 
\end{table}

When Alcubierre proposed the warp drive theory, he came across one of the main caveats: the need for negative energy densities to create a warp bubble and the violation of weak and dominant energy conditions. Alcubierre found this result by contracting the Einstein tensor $G^{\mu\nu}$ with the covariant velocity of the Eulerian observers 
\be
u_\mu = (-1,0,0,0),
\ee
and making use of the Einstein equations to find the energy density
\be
\rho = \frac{1}{8\pi}G^{\mu\nu} u_\mu u _\nu = \frac{G^{00}}{8\pi} =  T^{00}.
\ee
The same result can be found with the expression below
\begin{align}
T_{\alpha\beta} u^\alpha u^\beta &= \frac{1}{8\pi}G_{\alpha\beta} u^\alpha u^\beta
\\& = \frac{1}{8\pi}\left(G_{00} - 2\beta G_{01} + \beta^2 G_{11}\right),
\end{align}
which results in the following expression
\be
\rho = -\frac{1}{32\pi} \left[\left( \frac{\partial \beta}{\partial y} \right)^2 + \left( \frac{\partial \beta}{\partial z} \right)^2 \right] \leq 0,
\ee
The above relation is non-positive everywhere. However, Alcubierre did not account for zero or complex-valued energy densities \index[topics]{complex-valued energy density} because of his choice of the shift vector 
\be
\beta(t,x,y,z) = - v_s(t) f\left[r_s(t,x,y,z)\right].
\ee
and the real-valued regulating function\index[topics]{regulating function}. 

\subsection{Warp drive spacetime curvature}

A Burgers-type equation gives rise to a vacuum Einstein equation solution considering the warp drive coupled to dust. This is a significant result that needs more inspection since it suggests that this is an interior solution for the geometry of the warp drive, considering the warp bubble is formed by three regions: one interior, one exterior, and one warped region between the other two with limited thickness and delimited by the warp bubble walls. The warp drive may be only part of an exterior, possibly globally flat spacetime. This section discusses only the local flatness regarding the boundary conditions found for the Einstein equation solutions and the limitations on the shift vector. Since $\beta = \beta(t,x)$ for the vacuum solution of Einstein equations in a warp drive background, it is possible to use this ansatz to discuss the curvature of the warp drive spacetime by inspecting the components of the Riemann and Ricci tensors. With this limit condition on the shift vector, the only non-zero Riemann tensor components are
\be
R_{\phantom{t}t t x}^{t\phantom{t}\phantom{t}\phantom{x}} =  \beta \, R_{\phantom{t}x t }^{t\phantom{t}\phantom{t}\phantom{x}} = \beta \, \frac{\partial}{\partial x}\left[
\frac{\partial \beta}{\partial t}
+ \frac{1}{2} \frac{\partial(\beta^2)}{\partial x}\right] \,,
\ee
and the only non-zero Ricci components are 
\be
R_{00} = \left(\beta^2 - 1\right)\frac{\partial}{\partial x}\left[\frac{\partial\beta}{\partial t} + \frac{1}{2}\frac{\partial}{\partial x}\left(\beta^2\right)\right] \,,
\label{newricci00}
\ee
\be
R_{11} = \frac{\partial}{\partial x}\left[\frac{\partial\beta}{\partial t} + \frac{1}{2}\frac{\partial}{\partial x}\left(\beta^2\right)\right] \,.
\label{newricci11}
\ee
 Equations \eqref{newricci00} and \eqref{newricci11} can be summarized as
\be
R_{00} = (\beta^2 - 1) R_{11} \,.
\ee
Inserting the Burgers-type equation solution vanishes the Ricci tensor components $R_{00}$ and $R_{11}$, which shows that this spacetime is flat only locally. The connection between the warp drive and shock waves may be a new vacuum solution for the Einstein equations. These results suggest that a deeper investigation of the warp drive employing the junction conditions is necessary, considering the Minkowski as the exterior spacetime and the warp drive as the interior boundary. Also, this investigation may shed clarity on the horizon problem of the warp bubble formation, and the Burgers-type equations strongly suggest that the warp bubble may result from different spacetime discontinuities.

\subsection{Discussion}

A solution for the Einstein equations is presented for the dust with a warp drive background, where the matter density is null. These solutions connect the warp drive to shock waves by a Burgers-type equation. The right-hand side of the Burgers-type equation is composed by a function that only depends on the time coordinate $h = h(t)$, an arbitrary function determined by boundary and initial conditions. Two possibilities are considered regarding this equation: If $h(t) = 0$, the well-known inviscid Burgers equation is recovered. If $h(t) \neq 0$, this equation becomes a viscous Burgers-type equation under certain limitations for a dissipative system with a diffusion coefficient that depends only on time coordinates. This result indicates that the function $h(t)$ can act as a source term that originates the shock wave and, consequently, the warp bubble. However, this mechanism must be inspected more deeply since coupling the dust to a warp drive background resulted in a vacuum solution. This means that dust does not have enough momentum and energy to create a warp bubble, but shock waves appear remarkably. It is plausible that the warp bubble may be interpreted as a shock wave from classical fluid dynamics, with zero matter density as a new vacuum solution for the Einstein equations. 

Another relevant result is that the shift vector only depends on $(t,x)$ coordinates, so its dynamics are preserved in the other spatial directions, which makes the warp bubble change form in only one direction. The Riemann and Ricci tensors vanish with these results, but this is not a global result, meaning that this spacetime is globally flat but only locally flat in a particular region. The warp drive has a local isomorphism with Minkowski spacetime since it is asymptotically flat, as first proposed by Alcubierre; when the warp bubble walls become very thin, the form function behaves as a step function, making the warp drive spacetime flat outside the warp bubble and in its interior. Still, this warp drive spacetime is not globally flat since the Ricci and Riemann tensors only vanish for certain conditions where the vacuum solutions are satisfied.

\section{Perfect fluid}

Four solutions of the Einstein equations for the perfect fluid and the warp drive are presented in this chapter. Two of those solutions connect warp drive and shock waves via Burgers-type equations identical to dust. The other two solutions are similar, with the difference being the spacetime coordinate dependency. The shift vector can be complex-valued, with a positive real part allowing the warp drive to respect the energy conditions, resulting in boundary conditions for the shift vector. 

The first section presents the solutions of the Einstein equations when the perfect fluid is coupled to the warp drive spacetime. All conditions and results for the four sets of solutions are discussed in detail, with a summarizing table at the end of the section. The second section discusses the energy conditions for the perfect fluid in a warp drive background, which is a different approach than the one made by Alcubierre in his work, considering the geometrical aspect first to conclude that weak and dominant energy conditions are violated. This matter-energy approach highlights the geometrical needs for a possible warp drive metric without energy condition violation. Limiting conditions relating to the geometrical aspects of the warp drive, the shift vector, and the source contents are presented considering the four sets of solutions for the Einstein equations. The third section has the same objective as the second one: essential relations between the geometrical and matter content aspects are presented when there is an imposition of null divergence of the energy-momentum tensor and the four sets of solutions to the Einstein equations. The fourth section discusses the results.

\subsection{Solving Einstein equations}

The Einstein tensor components are given by the set of equations from Eq.\,\eqref{et00} to Eq.\,\eqref{et33}, and the important expressions used to solve the Einstein equations for different matter sources are given by Eqs.\,\eqref{burgerseq}, \eqref{expet1}, \eqref{gradpress}, \eqref{eqstaterel}. 

From Eq.\,\eqref{eqstaterel} it is possible to find the state equation for the perfect fluid, and the warp drive is found to be
\be
p = 3 \mu \,.
\label{eqstateeq}
\ee
from the Einstein tensor component $G_{23} = 8 \pi T_{23} = 0$. Two case conditions regarding derivatives of the shift vector are found for the perfect fluid and the warp drive, as follows,
\begin{align}
&1) \ \frac{\partial \beta}{\partial z} = 0\,,  
\\[10pt]
&2) \ \frac{\partial \beta}{\partial y} = 0 \,.
\end{align}
The above conditions lead to two different solutions for this configuration regarding the perfect fluid. Those two solutions are discussed and separated as Case 1 and Case 2.

\ni\textbf{Case 1) $\displaystyle{\frac{\partial \beta}{\partial z} = 0}$}

The following set of equations obtained from Einstein's equations manipulations determine the first set of solutions for the perfect fluid and the warp drive geometry
\begin{eqnarray} 
&& p = 3 \mu,
\label{eepf01} \\[10pt]
&& \left(\frac{\partial \beta}{\partial y}\right)^2 
= - \frac{32\pi}{3} p = - 32 \pi \mu\,,
\label{eepf02} \\[10pt]
&& \frac{\partial^2 \beta}{\partial y^2} = 0\,,
\label{eepf03} \\[10pt]
&& \frac{\partial^2 \beta}{\partial x\partial y}=0\,,
\label{eepf04} \\[10pt]
&& 2 \frac{\partial \beta}{\partial y} \, 
\frac{\partial \beta}{\partial x} 
+ \frac{\partial^2 \beta}{\partial t \partial y}=0 \,,
\label{eepf05} \\[10pt]
&& \frac{\partial}{\partial x}\left[
\frac{\partial \beta}{\partial t}
+ \frac{1}{2} \frac{\partial}{\partial x} (\beta^2)\right]
= - \frac{64\pi}{3} p \,= - 64 \pi \mu \,,
\label{eepf06} \\[10pt]
&& \frac{\partial}{\partial x}\left[
\frac{\partial \beta}{\partial t}
+ \frac{1}{2} \frac{\partial}{\partial x} (\beta^2)
\right] = - \frac{128\pi}{3} p = - 128\pi\mu\,.
\label{eepf07}
\end{eqnarray}
Eq.\,\eqref{eepf01} is the state equation found for the perfect fluid and the warp drive, and the other equations help to solve the Einstein equations. Case 1 unfolds in the two following subsets of solutions.

\ni\textbf{Case 1a:} $\left[\displaystyle{\frac{\partial \beta}{\partial z} = 0} \ \text{and} \ \displaystyle{\frac{\partial\beta}{\partial x} = 0}\right]$ 

\ni are obtained by imposing the sub-conditions and placing them into the condition found for Case 1. The set of equations from Eq.\,\eqref{eepf01} to Eq.\,\eqref{eepf07} are simplified by the conditions for Case 1a, and given by expressions below
\be 
p = 3 \mu,
\ee
\be
\frac{\partial \beta}{\partial y} =\pm\sqrt{- 32 \pi \mu}.
\ee
The above equation shows derivatives of the shift vector that can be readily integrated if the matter density is assumed to be constant, leading to the following result
\be
\beta(t,y) = \pm\sqrt{- 32 \pi \mu} \, y + g(t) \,,
\label{beta1a}
\ee
where $g(t)$ is an integration constant. This function depends only on time and can be determined by boundary and initial conditions. It must be noted that for a real-valued solution for the shift vector, the matter density must be of a negative value. For a non-constant matter density, $\beta$ may assume real values in regions where $\mu$ is negative or complex values when $\mu$ is positive. Further analysis of this point is outside the scope of this thesis.

\ni\textbf{Case 1b:} $\left[\displaystyle{\frac{\partial \beta}{\partial z} = 0} \ \text{and} \ \displaystyle{\frac{\partial\beta}{\partial y} = 0}\right]$ 

\ni The second set of solutions, has vanishing pressures since the condition $\partial \beta/\partial y = 0$ is imposed from Einstein equations components. This subcase recovers the same vacuum shock waves type solution found for the dust in Chap.\,4 when coupled to the warp drive geometry, leading to null pressure and matter density
\be 
p = 0 = \mu,
\ee 
\be
\frac{\partial}{\partial x}\left[\frac{\partial \beta}{\partial t} + \frac{1}{2} \frac{\partial}{\partial x} (\beta^2) \right] = 0\,.
\label{neweepf07}
\ee
Or equivalently, as found in Chap.\,4, a Burgers-type equation with source $h(t)$ that is dependent only on time, and this being a vacuum solution to Einstein equations
\be
\frac{\partial \beta}{\partial t}
+ \frac{1}{2}\frac{\partial}{\partial x} (\beta^2) = h(t)\,.   \label{pfadvec4}
\ee
Next, we proceed to analyze the second case solutions

\ni\textbf{Case 2) $\displaystyle{\frac{\partial \beta}{\partial y} = 0}$}

The set of equations that solve Einstein equations are
\begin{align} 
&p = 3 \mu,
\label{eepf01c2} \\[10pt]
&\left(\frac{\partial \beta}{\partial z}\right)^2 
= - \frac{32\pi}{3} p,
\label{eepf02c2} \\[10pt]
&\frac{\partial^2 \beta}{\partial z^2} = 0,
\label{eepf03c2} \\[10pt]
&\frac{\partial^2 \beta}{\partial x\partial z} = 0,
\label{eepf04c2} \\[10pt]
&2 \frac{\partial \beta}{\partial z} \, 
\frac{\partial \beta}{\partial x} 
+ \frac{\partial^2 \beta}{\partial t \partial z} = 0,
\label{eepf05c2} \\[10pt]
&\frac{\partial}{\partial x}\left[\frac{\partial \beta}{\partial t}
+ \frac{1}{2} \frac{\partial}{\partial x} (\beta^2)\right] = - \frac{64\pi}{3} p\,,
\label{eepf06c2} \\[10pt]  
&\frac{\partial}{\partial x}\left[\frac{\partial \beta}{\partial t} + \frac{1}{2} \frac{\partial}{\partial x} (\beta^2) \right] = - \frac{128\pi}{3} p\,,
\label{eepf07c2}
\end{align}
where Eq.\,\eqref{eepf01c2} is the state equation for the perfect fluid coupled to the warp drive geometry. The other equations above help solve Einstein's equations. Case 2 also unfolds in two subcases (2a and 2b), as discussed below. 

\ni\textbf{Case 2a:} $\left[\displaystyle{\frac{\partial \beta}{\partial y} = 0} \ \text{and} \ \displaystyle{\frac{\partial\beta}{\partial x} = 0}\right]$ 

\ni Imposing the conditions for the subcase 2a in Einstein equations results in the following equivalent set of equations where $\mu$ must be negative or zero for a non-complex solution considering the shift vector for Einstein equations.
\begin{align} 
&p = 3 \mu \,,
\label{p3mu} \\[10pt]
&\frac{\partial \beta}{\partial z}= \pm \,\sqrt{- 32 \pi \mu}\,.
\label{bbb1}
\end{align}
If the matter density $\mu$ is assumed constant, the above expressions can be integrated 
\be
\beta(z,t) = \pm\sqrt{- 32 \pi \mu} \, z + w(t)\,,
\label{beta2a}
\ee
where $w(t)$ is an arbitrary constant of integration, due to the integral on the $z$ coordinate. This function must be determined by boundary and initial conditions. This solution is similar to the one found in Case 1a, where there is a dependency on the sign of the matter density for the shift vector to possess real or complex values. As mentioned in Case 1a, it is out of the scope of this thesis to discuss general solutions of these equations considering a non-constant matter density. Next, the second subcase for Case 2 is analyzed.

\ni\textbf{Case 2b:} $\left[\displaystyle{\frac{\partial \beta}{\partial y} = 0} \ \text{and} \ \displaystyle{\frac{\partial\beta}{\partial z} = 0}\right]$ 
 
 Imposing the conditions for the second subcase in Einstein equations results for Case 2b to recover the same vacuum solutions found in Chap.\,4 and Case 1b, with the following
\be 
p = 3 \mu = 0,
\ee 
\be
\frac{\partial \beta}{\partial t}
+ \frac{1}{2}\frac{\partial}{\partial x} (\beta^2) = h(t)\,. 
\ee

As seen above, Cases 1b and 2b present the same family of solutions, connecting the vacuum solutions of Einstein equations for the warp drive geometry to shock waves via a Burger-type equation. These solutions are the same found in Chap.\,4 for dust. 

Positive values of the matter density imply that the solutions for Cases 1a and 2a result in complex-valued functions for the shift vector. For real-valued shift vector functions, the matter density must be negative. These two cases are similar but differ on the dependency of spacetime coordinates. Case 1a has a solution where the shift vector is a function of $(t,y)$ coordinates, and Case 2a has a solution where the shift vector is a function of $(t,z)$ spacetime coordinates. Both cases share the same state equation
\be
p = 3\mu \,.
\ee 

A complex-valued shift vector analysis is out of the scope of this thesis, so it is clear that negative matter density is necessary for creating a warp bubble concerning this warp drive spacetime. It is also out of the scope of this thesis to analyze a non-constant matter density. 

The following section analyzes the energy conditions for each solution found: Cases 1a and 1b and Cases 2a and 2b. Table \ref{tab1} summarizes all the found case solutions.

\begin{table}[!ht]
\centering
\begin{tabular}{| m{3cm} | m{3cm} | m{8cm} |}
\hline 
Case & Condition & Results \\ 
\hline 
\multirow{2}{*}{$1) \
\displaystyle{\frac{\partial \beta}{\partial z} = 0}$}
& 
$1a) \ \displaystyle{\frac{\partial \beta}{\partial x} = 0}$
&
$\begin{array} {ll} 
p = 3 \mu \\ [6pt]
\beta = \beta(y,t)\\ [6pt]
\displaystyle{\frac{\partial \beta}{\partial y} 
= \pm\sqrt{- 32 \pi \mu}} \\ [8pt]
\end{array}$ \\ [28pt]
\cline{2-3}   
&
$1b) \ \displaystyle{\frac{\partial \beta}{\partial y} = 0}$
&
$\begin{array} {ll} 
p = 3 \mu = 0 \\ [6pt]
\beta = \beta(x,t)\\ [6pt]
\displaystyle{\frac{\partial \beta}{\partial t} 
+ \frac{1}{2} \frac{\partial}{\partial x}(\beta^2)
= h(t)} \\ [8pt]
\rightarrow\mbox{this is the dust solution of Ref.\ \cite{nos1}} \\ [4pt]
\end{array}$ \\ [28pt] 
\hline 

\multirow{2}{*}{$2) \
\displaystyle{\frac{\partial \beta}{\partial y} = 0}$}
& 
$2a) \ \displaystyle{\frac{\partial \beta}{\partial x} = 0}$
&
$\begin{array} {ll} 
p = 3 \mu \\ [6pt]
\beta = \beta(z,t)\\ [6pt]
\displaystyle{\frac{\partial \beta}{\partial z} 
= \pm \, \sqrt{-32 \pi\mu}}\\ [6pt]
\beta = \beta(z,t)\\ [6pt]
\displaystyle{\frac{\partial \beta}{\partial z} 
= \pm \, \sqrt{\pm 96 \pi\mu}}\\ [8pt]
\end{array}$ \\ [28pt]
\cline{2-3}   
&
$2b) \ \displaystyle{\frac{\partial \beta}{\partial z} = 0}$
&
$\begin{array} {ll} 
p = 3 \mu = 0 \\ [6pt]
\beta = \beta(x,t)\\ [6pt]
\displaystyle{\frac{\partial \beta}{\partial t} 
+ \frac{1}{2} \frac{\partial}{\partial x}(\beta^2)
= h(t)} \\ [6pt]
\rightarrow\mbox{this is the dust solution of Ref.\ \cite{nos1}} \\ [4pt]
\end{array}$ \\ [28pt] 
\hline 
\end{tabular}
\caption{Summary of all solutions of the Einstein equation with the warp drive metric coupled to the perfect fluid as the mass-energy source.}
\label{tab1}
\end{table}

\subsection{Energy conditions}

This section presents the calculation for all the energy conditions considering the perfect fluid coupled to the warp drive spacetime. The inequalities are calculated considering all four case solutions (1a, 1b, 2a, 2b) found in the previous section. Expressions relating the matter density $\mu$, the static pressure $p$, and the shift vector are found, and the possibility of a warp drive not violating energy conditions is discussed. Since Cases 1b and 2b are vacuum solutions to the Einstein equations, the energy conditions are immediately satisfied. For Cases 1a and 2a, the Weak, Dominant, and Strong Energy Conditions are satisfied if the static pressure is positive. The null energy condition is satisfied if the pressure is positive and large enough to surpass the negative matter density. These results for the energy conditions do not differ much from other fluid solutions for the Einstein equation since the state equations found for the warp drive and the perfect fluid do not depend on geometric terms such as the shift vector, but only on matter-energy terms such as pressure.

\subsubsection{Weak energy condition} 

The weak energy condition requires that at each point in spacetime,
\be
T_{\alpha \sigma} \, u^\alpha u^\sigma \geq 0.    
\label{wec111pf}
\ee
This is valid for any timelike vector $\textbf{u} \, (u_\alpha u^\alpha < 0)$ and null vector $\textbf{k} \, (k_\alpha k^\alpha = 0)$. For an observer with a unit tangent vector at some point in spacetime, the local energy density measured by any observer is non-negative \cite{HawkingEllis1973}. Considering the same Eulerian observers that Alcubierre considered \cite{Alcubierre1994} with 4-velocity given by
\begin{align}
u_\alpha &= (- 1,0,0,0)\,, \\
u^\alpha &= g^{\alpha\beta} u_\beta =  (1, \beta, 0, 0)\,, 
\end{align}
and the perfect fluid energy-momentum tensor leads to the following result
\be
T_{\alpha \sigma} \, u^\alpha u^\sigma = \mu \,.
\label{eq88pf}
\ee
For cases 1a and 2a, the state equation is given by $p = 3\mu$, so the Weak Energy Condition\index[topics]{weak energy condition} is satisfied if the pressure and the matter density are positive, implying that, in this case, the shift vector is a complex-valued function. For cases 2a and 2b, the vacuum shock wave solution with the Burgers-type equation is immediately satisfied.

\subsubsection{Dominant energy condition}

The dominant energy conditions state that for every timelike vector $u_a$, the following two inequalities must be satisfied
\be
T^{\alpha \beta} \, u_\alpha u_\beta \geq 0, \quad \text{and} \quad 
F^\alpha  F_\alpha  \leq 0, 
\ee
where $F^\alpha = T^{\alpha \beta} u_\beta$ is a non-spacelike vector, so for any observer, the local energy density is non-negative, and the local energy flow vector is non-spacelike. Hawking and Ellis \cite{HawkingEllis1973} suggested that this condition must hold for all known forms of matter and that it should be the case in all situations.
\be
T^{00} \geq |T^{ab}|, \ \text{for each} \ a, b\,.
\ee
For the perfect fluid in warp drive geometry
\be
T^{\alpha \beta} \, u_\alpha u_\beta = \mu \geq 0\,,
\label{dec1pf}
\ee 
and the second inequality $F^\alpha F_\alpha \leq 0$ yields,
\be
(T^{\alpha \sigma} \, u_\sigma)(T_{\alpha\sigma} \, u^\sigma) = - p^2 \leq 0\,.
\label{dec2pf}
\ee
For the solutions 1a and 2a found for both the perfect fluid and warp drive, (see Table \ref{tab1}, Eq.\,\eqref{dec1pf}) depend on the sign of $\mu$, but Eq.\,\eqref{dec2pf} is always satisfied. A positively valued matter density is enough for the dominant energy condition\index[topics]{dominant energy condition} to be satisfied. However, this also means a complex result for the shift vector $\beta$. For cases 1b and 2b, the vacuum solution found in Chap.\,4 is recovered, and the DEC is trivially satisfied.

\subsubsection{Strong energy condition}

For any timelike vector $u^\alpha$, the strong energy condition is
\be
\left(T_{\alpha \beta} - \frac{1}{2}T \, g_{\alpha \beta} \right) 
u^\alpha u^\beta \geq 0 \,,
\ee
This requirement is stronger than the weak energy condition. The Einstein equations imply that gravity is always attractive. For a perfect fluid, the warp drive, and the Eulerian observers, the strong energy condition\index[topics]{strong energy condition} is then given by the following expression
\be
\left(T_{\alpha\sigma}-\frac{1}{2}T \, g_{\alpha\sigma}\right) u^\alpha u^\sigma = \frac{5}{3}p\,.
\label{secaaapf}
\ee
The strong energy conditions can then be calculated for solutions 1a and 2a for the perfect fluid and the warp drive geometry, see (Table\,\ref{tab1}), and are satisfied if Eq.\eqref{secaaapf} is greater than or equal to zero, meaning if the pressure is positively valued, which leads to complex-valued $\beta$ functions. Cases 1b and 2b are just the vacuum solution found in Chap.\,4, immediately satisfying the strong energy condition.

\subsubsection{Null energy condition}

The strong and weak energy conditions must be satisfied in the limit of the observers with 4-velocity $\textbf{k}$, which is a null vector. To satisfy the Null Energy Condition, the energy-momentum tensor must follow the inequality below, 
\be
T_{\alpha \sigma} \, k^\alpha k^\sigma \geq 0, \ \text{for any null vector} \ k^\alpha \,.
\label{neceqcondipf}
\ee
To calculate the Null Energy Condition, assume the null vector below,
\be
k^\alpha = (a,b,0,0) \,,    
\label{neck1pf}
\ee
$a$ and $b$ can be obtained by imposing the null condition $k_\alpha k^\alpha = 0$, 
\be
\frac{a}{b} = \frac{1}{\beta \pm 1}\,,
\label{necrootspf}
\ee
and the Null Energy Condition reads
\be
T_{\alpha \sigma} \, k^\alpha k^\sigma = 
\frac{b^2}{(\beta \pm 1)^2} \left(\mu + p\right) \geq 0 \,.
\ee
Cases 1a and 2a for the perfect fluid coupled to the warp drive geometry, see Table \eqref{tab1}, have the state equation $p = 3\mu$, so the Null Energy Condition \index[topics]{null energy condition} is satisfied if the pressure and the matter density are positive, which means complex-valued solutions for the shift vector. Cases 1b and 2b are the vacuum solutions found in Chap.\,4, and they immediately satisfy the null energy condition.

Table \eqref{tab3} summarizes the results of all energy conditions for the perfect fluid matter source.

\begin{table}[ht!]
\centering
\begin{tabular}{| m{4.5cm} | m{4.0cm} | m{5.0cm} |}
\hline Cases 
& Energy Conditions
& Results \\ \hline  
1a and 2a & \vspace{8pt} Weak &
$\begin{array} {ll} \\ [-6pt]
\displaystyle{p \geq 0} \\ [8pt]
\end{array}$ \\ [8pt] \cline{2-3}   
& \vspace{8pt} Dominant &
$\begin{array} {ll} \\ [-6pt]
\displaystyle{p \geq 0} \\ [8pt]
\end{array}$ \\ [8pt]
\cline{2-3} & \vspace{8pt} Strong &
$\begin{array} {ll} \\ [-6pt]
\displaystyle{p} \geq 0 \\ [8pt]
\end{array}$ \\ [8pt]
\cline{2-3} & \vspace{8pt} Null &
$\begin{array} {ll} \\ [-6pt]
\displaystyle{\mu + p \geq 0} \\ [8pt]
\end{array}$ \\ [8pt]
\hline 
1b and 2b & \vspace{8pt}  Weak &
$\begin{array} {ll} \\ [-6pt]
\text{immediately satisfied.} \\ [8pt]
\end{array}$ \\ [8pt]
\cline{2-3} & \vspace{8pt} Dominant &
$\begin{array} {ll} \\ [-6pt]
\text{immediately satisfied.} \\ [8pt]
\end{array}$ \\ [8pt]
\cline{2-3} & \vspace{8pt} Strong &
$\begin{array} {ll} \\ [-6pt]
\text{immediately satisfied.} \\ [8pt]
\end{array}$ \\ [8pt]
\cline{2-3} & \vspace{8pt} Null &
$\begin{array} {ll} \\ [-6pt]
\text{immediately satisfied.} \\ [8pt]
\end{array}$ \\ [8pt]
\cline{2-3} \hline
\end{tabular}
\caption{Summary of all energy conditions results for the perfect fluid coupled with warp drive geometry.}
\label{tab3}
\end{table}

In summary, all energy conditions can be satisfied for Eulerian observers as long as the matter density is considered positive. However, this would lead to complex-valued solutions for the shift vector considering cases 1a and 2a. Even though complex functions are often used in physics, a deeper evaluation of this result is needed for further work. 

This thesis presents the solutions for the simple matter-energy sources, but practical applications and possible models for experimental results are not discussed. It is important to state that the fact that the energy conditions can be satisfied for one class of observers, in this particular example for Eulerian observers, does not mean that the conditions will always be fulfilled for other classes of observers. The aim of these calculations is not to demonstrate physical warp drives since that it would require general theorems proving the warp drive would always satisfy energy conditions. It is demonstrated that limiting theorems regarding warp drive and energy conditions published since 1994 are not in a strong form since they lack boundary conditions. This work shows how warp drives do not always violate energy conditions. 

In the next section, the null divergence of the energy-momentum tensor is imposed, and boundary conditions are found, leading to expressions relating to the matter density, the static pressure, and the shift vector.

\subsection{Divergence of energy-momentum tensor}

An important aspect of fluids in general relativity is the null divergence of the energy-momentum tensor, which connects it to energy conservation and other physical properties. Calculating the divergence given by the following expression
\be
{T^{\alpha \sigma}}_{;\sigma}=0
\ee
for the perfect fluid energy-momentum tensor and imposing the null condition for each of the solutions found (cases 1a, 1b, 2a, 2b), the Einstein equations lead to the results discussed below. The leading equations for the null divergence of the perfect fluid energy-momentum tensor are
\begin{align}
- \frac{\partial \mu}{\partial t}-\beta \left(\frac{\partial p}{\partial x} + \frac{\partial\mu}{\partial x}\right)&= 0\,,
\label{pf1adivT0} \\[2pt]
\frac{\partial p}{\partial x}&= 0\,,
\label{pf1adivT1} \\[2pt]
\frac{\partial p}{\partial y}&= 0\,,
\label{pf1adivT2} \\[2pt]
\frac{\partial p}{\partial z}&= 0\,,
\label{pf1adivT3}
\end{align}
so the pressure does not depend on the spatial coordinates, and Eq.\,\eqref{pf1adivT0} reduces to
\be
\frac{\partial\mu}{\partial t} + \beta\frac{\partial\mu}{\partial x} = 0\,,
\label{divpf0}
\ee 
which is the continuity equation, where $\mu$ plays the role of the fluid density, and $\beta$ is the flow velocity vector field. It is worth mentioning that the fluid has incompressible flow for a constant density. All the partial derivatives of $\beta$ in terms of the spatial coordinates vanish. The flow velocity vector field has null divergence, this being a classical fluid dynamics scenario, and the local volume dilation rate is zero.

\ni\textbf{Case 1a: $\bm{\left[\displaystyle\frac{\partial \beta}
{\partial z} = 0\right.}$ and $\bm{\left.\displaystyle\frac{\partial
\beta}{\partial x} = 0\right]}$}\label{sit1a}

\ni For this case solution,  this state equation is given by $p = 3 \mu$, and from Eqs.\,\eqref{pf1adivT1}, \eqref{pf1adivT2}, \eqref{pf1adivT3} it implies that the pressure may depend only on time, but it is constant regarding spatial spacetime coordinates. Using the state equation and Eq.\,\eqref{divpf0} implies
\be
\frac{\partial \mu}{\partial t} = 0 \,,
\ee
demonstrating that both pressure and matter density are constant.

\ni \textbf{Case 2a: $\bm{\left[\displaystyle\frac{\partial \beta}{\partial y} = 0\right.}$ and $\bm{\left.\displaystyle\frac{\partial\beta}{\partial x} = 0\right]}$}\label{sit2a}

\ni The results for this case are identical to Case 1a. 

\ni\textbf{Cases 1b and 2b: $\bm{\left[\displaystyle\frac{\partial \beta}
{\partial y} = 0\right.}$ and $\bm{\left.\displaystyle\frac{\partial
\beta}{\partial z} = 0\right]}$}\label{1b2b}

\ni Solution $\mu = p = 0$ For this case, the null divergence is immediate.

\subsection{Discussion}

Coupling the warp drive geometry to the perfect fluid led to four solutions to the Einstein equations, named 1a, 1b, 2a, and 2b. Cases 1b and 2b are the same as vacuum solutions found for the dust in Chap.\,4. A perfect fluid source can also lead to a vacuum solution, connecting the warp drive to shock waves via a Burgers-type equation. Cases 1a and 2a share the same equation of state $p = 3 \mu$, but with different spacetime coordinate dependencies since $\beta = \beta(y,t)$ and $\beta = \beta(z,t)$. For $\beta$ to be a real-valued function, $\mu$ must be negative in both cases. For cases 1a and 2a, if the matter density is positive, the shift vector takes a complex-valued form of solution, not violating the energy conditions. Cases 1b and 2b are vacuum solutions that immediately satisfy the energy conditions. 

A complex-valued solution for $\beta$ might have physical meaning since, in the warp drive scenario, $\beta = v_s(t) f(r_s)$ determines the bubble's velocity and shape. A complex velocity has no physical meaning, but a complex regulating function could be acceptable if we only consider its real part. Thus, a warp bubble with superluminal speeds could still be formed. Nevertheless, caution is required because if the result of integrating $\beta$ turns out to be purely imaginary, it is unclear what the bubble shape represented by an imaginary function means. It seems reasonable to presume that the warp bubble requires a perfect fluid with negative mass-energy density and $\beta$ as a real function. 

The energy conditions are readily satisfied for the Burgers-type equation solutions analogous to the one we found in Chap.\,4. However, for the other solutions, the energy conditions requirements require both pressure and matter density to be positive. It should be noted that finding specific cases of warp drives not violating energy conditions for Eulerian observers it does not mean that they are physical warp drives. Still, it is a relevant result since it makes limiting theorems regarding classes of warp drives violating all energy conditions weaker in the mathematical sense. The imposition of the energy-momentum tensor null divergence implies that matter density and fluid pressure are constant when coupled to the warp drive. Perhaps more general fluids with dissipative terms, heat flow, and momentum flow could result in more general solutions to the Einstein equations considering warp drive geometry. Since the shift vector is of complex-valued form, one could also think about defining the total mass-energy density as follows
\be
\mu(t,x^j) = \mu^+ + a(t,x^j) \mu^- \leq 0 \ , \ x^j=y,z \,, 
\label{mattnegpos}
\ee
where $\mu^+$ is the positive portion of the matter density of the perfect fluid and $\mu^-$ its negative portion that would allow the warp bubble to exist. Where $a(t,x^j)$ would be a regulating function related to the bubble's shape and location. Remembering that $x^j = y$ for Case 1a and $x^j = z$ for Case 2a, since $\mu(t,x^j)$ must be negative, there would be a restriction for the positive and negative portions of the matter density in Eq.\,\eqref{mattnegpos}.

The results above imply that a perfect fluid source generates more complex solutions for Einstein's equations than dust when coupled to warp drive spacetime. This configuration even respects the energy conditions for a complex-valued shift vector. At this point, it is reasonable to think that sources with more complex physical properties can give rise to interesting Einstein equation solutions considering classes of warp drive geometry. Yet, we may still be far from feasible physical warp drives.
\section{Anisotropic fluid}

As an advance on Einstein equation's solution to the warp drive beyond dust and the perfect fluid, an anisotropic fluid is proposed, with five different parameters: $A, B, C$ being anisotropic static pressures, $p$, a static intrinsic pressure related to the energy density, and $D$, a momentum and energy flow density parameter. In matrix form, this source is
\be
T_{\alpha \sigma} = 
\begin{pmatrix} 
\mu + \beta^2 p & - \beta D & 0  & 0  \\ 
- \beta D     & A         & 0  & 0  \\ 
0             & 0         & B  & 0  \\ 
0             & 0         & 0  & C 
\end{pmatrix}.
\label{71}
\ee
Four solutions to the Einstein equations are presented. Two solutions are similar to the dust, connecting the warp drive to shock waves via a Burgers-type equation. The other two solutions lead to possible complex-value solutions for the shift vector. The energy conditions are analyzed, and constraints are found between the pressure parameters, the shift vector, and the matter density for the warp drive solutions to satisfy the inequalities of energy conditions. The null divergence of the energy-momentum tensor is also investigated, and all solutions are used as input to find expressions relating to the free parameters of the anisotropic fluid. Even with negative energy density, 
\be
\rho = T_{\mu\nu} u^\mu u^\nu < 0,
\ee
is possible to construct solutions where the energy conditions are satisfied; this is acquired by considering complex-valued solutions for the shift vector. Even with five free parameters $A, B, C, D, p$, obtaining well-behaved warp bubbles is still impossible.

This chapter is organized as follows: the first section presents the Einstein equation solutions to the anisotropic quasi-perfect fluid. Section two discusses the validity of energy conditions. Section three presents some conditions regarding the validity of the null divergence of the energy-momentum tensor.

\subsection{Solving Einstein equations}

The Einstein tensor components are given by the set of equations from Eq.\,\eqref{et00} to Eq.\,\eqref{et33}, and the important expressions used to solve the Einstein equations for different matter sources are given by equations \eqref{burgerseq}, \eqref{expet1}, \eqref{gradpress}, \eqref{eqstaterel}. From Eq.\,\eqref{eqstaterel}
\be
T_{00} + 2 \beta T_{01} + \frac{1}{3}(3\beta^2 - 1)T_{11} = 0 
\ee
which leads to the following state equation relating the static pressures $A, B, C, p$, the momentum flow $D$, the shift vector $\beta$, and the matter density $\mu$
\be 
\beta^2 (2D - A - p) + \frac{A}{3} = \mu.
\label{set01}
\ee
Since the anisotropic fluid has different pressures, it is not possible to find a traditional state equation of the form
\be
p = \omega \, \mu \,,
\ee
but it is interesting that Eq.\,\eqref{set01} implies an explicit function of the matter density as a function of the fluid pressure and the shift vector. 

Regarding Einstein equation solutions, the anisotropic fluid also leads to four possible solutions, which are named Case 1 and Case 2, with the following conditions stated below
\be
\text{Case} 1) \ \ \frac{\partial\beta}{\partial z} = 0,
\ee
\be
\text{Case} 2) \ \ \frac{\partial\beta}{\partial y} = 0,
\ee
Each of these cases leads to two subcases with different conditions and results, which will be analyzed below. The relevant Einstein equations components are expressed by 
\be
\frac{\partial^2 \beta}{\partial x \partial y} = 0,
\label{set02}
\ee
\be
\frac{\partial^2 \beta}{\partial x \partial z}  = 0,
\label{set03} 
\ee
\be
\left(\frac{\partial \beta}{\partial y}\right)^2 + \left(\frac{\partial
\beta}{\partial z}\right)^2 = - \frac{32 \pi}{3} A,
\label{set04} 
\ee
\be
\left(\frac{\partial \beta}{\partial y}\right)^2 - \left(\frac{\partial
\beta}{\partial z}\right)^2 = 16 \pi (C-B),
\label{set05} 
\ee
\be
2 \frac{\partial \beta}{\partial y} \, \frac{\partial \beta}{\partial x} + \frac{\partial^2 \beta}{\partial t \partial y} = 0,
\label{set06}
\ee
\be
2 \frac{\partial \beta}{\partial z} \, \frac{\partial \beta}{\partial x} + \frac{\partial^2 \beta}{\partial t \partial z} = 0,
\label{set07}
\ee
\be
\frac{\partial \beta}{\partial z} \, \frac{\partial \beta}{\partial y} = 0,
\label{set08}
\ee
\be
\frac{\partial}{\partial x}\left[ \frac{\partial \beta}{\partial t} + \frac{1}{2}\frac{\partial}{\partial x}(\beta^2) \right] = - 32 \pi (B + C),
\label{set09}
\ee
\be
\frac{\partial^2 \beta}{\partial y^2} + \frac{\partial^2 \beta}{\partial z^2} = 16 \pi \beta (A-D).
\label{set10}
\ee
Eq.\,\eqref{set08} implies two conditions on the values for $\beta$,
\begin{align}
\frac{\partial \beta}{\partial z} &= 0, \\  
\frac{\partial \beta}{\partial y} &= 0,
\end{align}
which results in two solutions, cases 1 and 2, respectively. Starting with Case 1, and Eq.\,\eqref{set02} implies that either 
\begin{align}
\frac{\partial\beta}{\partial x} &= 0, \\  \frac{\partial\beta}{\partial y} &= 0 \,,
\end{align}
resulting in Case 1a and Case 1b. For Case 2, and from Eq.\,\eqref{set03}, it follows that
\begin{align}
\frac{\partial\beta}{\partial x} &= 0, \\ 
\frac{\partial\beta}{\partial z} &= 0,
\end{align}
resulting in cases 2a and 2b. All four cases will be further analyzed as follows.

\ni\textbf{Case 1a:} $\left[\displaystyle{\frac{\partial \beta}{\partial z} = 0} \ \text{and} \ \displaystyle{\frac{\partial\beta}{\partial x} = 0}\right]$ 

For this case, the leading equations are taken from Eq.\,\eqref{set02} to Eq.\,\eqref{set10}, and after inclusion of the results stated for Case 1a, the leading equations become
\be 
\beta^2 (2D - A - p) + \frac{A}{3} = \mu,
\ee
\be
B = - C = \frac{A}{3}\,,
\ee
\be
\frac{\partial^2 \beta}{\partial y^2} = 16 \pi \beta (A-D),
\ee
\be
\left(\frac{\partial \beta}{\partial y}\right)^2 = - 32 \pi C,
\ee
and the energy-momentum tensor becomes, in matrix form, the following
\be
T_{\alpha \sigma} = 
\begin{pmatrix} 
\beta^2 (2D-A) + A/3 & - \beta D & 0  & 0  \\ 
- \beta D     & A         & 0  & 0  \\ 
0             & 0         & A/3  & 0  \\ 
0             & 0         & 0  & -A/3 
\end{pmatrix}.
\label{emtc1a}
\ee   
The sign of $T_{00}$ depends on how the shift vector relates to other anisotropic fluid pressures. Assuming that the energy density is negative, $T_{00} < 0$, requires that
\be
\beta^2 > \frac{A}{3(A-2D)}\,,
\label{ineq1b}
\ee
with the constraints that $A \neq 2D$ to avoid points where we do not have fixed the inequality. If $A > 0$, and $A > 2D$ or $A < 0$, and $A < 2D$, then $\beta$ can be considered of real-valued form. Otherwise, $\beta$ shall be considered complex-valued for the inequality in Eq.\,\eqref{ineq1b} to be analyzed. For superluminal cases, and imposing that $\beta^2 > 1$ in Eq.\,\eqref{ineq1b}, 
\be
\frac{A}{3(A-2D)} > 1,
\ee
it is necessary that $A < 3 D$. Defining a parameter $k^2$, which expresses how much greater than the speed of light $\beta$ is,  via the following equation
\be
k^2 \equiv \frac{A}{3(A-2D)},
\ee
gives a relation between $A$ and $D$, described by the following expression
\be
A =  \frac{6k^2}{3k^2 - 1} D.
\label{express1}
\ee
So, for example, if the shift vector equals the speed of light $(k = 1)$, then 
\be
A = 3 D \,.
\ee 
In the limit of a very large value of the shift vector $\beta$ in the superluminal regime, where the factor $k \to \infty$, then Eq.\,\eqref{express1} has the following limit value
\be
\lim_{k \to \infty} A = 2 D.
\ee
Hence, even for unbounded values of the shift-vector, in the superluminal regime, the anisotropic fluid presents bounded relative values between its pressures $A$ and $D$. Remember that those expressions are valid only considering negative values for $T_{00}$. Next, the second solution to the Einstein equations is presented.

\ni\textbf{Case 1b:} $\left[\displaystyle{\frac{\partial \beta}{\partial z} = 0} \ \text{and} \ \displaystyle{\frac{\partial\beta}{\partial y} = 0}\right]$ 

For Case 1b, where $\partial \beta/\partial z = 0$, and the condition $\partial\beta/\partial y = 0$ is required, the equations that determine this solution to the Einstein equations are
\be
\mu = - \beta^2 p,
\ee
\be
B = C,
\ee
\be
A = 0
\ee
\be
\frac{\partial}{\partial x}\left[ \frac{\partial \beta}{\partial t}
+ \frac{1}{2}\frac{\partial}{\partial x}(\beta^2) \right] = - 64 \pi B \,. 
\label{c1bset04}
\ee
Case 1b is similar to the vacuum solution found for dust in Chap.\,4 and to the solution found for the perfect fluid, discussed in the last chapter. The difference is, in this case, for it to be considered a vacuum solution, it is necessary that $B = 0$, otherwise, it leads to the unphysical solution 
\be
T_{\alpha \sigma} = 
\begin{pmatrix} 
0 & 0 & 0  & 0  \\ 
0 & 0  & 0  & 0  \\ 
0 & 0  & B  & 0  \\ 
0 & 0  & 0  & B 
\end{pmatrix},
\label{emtc1b}
\ee
where there are only two components of static pressure equal to $B$. We have no flow of momentum $D = 0$, and one of the static pressures is zero, $A = 0$. An equation of state $p = -\frac{1}{\beta^2}\mu$ may lead to negative pressure $p$ when positive matter density is considered. No energy density exists since $\rho = T_{\mu\nu} u^\mu u^\nu = 0$. Case 2 unfolds in two subcases, 
\be
2a) \ \ \frac{\partial\beta}{\partial y} = 0 \ \ \text{and} \ \
\frac{\partial\beta}{\partial x} = 0,
\ee
\be
2b) \ \ \frac{\partial\beta}{\partial y} = 0 \ \ \text{and} \ \
\frac{\partial\beta}{\partial z} = 0.
\ee

\ni\textbf{Case 2a:} $\left[\displaystyle{\frac{\partial \beta}{\partial y} = 0} \ \text{and} \ \displaystyle{\frac{\partial\beta}{\partial x} = 0}\right]$ 

For Case 2a, the leading equations are
\be
\beta^2(2D - A - p) + \frac{A}{3} = \mu,
\ee
\be
\left(\frac{\partial \beta}{\partial z}\right)^2 = - \frac{32 \pi}{3} A,
\ee
\be
A = 3B = -3C,
\ee
\be
\frac{\partial^2 \beta}{\partial z^2} = 16 \pi \beta (A-D).
\ee
The energy-momentum tensor, in matrix form, for this case, is 
\be
T_{\alpha \sigma} = 
\begin{pmatrix} 
\beta^2 (2D-A) + A/3 & - \beta D & 0  & 0  \\ 
- \beta D     & A         & 0  & 0  \\ 
0             & 0         & - A/3  & 0  \\ 
0             & 0         & 0  & A/3 
\end{pmatrix},
\label{emtc2a}
\ee
these results are similar to Case 1a (see Eq.\,\eqref{emtc1a} for comparison). The conditions for negative values of energy density, $T_{00} < 0$, are the same for the ones found for Case 1a, 
\be
\beta^2 > \frac{A}{3(A-2D)}.
\ee
Superluminal speed can be attained under the condition
\be
A < 3D \,,
\ee
and for the solutions where the shift vector $\beta$ is real-valued, it is necessary that
\be
A < 2D < 0.
\ee
Again, $A = 2D$ is a singular value when analyzing the sign of $T_{00}$. Case 2a is very similar to Case 1a, with the difference that for Case 1a, the shift vector is a function of $(t,y)$, and for Case 2a, the shift vector is a function of $(t,z)$. Also, the form of the energy-momentum tensor presents a sign exchange between pressures $B$ and $C$ in both solutions. For Case 1a, 
\be
B = A/3 = - D\,,
\ee
and for Case 2a 
\be
B = - A/3 = - D.
\ee
Next, solution 2b is discussed.

\ni\textbf{Case 2b:} $\left[\displaystyle{\frac{\partial \beta}{\partial y} = 0} \ \text{and} \ \displaystyle{\frac{\partial\beta}{\partial z} = 0}\right]$ 

The equations that solve Einstein equations are identical to the equations found for Case 1b
\be 
- p \beta^2 = \mu,
\ee 
\be
A = D = 0,
\ee
\be
B = C,
\ee 
\be
\frac{\partial}{\partial x}\left[ \frac{\partial \beta}{\partial t} + \frac{1}{2}\frac{\partial}{\partial x}(\beta^2) \right] =  - 64 \pi B,
\label{aniso2b}
\ee
If the pressure $B = 0$ in Eq.\,\eqref{aniso2b}, then the same result found in Chap.\,4 is recovered, a vacuum solution to the Einstein equations connecting the warp drive to shock-waves via a Burgers-type equation. This is also the case for solution 1b.  

In summary, four solutions (1a, 1b, 2a, 2b) are obtained for the anisotropic fluid \index[topics]{anisotropic fluid} coupled with the warp drive spacetime. Cases 1a and 2a are similar, with the main difference being spacetime coordinate dependency for the shift vector. For Case 1a, the shift vector is a function of $(t,y)$ coordinates; for Case 2a, the shift vector is a function of $(t,z)$ coordinates. 

The values of pressure $A$ and the momentum flow $D$ are essential in determining whether the shift vector $\beta$ is of real or complex value for both Case 1a and Case 2a when considering the energy-momentum tensor component $T_{00}$ as negative. 

Cases 1a and 2a also share the same state equation for the anisotropic fluid but with opposing signs for the pressures $B$ and $C$. Cases 1b and 2b are identical and were dismissed as nonphysical due to the energy-momentum configuration, with zero energy density, and the state equation
\be
\mu = - \beta^2 p,
\ee
and only two non-zero components as static pressures. It is relevant to mention that even with more free parameters in the energy-momentum tensor, the anisotropic fluid was insufficient to generate more tangible physical solutions, with $\beta$ not being so restricted to regions of the warp drive. Still, the constraints in all the subcases suggest that the warp bubble may be constructed as a piecewise solution of Einstein equations.

Table \ref{table_summary} summarizes the Einstein equations solutions obtained for the anisotropic (parametrized) fluid and the warp drive.

\begin{table}[!ht]
\centering
\begin{tabular}{| m{3cm} | m{3cm} | m{7cm} |}
\hline 
Case & Conditions & Results \\ 
\hline 
\multirow{2}{*}{$1) \
\displaystyle{\frac{\partial \beta}{\partial z} = 0}$}
& 
$1a) \ \displaystyle{\frac{\partial \beta}{\partial x} = 0}$
&
$\begin{array} {ll} 
\\ [-10pt]
\displaystyle{\mu = \beta^2(2D-A-p) + \frac{A}{3}} \\ [7pt]
\beta = \beta(t,y)\\ [7pt]
\displaystyle{B = - C = \frac{A}{3}} \\ [7pt]
\displaystyle{\left(\frac{\partial \beta}{\partial y}
\right)^2 = 32 \pi C},  \\ [10pt]
\displaystyle{\frac{\partial^2\beta}{\partial y^2} = 16 \pi \beta (A-D)}
\\[7pt]
\end{array}$ \\
\cline{2-3}   
&
$1b) \ \displaystyle{\frac{\partial \beta}{\partial y} = 0}$
&
$\begin{array} {ll} 
\\ [-10pt]
\displaystyle{\mu = - \beta^2p}\\ [7pt]
\beta = \beta(t,x)\\ [7pt]
\displaystyle{B = C}  \\ [5pt]
\displaystyle{A = D = 0} \\ [7pt]
\displaystyle{\frac{\partial}{\partial x}\left[
\frac{\partial \beta}{\partial t} 
+ \frac{1}{2} \frac{\partial}{\partial x}(\beta^2)
\right] = - 64 \pi B} \\[8pt]
\rightarrow\mbox{solution \textit{dismissed} as nonphysical} \\
\end{array}$ \\
\hline

\multirow{2}{*}{$2) \
\displaystyle{\frac{\partial \beta}{\partial y} = 0}$}
& 
$2a) \ \displaystyle{\frac{\partial \beta}{\partial x} = 0}$
&
$\begin{array} {ll} 
\\ [-10pt]
\displaystyle{\mu = \beta^2(2D-A-p) + \frac{A}{3}} \\ [7pt]
\beta = \beta(t,z)\\ [7pt]
\displaystyle{B = - C = \frac{A}{3}}\\ [7pt]
\displaystyle{\left(\frac{\partial \beta}{\partial z}
\right)^2 = 32 \pi C}  \\ [9pt]
\displaystyle{\frac{\partial^2\beta}{\partial z^2} =
16 \pi \beta (A-D)}
\\[7pt]
\end{array}$ \\
\cline{2-3}   
&
$2b) \ \displaystyle{\frac{\partial \beta}{\partial z} = 0}$
&
$\begin{array} {ll} \\ [-10pt]
\displaystyle{\mu = - \beta^2p}\\ [7pt]
\beta = \beta(t,x)\\ [7pt]
\displaystyle{B = C}  \\ [5pt]
\displaystyle{A = D = 0} \\ [7pt]
\displaystyle{\frac{\partial}{\partial x}\left[
\frac{\partial \beta}{\partial t} 
+ \frac{1}{2} \frac{\partial}{\partial x}(\beta^2)
\right] = - 64 \pi B} \\[8pt]
\rightarrow\mbox{solution \textit{dismissed} as nonphysical} 
\end{array}$ \\  
\hline 
\end{tabular}
\caption{Summary of all solutions of the Einstein equations for the warp drive metric with the anisotropic fluid as source.}
\label{table_summary}
\end{table}

\subsection{Energy conditions}

This section shows the calculation for all energy conditions for the anisotropic fluid coupled to the warp drive spacetime. The inequalities are calculated considering all four solutions (1a, 1b, 2a, 2b) found in the previous section. Expressions connecting the matter density $\mu$, pressures $A, B, C, D$, and the shift vector are found, and the possibility that the warp drive does not violate the energy conditions is discussed. Since the anisotropic fluid has four different pressures and momentum flow, the energy conditions inequalities present more options regarding their fulfillment. For Cases 1a and 2a, lower and upper bounds are found for the shift vector to satisfy the dominant energy condition. The weak energy condition is satisfied for these cases if $A \geq 0$. The dominant energy condition also requires $A \geq 0$, but with the addition of an upper bound for the shift vector. The strong energy condition is not violated as long as $A \geq 0$. The null energy condition is not violated if the shift vector does not assume specific values in a limited range, depending on the pressures. Since Cases 1b and 2b are vacuum solutions to the Einstein equations when $B = 0$, the energy conditions are immediately satisfied for this scenario. The strong energy condition is only satisfied if $B = 0$ for cases 1b and 2b.

\subsubsection{Weak energy condition} 

Considering the Eulerian observers with 4-velocity given by
\be
u^\alpha = (1, \beta, 0, 0), \ \ 
u_\alpha = (- 1,0,0,0),
\ee
and the anisotropic fluid energy-momentum tensor shown in Eq.\,\eqref{emtanisofluid} yields 
\be
\rho = T_{\alpha \sigma} \, u^\alpha u^\sigma = \mu + \beta^2(p - 2 D + A). 
\label{eq88}
\ee
For cases 1a and 2a shown in Table \ref{table_summary}, the state equation for the anisotropic fluid can be found using Eq.\,\eqref{eqstaterel}, being equal to  
\be
\mu = \beta^2(2D - A - p) + \frac{A}{3}
\label{eqstateaux}
\ee
inserting the equation of state in Eq.\,\eqref{eqstateaux} into Eq.\,\eqref{eq88} leads to the following expression
\be
T_{\alpha\sigma}\, u^\alpha u^\sigma=\frac{A}{3} \geq 0.
\label{wec}
\ee
For the weak energy condition to be satisfied, the pressure $A$ must be positive, and for real-valued forms for $\beta$ as shown in the condition given by Eq.\,\eqref{ineq1b} means that $A > 2D$. There is a superluminal regime with weak energy conditions being satisfied if
\be
2D < A < 3D.
\ee 
If Eq.\,\eqref{wec} holds true, a condition for positive matter density $\mu$ is given by
\be
A + p \leq 2D.
\ee
For cases 1b and 2b, where a Burgers-type equation solves the Einstein equations, as shown in Table \ref{table_summary}, the weak energy condition is trivially satisfied since $A = D = 0$.

\subsubsection{Dominant energy condition}

The first condition of the dominant energy condition calculated for the anisotropic fluid energy-momentum tensor is
\be
T^{\alpha \beta} \, u_\alpha u_\beta = 
T^{00} = \beta^2(A - 2D + p) + \mu \geq 0,
\ee 
the second inequality $F^\alpha F_\alpha \leq 0$ yields
\begin{align}
\nonumber (T^{\alpha \beta} \, u_\beta) (T_{\alpha \beta} \, u^\beta)
=& - \mu^2 - A^2 \beta^4 - \beta^4 p^2 + A^2 \beta^2 
- (4 \beta^4 - \beta^2) D^2
\\
\nonumber &+2 (2 A \beta^4 - A \beta^2) D - 2 (A \beta^2 - 2 \beta^2 D)
\mu 
\\ 
&-2 (A \beta^4 - 2 \beta^4 D + \beta^2 \mu) p \leq 0 \,.
\label{anisodec}
\end{align}
Inserting Eq.\,\eqref{anisodec} in the equations found for cases 1a and 2a, see Table \ref{table_summary}, results in the following expression for the first and second conditions to the dominant energy
\be
T^{\alpha \beta} \, u_\alpha u_\beta = \frac{A}{3} \geq 0,
\label{anisodec1}
\ee
\be
(T^{\alpha \beta} \, u_\beta) (T_{\alpha \beta} \, u^\beta) =
\left[(A-D)\beta+\frac{A}{3}\right]\left[(A-D)\beta-\frac{A}{3}
\right] \leq 0.
\label{anisodec2}
\ee
Eq.\,\eqref{anisodec1} is the same as the weak energy condition from Eq.\,\eqref{wec}. Eq.\,\eqref{anisodec2} then reads
\be
\left|\beta\right| \leq \frac{1}{3} \left|\frac{A}{D-A}\right|.
\label{dec5}
\ee
Eq.\,\eqref{dec5} means that $\beta$ possesses an upper bound, but it is possible that $\beta > 1$. The dominant energy condition is trivially satisfied for cases 1b and 2b since $A = D = 0$.

\subsubsection{Strong energy condition}

Using the general formula for the anisotropic fluid energy-momentum tensor and the formula for the strong energy condition, the following expression is obtained 
\be
\left(T_{\alpha\sigma}-\frac{1}{2}T \, g_{\alpha\sigma}\right) 
u^\alpha u^\sigma = \frac{2A}{3} + \frac{(B+C)}{2}  \geq 0.
\label{secaaa}
\ee
Using the expressions and limiting conditions for the energy-momentum tensor found for Case 1a and Case 2a, see Table \ref{table_summary}, and inserting those conditions in Eq.\,\eqref{secaaa} results in the following expression
\be
\left(T_{\alpha \sigma} - \frac{1}{2}T \, g_{\alpha \sigma} \right) 
u^\alpha u^\sigma = \frac{2}{3}A  \geq 0.
\ee
For the strong energy condition to be satisfied, it is only necessary that the pressure $A$ be non-negative. This is an interesting result since no assumption was made regarding the sign of $T_{00}$ or the necessity for negative matter density. Since Case 1b and Case 2b have null values for pressures $A$ and $D$ and $B = C$, the strong energy condition is immediately satisfied for these two solutions when $B = 0$.

\subsubsection{Null energy condition}

For the null energy condition, firstly, it is necessary to find a general null vector for the Eulerian observers, so considering a 4-velocity of the form
\be
k^\alpha = (a,b,0,0),    
\label{neck1}
\ee
and imposing the null condition for the 4-velocity, $k_\alpha k^\alpha = 0$, the following expression relates the two parameters $a, b$ of the 4-velocity for this observer
\be
\frac{a}{b} = \frac{1}{\beta \pm 1}.
\label{necroots}
\ee
Then, inserting the above expression in the 4-velocity $k_\alpha$ and contracting the energy-momentum tensor to find the expression for the null energy condition results in 
\be
T_{\alpha \sigma} \, k^\alpha k^\sigma =
a^2 \beta^2 p - 2 a b \beta D + b^2 A + a^2 \mu \geq 0.
\label{820}
\ee
Substituting Eq.\,\eqref{necroots} into Eq.\,\eqref{820} results in the following expressions regarding the anisotropic pressures $A, B, C$, the flow of momentum $D$, the intrinsic static pressure $p$, the shift vector $\beta$, and the two parameters $a, b$
\be
\frac{b^2}{(\beta + 1)^2} \, \left[\mu - \beta^2(2D - A - p) + 2\beta(A-D) + A\right] \geq 0,
\label{necin1}
\ee
\be
\frac{b^2}{(\beta - 1)^2} \, \left[\mu - \beta^2(2D - A - p) - 2\beta(A-D) + A\right] \geq 0.
\label{necin2}
\ee
Considering the equation of state for the anisotropic fluid found in Case 1a and Case 2a
\be
\mu = \beta^2 (2D - A - p) + \frac{A}{3},
\ee
and with the conditions found for the pressures of the anisotropic fluid solutions in Case 1a and Case 2a, see Table \eqref{table_summary}, the results in the following conditions
\be
\frac{b^2}{(\beta+1)^2}\,\left[\frac{4A}{3}+2\beta(A-D)\right]\geq 0,
\label{anisonec1}
\ee
\be
\frac{b^2}{(\beta-1)^2}\,\left[\frac{4A}{3}-2\beta(A-D)\right]\geq 0,
\label{anisonec2}
\ee
Eq.\,\eqref{anisonec1} and Eq.\,\eqref{anisonec2} can be simplified into the following conditions
\be
- \frac{2}{3} \frac{A}{D-A} \leq \beta \leq \frac{2}{3} \frac{A}{D-A},
\label{null}
\ee
\be
\beta \neq \pm 1,
\ee
with the constraint that the shift vector cannot assume the exact light speed $(\beta \neq \pm 1)$, since this is a singular value in equations \eqref{anisonec1} and \eqref{anisonec1}, which makes these inequalities not analytical. Also, there is a limiting condition associated with the anisotropic pressure $A$ and the momentum flow $D$ since they cannot assume the same value and since this is a singular value in Eq.\,\eqref{null}. For Case 1b and Case 2b, $A = D = 0$ implies that the null energy condition is immediately satisfied for these two solutions of Einstein equations for the anisotropic fluid and the warp drive. 

Table \ref{tab3ppf} summarizes the results of all energy conditions for the anisotropic fluid energy-momentum tensor coupled to warp drive. For the case solutions, refer to Table \ref{table_summary}

\begin{table}[h]
\centering
\begin{tabular}{| m{2.5cm} | m{3.5cm} | m{6.0cm} |}
\hline 
Cases & Energy Conditions & Results \\ \hline 
\multirow{4}{*}{1a and 2a} 
& Weak &
$\begin{array} {ll} \\ [-6pt]
\displaystyle{A \geq 0 } \\[8pt]
\end{array}$ \\[8pt]
\cline{2-3}   
& Dominant &
$\begin{array} {ll} \\[-6pt]
\displaystyle{A \geq 0} \ \ \text{and} \ \ 
\displaystyle{\left|\beta\right| \leq \frac{1}{3}
\left|\frac{A}{D-A}\right|} \\[12pt]
\end{array}$ \\[8pt]
\cline{2-3}   
& Strong &
$\begin{array} {ll} \\[-6pt]
\displaystyle{A} \geq 0 \\[8pt]
\end{array}$ \\[8pt]
\cline{2-3} 
& Null &
$\begin{array} {ll} \\[-6pt]
\displaystyle{\beta \geq \frac{2}{3} \frac{A}{D - A} \ \ \text{and}
\ \ \beta \neq - 1} \\[8pt]
\displaystyle{\beta \leq \frac{2}{3} \frac{A}{A - D} \ \ \text{and}
\ \ \beta \neq 1} \\[8pt]
\end{array}$ \\[8pt]
\hline
\multirow{4}{*}{1b and 2b} 
& Weak &
$\begin{array} {ll} \\[-6pt]
\text{immediately satisfied} \\ [8pt]
\end{array}$ \\ [8pt]
\cline{2-3} 
& Dominant &
$\begin{array} {ll} \\[-6pt]
\text{immediately satisfied.} \\ [8pt]
\end{array}$ \\ [2pt]
\cline{2-3}   
& Strong &
$\begin{array} {ll} \\[-6pt]
B = 0  \\ [2pt]
\end{array}$ \\[8pt]
\cline{2-3} 
& Null &
$\begin{array} {ll} \\[-6pt]
\text{immediately satisfied.} \\ [8pt]
\end{array}$ \\ [8pt]
\cline{2-3}
\hline
\end{tabular}
\caption{Summary of all energy conditions results for the anisotropic fluid energy-momentum tensor coupled to the warp drive spacetime.}
\label{tab3ppf}
\end{table}

It is possible to find solutions to Einstein equations for the warp drive coupled to a source of energy-momentum tensor that satisfy the energy conditions for particular classes of observers, even for negative values of $T_{00}$. Cases 1b and 2b are vacuum solutions when the anisotropic pressure $B = 0$, and all of the energy conditions are immediately satisfied. 

The Strong Energy Condition is only satisfied if $B = 0$ for cases 1b and 2b. Considering the other two solutions, Case 1a and Case 2a, lower and upper bounds are found for the shift vector to satisfy the dominant energy condition. There are also limiting conditions, such as $A \neq D$ and $A \geq 0$. The weak energy condition is satisfied for cases 1a and 2a if $A \geq 0$. The Strong Energy Condition is not violated as long as $A \geq 0$. The null energy condition is not violated if the shift vector does not assume specific values in a limited range depending on the pressure $A$ and the momentum flow $D$. Also, for the null energy condition, the shift vector cannot assume the exact speed of light $(\beta \neq \pm 1)$. In light of this result for the anisotropic fluid and the one found for the perfect fluid, it is clear that better validity conditions are necessary when stating limiting theorems regarding the violation of energy conditions for classes of warp drive. Energy conditions are not violated if one considers vacuum solutions for the warp drive. 

\subsection{Divergence of energy-momentum tensor}

This section imposes the null divergence for the anisotropic fluid energy-momentum tensor via the following expression
\be
{T^{\alpha \sigma}}_{;\sigma} = 0\,,
\ee 
as the null divergence for the energy-momentum tensor results in the following 
\begin{align} 
\nonumber
- \frac{\partial \beta}{\partial x} (D + \mu) 
- \frac{\partial \mu}{\partial t} 
- \beta \left[\frac{\partial D}{\partial x} 
+ \frac{\partial \mu}{\partial x} 
+ \frac{\partial \beta}{\partial t}(2p + A - 3D)\right]& \\ 
+ \beta^2\left[\frac{\partial D}{\partial t} 
- \frac{\partial p}{\partial t} 
+ 3\frac{\partial \beta}{\partial x}(D-p)\right]
+ \beta^3 \left(\frac{\partial D}{\partial x} 
- \frac{\partial p}{\partial x}\right)&= 0,
\label{divT0} \\[2pt]
\frac{\partial A}{\partial x} 
+ \frac{\partial \beta}{\partial t} (D-A)
+ \beta \left[ 3 \frac{\partial \beta}{\partial x} (D-A)
+ \frac{\partial D}{\partial t} 
- \frac{\partial A}{\partial t}\right] 
+ \beta^2 \left(\frac{\partial D}{\partial x}
- \frac{\partial A}{\partial x}\right) &= 0,
\label{divT1} \\[2pt]
\frac{\partial B}{\partial y} +
\beta \frac{\partial \beta}{\partial y} (D-A) &= 0,
\label{divT2} \\[2pt]
\frac{\partial C}{\partial z} +
\beta \frac{\partial \beta}{\partial z} (D-A) &= 0.
\label{divT3}
\end{align}
The expressions above will be further analyzed considering the conditions found for the Einstein equation solutions for the anisotropic fluid and the warp drive. The conditions used in this analysis are found for Cases 1a, 1b, 2a, and 2b; see Table \eqref{table_summary}.

\ni\textbf{Case 1a: $\bm{\left[\displaystyle\frac{\partial \beta}
{\partial z} = 0\right.}$ and $\bm{\left.\displaystyle\frac{\partial
\beta}{\partial x} = 0\right]}$}

\ni Eq.\,\eqref{divT0} to Eq.\,\eqref{divT3} are reduced to the following expressions regarding the conditions and results found for Case 1a 
\begin{align}
\nonumber
- \frac{\partial \mu}{\partial t} - \beta \left[\frac{\partial D}
{\partial x}+\frac{\partial \mu}{\partial x}+\frac{\partial \beta}
{\partial t}(2p + A - 3D)\right]+\beta^2 \left(\frac{\partial D}
{\partial t}-\frac{\partial p}{\partial t}\right)& \\
+ \beta^3 \left(\frac{\partial D}{\partial x} - \frac{\partial p}
{\partial x}\right)&= 0,
\label{npf1adivT0} \\[2pt]
\frac{\partial A}{\partial x}+\frac{\partial \beta}{\partial t}(D-A)
+ \beta \left(\frac{\partial D}{\partial t}-\frac{\partial A}
{\partial t} \right)+\beta^2 \left(\frac{\partial D}{\partial x}
- \frac{\partial A}{\partial x}\right)&= 0, \label{npf1adivT1} 
\end{align}
\be
\frac{\partial B}{\partial y} + \beta \frac{\partial \beta}{\partial y} (D-A)= 0,
\label{npf1adivT2}
\ee
\be
\frac{\partial C}{\partial z}= 0.
\label{npf1adivT3}
\ee
These four equations and the results found for Case 1a, shown in Table \ref{table_summary}, mean an over-determined system for the anisotropic fluid six parameters $\mu, p, A, B, C, D$ with more limiting and boundary conditions than the ones found for the perfect fluid. The only immediate condition is that the anisotropic pressures $B = - C$ do not depend on the $z$ coordinate, from Eq.\,\eqref{npf1adivT3}. The other partial equations obtained from the null divergence imposition must be solved analytically or numerically, considering the expressions found for the Einstein equations for Case 1a to fully determine if the warp drive is physical and does not violate energy conditions. Those equations are not fully solved in this work. Next, due to its similarity to Case 1a, Case 2a is discussed.

\ni\textbf{Case 2a: $\bm{\left[\displaystyle\frac{\partial \beta}
{\partial y} = 0\right.}$ and $\bm{\left.\displaystyle\frac{\partial
\beta}{\partial x} = 0\right]}$}

\ni Eqs\,\eqref{divT0} to \eqref{divT3} simplify to the following
expressions
\begin{align}
\nonumber
- \frac{\partial \mu}{\partial t}-\beta \left[\frac{\partial D}
{\partial x}+\frac{\partial \mu}{\partial x}+\frac{\partial \beta}
{\partial t}(2p + A - 3D)\right]& \\
+ \beta^2\left(\frac{\partial D}{\partial t}-\frac{\partial p}
{\partial t}\right)+\beta^3 \left(\frac{\partial D}{\partial x} 
- \frac{\partial p}{\partial x}\right)&= 0,
\label{npf2adivT0} 
\end{align}
\be
\frac{\partial A}{\partial x}+\frac{\partial \beta}{\partial t}(D-A)
+ \beta \left(\frac{\partial D}{\partial t}-\frac{\partial A}
{\partial t} \right)+\beta^2 \left(\frac{\partial D}{\partial x}
- \frac{\partial A}{\partial x}\right)= 0,
\label{npf2adivT1} 
\ee
\be
\frac{\partial B}{\partial y} = 0,
\label{npf2adivT2}
\ee
\be
\frac{\partial C}{\partial z}+\beta \frac{\partial \beta} {\partial z} (D-A) = 0 \,.
\label{npf2adivT3}
\ee
 Case 2a is similar to Case 1a, but has  $\beta = \beta(t,z)$ and Case 1a has $\beta = \beta(t,y)$. All of the above equations must be solved analytically or numerically.

\ni\textbf{Cases 1b and 2b: $\bm{\left[\displaystyle\frac{\partial \beta} {\partial y} = 0\right.}$ and $\bm{\left.\displaystyle\frac{\partial \beta}{\partial z} = 0\right]}$}

\ni Considering Case 1b and Case 2b, where $A=D=0$, $B=C$, and $p =- \mu/\beta^2$ leads to 
\begin{align}
-\frac{\partial \mu}{\partial t}-\beta \left(\frac{\partial \mu}
{\partial x}+2p \frac{\partial \beta}{\partial t}\right)-\beta^2
\frac{\partial p}{\partial t}-\beta^3 \frac{\partial p}{\partial x}
&= 0,
\label{newarp driveivT0}
\\[2pt]
\frac{\partial B}{\partial y}&= 0,
\label{newarp driveivT2}
\\[2pt]
\frac{\partial C}{\partial z}=\frac{\partial B}{\partial z}&= 0.
\label{newarp driveivT3}
\end{align}
Eq.\,\eqref{newarp driveivT2} and Eq.\,\eqref{newarp driveivT3} imply that $B$ does not
depend on $y$ nor $z$. Inserting the state equation $p =- \mu/\beta^2$ into Eq.\,\eqref{newarp driveivT0} yields
\be
p \beta^2 \frac{\partial \beta}{\partial x} = 0.
\ee
This expression can be separated into three subcases: $\beta = 0$,
$p = 0$, $\partial \beta/\partial x = 0$, which are discussed below

\ni\textbf{{Subcase 1:} $\left[\,\beta=0\,\right]$}

This condition reduces the warp drive metric into Minkowski. 
 
\ni\textbf{{Subcase 2:} $\left[\,B=0\,\right]$}

This means that all components of the anisotropic fluid energy-momentum tensor are zero, and even though there remains an equation of state $\mu = - \beta^2 p$ remains, all the components are zero, leading to the vacuum solution found for the dust solution \cite{nos1}.
 
\ni\textbf{{Subcase 3:} $\left[\,p=0\,\right]$}

Considering the equation of state $\mu = - \beta^2 p$ then the
matter density is also zero, leading to the vacuum solution found in Chap.\,4 for dust.

\subsection{Discussion}

Considering the perfect fluid as a source for the warp drive raised the need for a more complex form of energy momentum that could create a stable warp bubble with apparent negative energy density. The anisotropic fluid was an attempt, with four free parameters $A, B, C, D$, to solve the Einstein equations with more degrees of freedom. Those parameters are the anisotropic pressure of the fluid. Four solutions for the anisotropic fluid were found: Case 1 (1a and 1b) and Case 2 (2a and 2b). Cases 1a and 2a are the only physically plausible solutions for the anisotropic fluid as matter-energy source. 

Cases 1a and 2a satisfied all energy conditions if the pressure $A > 0$ and the shift vector $\beta$ is bounded by values $\pm 2 A/3(D-A)$, and it can not assume the values $\pm 1$ for the null energy condition. So, in this case, there is no luminal regime for the anisotropic perfect fluid; only subluminal and superluminal regimes are used. As for the dominant energy condition, the shift vector is bounded by $\pm A/3(D-A)$ values. So, if we want both Dominant and null energy conditions to be satisfied, the shift vector must respect the lower bound set by the dominant conditions. The shift vector only depends on $(t,y)$ coordinates for the Case 1a, and depends on $(t,z)$ for the Case 2a, despite this difference, the two sets of solutions are very similar in their form. Both depend on positive signs of the pressure $C$, and that $A > D$ for the shift vector to be a real-valued function. However, the solutions are not unique.

Cases 1b and 2b were dismissed as nonphysical due to the configuration of the energy-momentum tensor, with only two pressure components, zero energy density with a state equation given by $p =- \mu/\beta^2$. This is curious, since a positive static pressure would mean negative matter density in the same proportion, causing the energy density to be zero. The pressures $A$ and $D$ are both zero in this solution. The same Burgers-type equation was also found in this case. However, the right-hand side of the equation is no longer a function only of the time coordinate. Still, the pressure $B$ depends on the whole set of warp drive spacetime coordinates; in this case, the coordinates $y$ and $z$ would be constants since the shift vector only depends on $(t,x)$. In case $B = 0$, the result found in Chap.\,4 is recovered, connecting a vacuum solution to Einstein equations for the warp drive to shock-waves via a Burgers-type equation. All energy conditions are trivially satisfied for cases 1b and 2b, except the strong energy condition, which is satisfied only in vacuum, when the pressure $B$ equals zero.

\section{Charged dust}

This chapter analyzes the solutions of the Einstein equations with a charged dust energy-momentum tensor as source. Several regimes for different warp velocities are discussed. The Einstein equations with the cosmological constant are written, and all solutions with energy-momentum tensor components for the electromagnetic fields generated by charged dust are presented, as well as the respective energy conditions. 

The results show an interplay between the energy conditions and the electromagnetic field. In some cases, the former can be satisfied by both positive and negative matter density. In other cases, the Dominant and null energy conditions\index[topics]{null energy condition} are violated. A result connecting the electric energy density with the cosmological constant\index[topics]{cosmological constant} is also presented, as are the effects of the electromagnetic field on bubble dynamics.

One of the primary outcomes from this analysis is that, depending on the strength of the electric and magnetic field, it is possible to respect the energy conditions and obtain superluminal speed where $\beta > 1$. 

Since this energy-momentum tensor is quite cumbersome in finding simple solutions, a specific choice of the electromagnetic field is presented and analyzed. The main result is that the energy density for the warp drive might be negative, depending on the strength of the electric and magnetic fields, and that the matter density has to be negative. For subluminal and luminal regimes, it is also possible to obtain apparent negative energy density depending on the strength of the electromagnetic fields.

\subsection{The energy-momentum and Einstein tensor components}

Considering a warp drive spacetime background, the energy-momentum tensor components have extra $\beta$-correction terms compared to the Minkowski background. In this scenario, the warp drive interacts with electromagnetic fields \index[topics]{electromagnetic field} as seen from Eqs.\,\eqref{eemt00} to \eqref{eemt33}. The $T_{00}$ term given by Eq.\,\eqref{eemt00} is a $\beta$-fourth degree polynomial, where there are different regimes for the warp bubble speed, where $O(\beta)$ corrections can be subluminal, and luminal regimes, $O(\beta^2)$ and higher are superluminal regimes
\be
T_{00}(\beta) = \omega_{(0)} + \omega_{(1)} \beta + \omega_{(2)} \beta^2  + \omega_{(3)} \beta^3 + \omega_{(4)} \beta^4,
\label{t00energy}
\ee
where the coefficients $\omega_{(k)}$ are given by the expressions 
\be
\omega_{(0)} = \mu + \frac{1}{2}(B^2 + E^2),
\label{t00omega0}
\ee
\be
\omega_{(1)} = B_3 E_2 - B_2 E_3,
\label{t00omega1}
\ee
\be
\omega_{(2)} = \frac{1}{2}\left(B^2 - 2B_1^2 - E^2 - E_3^2\right),
\label{t00omega2}
\ee
\be
\omega_{(3)} = B_2 E_3 - B_3 E_2\,,
\label{t00omega3}
\ee
\be
\omega_{(4)} = \frac{1}{2}\left(E_2^2 + E_3^2\right).
\label{t00omega4}
\ee
For $\beta = 0$, the warp drive metric becomes Minkowski, and the only nonzero component of the $T_{00}$ term would be $\omega_{{}_0}$ as expected from Eq.\,\eqref{emtmatrix}. Assuming a weak warp drive that considers only shift vector $O(\beta)$ terms, the condition for $\beta$, where 
\be
T_{00} \approx \omega_{(0)} + \omega_{(1)} \beta, 
\label{betalow}
\ee
where the sign of the above equation depends on the shift vector and the electromagnetic field components. Assuming $T_{00} > 0$ in Eq.\,\eqref{betalow} leads to
\be
\beta < \frac{\mu}{B_2 E_3 - B_3 E_2}
+ \frac{1}{2}\frac{B^2 + E^2}{B_2 E_3 - B_3 E_2},
\label{t00lowbeta}
\ee
the shift vector has an upper bound limit, which can be very large if $B_2 E_3 \approx B_3 E_2$. Now, considering a usual warp drive, with higher orders of the shift vector $O(\beta^2)$, the conditions for $T_{00} > 0$ require that the terms $\omega_{(2)}$ and $\omega_{(3)}$ from Eqs.\,\eqref{t00omega2} and \eqref{t00omega3} be positive, so
\be
B_2^2 + B_3^2 \:\:>\:\: E_1^2 + E_2^2 + 2 E_3^2 + B_1^2.
\label{relfieldcomp2}
\ee
\be
B_2 E_3 - B_3 E_2 > 0\,,
\label{relfieldcomp3}
\ee
the term $\omega_{(4)}$ in Eq.\,\eqref{t00omega4} was neglected. Both Eqs\,\eqref{relfieldcomp2} and Eq.\,\eqref{relfieldcomp3} show connections between the electromagnetic field components in a way that $T_{00}$ be positive for the warp drive spacetime. After some algebraic manipulation of Einstein equations $G_{\mu\nu} = T_{\mu\nu}$, the following set of partial differential equations are obtained
\be
\frac{4}{3} \Lambda = T_{00} + 2 \beta T_{01} 
+ \left(\beta^2 - \frac{1}{3} \right)T_{11},	
\label{eqset1}
\ee
\be
\frac{1}{2}\frac{\partial^2 \beta}{\partial y^2} + 
\frac{1}{2}\frac{\partial^2 \beta}{\partial z^2} = 
T_{01} + \beta T_{11},
\label{eqset2}
\ee
\be
- \frac{3}{4}\left(\frac{\partial \beta}{\partial y}\right)^2 
- \frac{3}{4} \left(\frac{\partial \beta}{\partial z}\right)^2 
- \Lambda = T_{11},
\label{eqset3}
\ee
\be
\frac{1}{2}\left(\frac{\partial \beta}{\partial y}\right)^2 
- \frac{1}{2} \left(\frac{\partial \beta}{\partial z}\right)^2 
= T_{22} - T_{33},
\label{eqset4}
\ee
\be
- \frac{\partial}{\partial x} \left(\frac{\partial \beta}{\partial t}
+ \frac{1}{2} \frac{\partial}{\partial x} (\beta^2)\right) 
- 2 \Lambda = T_{33} + T_{22},
\label{eqset5}
\ee
\be
- \frac{1}{2}\frac{\partial^2 \beta}{\partial x \partial y} = 
T_{02} + \beta T_{12},
\label{eqset6}
\ee
\be
- \frac{1}{2}\frac{\partial^2 \beta}{\partial x \partial z} = 
T_{03} + \beta T_{13},
\label{eqset7}
\ee
\be
\frac{1}{2}\frac{\partial \beta}{\partial y}\frac{\partial \beta}{\partial z} = T_{23},
\label{eqset8}
\ee
\be
 - \frac{1}{4} \left(\frac{\partial\beta}{\partial y}\right)^2 
- \frac{1}{4} \left(\frac{\partial\beta}{\partial z}\right)^2 
+ \Lambda = T_{00} + 2 \beta T_{01} + \beta^2 T_{11}.
\label{eqset9}
\ee
The set of partial differential equations from Eq.\,\eqref{eqset1}
to Eq.\,\eqref{eqset9} are quite cumbersome to solve analytically, even considering the terms as constants. Next, the energy conditions are analyzed, and a particular configuration for the electromagnetic fields is proposed, as well as expressions relating the components of the electric and magnetic fields to the matter density and the shift vector. 

\subsection{Energy conditions} \label{engconds}

In this section, we will calculate the energy conditions for the \textit{electromagnetic energy-momentum tensor} (EEMT)\index[topics]{electromagnetic energy-momentum tensor} and check under what conditions they are satisfied, connecting the electromagnetic field components, the radiant matter density $\mu$, and the shift vector $\beta$. The Eulerian observers' 4-velocities were used to calculate the energy conditions
\be
u_\mu = (-1,0,0,0) \ \ , \ \ u^\mu = (-1,\beta,0,0),
\ee
which are indeed timelike since $u_\mu u^\mu = -1 < 0$, always.

\subsubsection{Weak energy conditions} 

For the weak energy condition, the electromagnetic energy-momentum tensor at each point of the spacetime must satisfy 
\be
T_{\alpha \sigma} \, u^\alpha u^\sigma \geq 0,    
\label{weccond}
\ee
where, for any timelike vector $\textbf{u} \, (u_\alpha u^\alpha < 0)$, and any zero vector $\textbf{k} \, (k_\alpha k^\alpha = 0)$, for an observer with unit tangent vector at a certain point of the spacetime, the local energy density measured by any observer is non-negative \cite{HawkingEllis1973}. For the electromagnetic energy-momentum tensor, the expression $T_{\alpha \sigma}\,u^\alpha u^\sigma$ is
\be
T_{\alpha \sigma} \, u^\alpha u^\sigma =  \frac{1}{2}\Big(E_2^2 + E_3^2\Big) \beta^2 - \Big(B_3 E_2 - B_2 E_3\Big) \beta + \frac{1}{2} \Big(B^2 + E^2\Big) + \mu.
\label{cdwec}
\ee
Eq.\,\eqref{cdwec} is a quadratic function of $\beta$. Imposing real-valued solutions for $T_{\alpha \sigma} u^\alpha u^\sigma \geq 0$ means that the discriminant of this quadratic inequality must be negative
\be
\Big(B_3 E_2 - B_2 E_3\Big)^2 - 2\Big(E_2^2 + E_3^2\Big)
\left[\frac{1}{2} \Big(B^2 + E^2\Big) + \mu\right] > 0,
\label{wec2}
\ee
which leads to
\be
0 < \mu < \frac{\left(B_3 E_2 - B_2 E_3\right)^2} {2\left(E_2^2 + E_3^2\right)} - \frac{1}{2}\Big(B^2 + E^2\Big),
\label{wec3}
\ee
which shows a condition between the electromagnetic components, 
\be
\Big(B_3 E_2 - B_2 E_3\Big)^2 > \Big(B^2 + E^2\Big)\Big(E_2^2 + E_3^2\Big),
\ee
the matter density must have a positive inferior minimum value since the right-hand side of Eq.\,\eqref{wec3} was assumed to be always positive. This tells us that the Weak Energy Condition may be satisfied by considering both positive and negative matter density. Solving Eq.\,\eqref{cdwec},
\be
\beta_\pm = \frac{B_3 E_2 - B_2 E_3}{E^2 - E_1^2} \pm \sqrt{\Bigg(\frac{B_3 E_2 - B_2 E_3}{E_2^2 + E_3^2}\Bigg)^2 -\frac{B^2+E^2+2\mu}{E_2^2 + E_3^2}},
\label{wec-sol}
\ee
where $\beta_- < \beta <\beta_+$, and considering a `weak' regime for the warp drive where we neglect $\beta^2 \approx 0$ in Eq.\ \eqref{cdwec}, then for the Weak Energy Condition to be valid, requires that
\be
\beta > \frac{1}{B_3 E_2 - B_2 E_3}\Bigg[\; \frac 12 \Big(B^2 + E^2\Big) + \mu \Bigg],
\label{wec4}
\ee
so the shift vector has a limiting minimum value condition. This is an interesting result since it implies that even for a lower order of $\beta$, the Weak Energy Condition does not limit the warp bubble speed. If $B_3 E_2 \rightarrow B_2 E_3$ the right-hand side of Eq.\,\eqref{wec4} assumes unlimited values, which could also be observed in low-strength electromagnetic fields and low density of radiant matter. 

\subsubsection{Dominant energy conditions}

For every timelike vector $u_a$, the following inequalities must be satisfied
\be
T^{\alpha \beta} u_\alpha u_\beta \geq 0, \quad \text{and} \quad F^\alpha  F_\alpha \leq 0, 
\ee
where $F^\alpha = T^{\alpha \beta} u_\beta$ is a non-spacelike vector, the local energy density appears non-negative, and the local energy flow vector is non-spacelike. On any orthonormal basis, the energy dominates the other components of the EMT,
\be
T^{00} \geq |T^{ab}|\,, \ \text{for each $a$, $b$}. 
\ee
Evaluating the first condition  $T^{\alpha \beta} u_\alpha u_\beta \geq 0$ for the dominant energy condition gives the same result as the weak energy condition seen in Eq.\,\eqref{wec2} and Eq.\,\eqref{wec3}, so this energy condition term can be satisfied by both positive and negative radiant matter density. Calculating the second condition for the dominant energy condition $F^\alpha F_\alpha$ leads to 
\be
F^\alpha  F_\alpha = \omega_{(4)} \beta^4 + \omega_{(3)} \beta^3 + \omega_{(2)} \beta^2 + \omega_{(1)} \beta + \omega_{(0)},
\label{cddec2}
\ee
where $\omega_{(0)},\ldots, \omega_{(4)}$ are implicit functions of spacetime coordinates and explicit functions in terms of the electromagnetic field components. The sub-indices of $\omega_{(k)}$'s are not tensor indices, and the terms $\omega_{(k)} \beta^k$ are not tensor contractions. The coefficients $\omega_{(k)}$ are
\be
\omega_{(4)} = \frac{1}{4}\left(E_2^2 + E_3^2\right)^2,
\ee
\be 
\omega_{(3)} = (E_2^2 + E_3^2)(E_2 B_3 - E_3 B_2),
\ee
\be
\omega_{(2)} = \frac{1}{2}(E_2^2 + E_3^2)(E^2 - B^2 - 2 \mu)
- (B_2 E_3 - B_3 E_2)^2,
\ee
\be 
\omega_{(1)} = (E_2 B_3 - E_3 B_2)(B^2 - E^2 + 2 \mu),
\ee
\begin{align}
\nonumber \omega_{(0)} &= - \frac{1}{4} (B^4 + E^2)
-  (E^2 + B^2) \mu
-  \mu^2
\\
\nonumber &- \frac{1}{2} (B_1 E_1 + B_2 E_2 + B_3 E_3)^2
+ \frac{1}{2} (B_1 E_2 - B_2 E_1)^2
\\
&+ \frac{1}{2} (B_1 E_3 - B_3 E_1)^2 
+ \frac{1}{2} (B_3^2 E_2^2 - B_2 E_3)^2.
\end{align}
Since $F^\alpha  F_\alpha$ is a fourth-order polynomial on $\beta$, it is rather challenging to calculate all the roots for this polynomial and find the general requirements for the dominant energy condition to be true. One way to impose the validity of this energy condition is to require that all the coefficients $\omega_{(k)}$ be of positive value.

\subsubsection{Strong energy conditions}

The strong energy condition is 
\be
\left(T_{\alpha \beta} - \frac{1}{2}T \, g_{\alpha \beta} \right) u^\alpha u^\beta \geq 0,
\label{sec}
\ee
for any timelike vector $u$. The strong energy condition in Eq.\,\eqref{sec} is
\be
 \left(E_2^2 + E_3^2\right) \beta^2 - 2 \left(B_3 E_2 - B_2 E_3\right) \beta + B^2 + E^2 + \mu \geq 0,
\label{sec2}
\ee
which is a quadratic $\beta$ inequation. For the strong energy condition to be satisfied, the discriminant of the above quadratic equation in $\beta$ must be negative,
\be
\left(B_3 E_2 - B_2 E_3\right)^2 - \left(E_2^2 + E_3^2\right)
\Big[B^2 + E^2 + \mu\Big] < 0\,,
\ee
which is similar to the expressions found for the weak and first dominant energy conditions, with a multiplication factor of 2
\be
\mu \leq \frac{\left(B_3 E_2 - B_2 E_3\right)^2}{\left(E_2^2 + E_3^2\right)} - \left(B^2 + E^2\right),
\label{sec3}
\ee
which is similar to the condition found in Eq.\,\eqref{wec3} for the weak energy condition. The matter density has an inferior bound, and it may be negative. But the strong energy condition could still be satisfied if $\mu$ is less than or equal to the right-hand side of the inequality in Eq.\,\eqref{sec3}.

\subsubsection{Null energy conditions}

The null energy conditions are satisfied for the null vector $\textbf{k}$ since the following inequality must be satisfied 
\be
T_{\alpha \sigma} k^\alpha k^\sigma \geq 0, \ \text{for any null 
vector} \ k^\alpha.
\ee
Assuming the following vector $k^\alpha = (a,b,0,0)$ where $a$ 
and $b$ can be determined by solving the equation $k_\alpha 
k^\alpha = 0$ for the null condition finding $a$ and $b$ respecting
\be
a = \frac{b}{\beta \pm 1}.
\label{abnullvecsolved}
\ee
so the calculation of $T_{\alpha \sigma} k^\alpha k^\sigma$ leads to
\be
T_{\alpha \sigma} \, k^\alpha k^\sigma =
T_{00} k^0 k^0 + 2 T_{01} k^0 k^1 + T_{11} k^1 k^1.
\label{nullcondcalc}
\ee
Substituting Eq.\,\eqref{abnullvecsolved}, $k^0 = a$, $k^1 = b$ into Eq.\,\eqref{nullcondcalc} results in
\be
T_{\alpha \sigma} \, k^\alpha k^\sigma = \omega_{(4)} \beta^4 + \omega_{(3)} \beta^3 + \omega_{(2)} \beta^2 + \omega_{(1)} \beta + \omega_{(0)},
\label{nec}
\ee
where the coefficients $\omega_{(k)}$ are also functions of $\beta$
\be
\omega_{(4)} = \frac{a^2}{2}(E_2^2 + E_3^2),
\ee
\be
\omega_{(3)} = a^2 (B_2 E_3 - B_3 E_2) - a b (E_2^2 + E_3^2),
\ee
\be
\omega_{(2)} = \frac{1}{2}\left[a^2 (B^2 - 2 B_1^2 - E_1^2) + 4 a b (B_3 E_2 - B_2 E_3) + (b^2 - 2 a^2) (E_2^2 + E_3^2)\right],
\ee
\be
\omega_{(1)} = a b (B_1^2 - B_2^2 - B_3^2 + E^2) + (a^2 - b^2) (B_3 E_2 - B_2 E_3),
\ee
\be
\omega_{(0)} = \frac{a^2}{2}(B^2 + E^2) + \frac{b^2}{2}(B^2 - 2B_1^2 + E^2 - 2 E_1^2) + 2 a b (B_2 E_3 - B_3 E_2) + a^2 \mu,
\ee
which are hard-working algebraic expressions concerning the null energy condition. The next step would be to solve them in a general way and to find the specific connections for the energy condition to be valid. Again, one simple way to analyze limiting conditions is to assume that the $\omega_{(k)}$ are constant and positive. We will not do such an analysis in this work. Instead, we will propose a particular scenario to simplify the analysis of Einstein equations, energy conditions, and the divergence of the electromagnetic energy-momentum tensor.

\subsection{Another expression of the energy-momentum tensor}

This section introduces a particular configuration of the electromagnetic energy-momentum tensor before analyzing the Einstein equations, where the electric and magnetic fields are orthogonal, \textit{i.e.} 
\be
E_i B^i = 0.
\label{plane-wave}
\ee
Assuming that $E_1$ is the only nonzero electric field component and $B_1 = 0$ from Eq.\,\eqref{plane-wave}. The nonzero components are 
\be
T_{00} = \mu + \frac{1}{2}(B^2 + E_1^2) + \frac{\beta^2}{2}\left(B^2 - E_1^2\right),
\label{simpeemt00}
\ee
\be
T_{01} = \frac{\beta}{2}\left(E_1^2 - B^2\right),
\label{simpeemt01}
\ee
\be
T_{02} = B_3 E_1, 
\label{simpeemt02}
\ee
\be
T_{03} = - B_2 E_1,
\label{simpeemt03}
\ee
\be
T_{11} = \frac{1}{2}\left(B^2 - E_1^2\right), 
\label{simpeemt11}
\ee
\be
T_{22} = \frac{1}{2} \left(B_3^2 - B_2^2 + E_1^2\right),
\label{simpeemt22} 
\ee
\be
T_{23} = - B_2 B_3,
\label{simpeemt23} 
\ee
\be
T_{33} =  \frac{1}{2}\left(B_2^2 - B_3^2 + E_1^2\right).
\label{simpeemt33}
\ee
Defining $B^2 = B_2^2 + B_3^2$ and $E^2 = E_1^2$ as the magnetic and electric field norms squared, we showed that both the weak and strong energy conditions could be satisfied in a general manner. Still, the dominant and null conditions require a hard-working calculation to show they are valid. We now proceed to analyze the energy conditions in this simplified scenario.

\subsubsection{Weak energy condition for the simplified choice}
 
Using the results obtained in the last section, together with the specific choice of the electromagnetic field, the weak energy condition from Eq.\,\eqref{cdwec} can be satisfied if 
\be
T_{\alpha \sigma} u^\alpha u^\sigma =  \frac{1}{2} \left(B^2 + E^2\right) + \mu \geq 0,
\label{simpwec}
\ee
The weak energy is still non-negative even with no radiant matter density. This energy condition could still be satisfied even with a negative matter density if the electromagnetic field strength $\left(B^2 + E_1^2\right)/2$ is larger than the matter density $\mu$ as seen in Eq.\,\eqref{simpwec}.

\subsubsection{Strong energy condition for the simplified choice}

From Eq.\,\eqref{sec2} the inequality simplifies to
\be
\left(T_{\alpha \beta} - \frac{1}{2}T g_{\alpha \beta} \right) u^\alpha u^\beta = B^2 + E^2 + \mu \geq 0,
\label{simpsec2}
\ee
since $B^2 + E^2$ is always positive, the strong energy condition is readily satisfied if the matter density $\mu$ is positive. For the negative matter density, the electromagnetic energy density $B^2 + E^2$ must be larger than $\mu$ to satisfy the strong energy condition. 

\subsubsection{Dominant energy condition for the simplified choice}

For the dominant energy inequality, the first condition $T^{\alpha \beta} u_\alpha u_\beta \geq 0$ gives the same result as the one from the weak energy inequality. With this simplified choice for the electromagnetic field, the second requirement for the dominant energy condition is
\be
F^\alpha  F_\alpha = -\mu^2 - (E^2 + B^2)\mu 
- \frac{1}{2} (E^2 - B^2)^2 - \frac{1}{2} E^2 B^2,
\label{simpdec2}
\ee
which is a quadratic function of the matter density $\mu$. The sign of this expression will depend on the value of the electromagnetic components. For real solutions, it is necessary  that
\be
2 E^2 B^2 \geq (E^2-B^2)^2.
\ee

\subsubsection{Null energy condition for the simplified choice}

Recovering the null energy condition from Eq.\,\eqref{nec}
\be
T_{\alpha \sigma} k^\alpha k^\sigma = \frac{a^2}{2}(B^2 - E^2)\beta^2 
+ a b (E^2 - B^2)\beta + \frac{1}{2}(a^2 + b^2)B^2 + \frac{1}{2}(a^2 - b^2)E^2 + a^2 \mu,
\label{simpnec}
\ee
neglecting that $a$ and $b$ are functions of $\beta$, using the ``ansatz" that Eq.\,\eqref{simpnec} is a quadratic function of $\beta$, and imposing that the discriminant of this equation must be positive or zero for real-valued solutions, the null energy condition might be satisfied for any $\beta$ if
\be
a^4 (E^2 - B^2)(\mu + B^2 + E^2) \geq 0.
\label{simpnec2}
\ee
considering $\mu > 0$, for the null energy condition to be satisfied, it is necessary that 
\be
E^2 - B^2 > 0 \,.
\ee 
Considering $\mu < 0$, there are two possible configurations
\be
E^2 + B^2 > \|\mu\| \, \ \text{and} \ E^2 > B^2,
\ee
\be
E^2 + B^2 < \|\mu\| \, \ \text{and} \ E^2 < B^2.
\ee

Table \ref{table:nonlin} summarizes all the requirements for the energy conditions to be valid simultaneously with the specific simplified choice of the electromagnetic field.

\begin{table}[h]
\caption{Summary for the energy conditions} 
\centering 
\begin{tabular}[c]{l @{\hspace{50pt}} l} 
\hline\hline 
Energy condition & Results \\ 
\hline\hline  \\[-12pt]
Weak         &  $\frac{1}{2} \left(B^2 + E^2\right) + \mu \geq 0$  \\ 
Strong       &  $B^2 + E^2 + \mu \geq 0$  \\
Dominant     &  $ E^2B^2\geq   (E^2-B^2)^2/2$ \\
Null         &  $ E^2 \geq B^2 \ , \ \mu>0$ \\ 
\hline 
\end{tabular}
\label{table:nonlin} 
\end{table}

For the simplified choice of the energy-momentum tensor, the new set of equations from Eq.\,\eqref{eqset1} to Eq.\,\eqref{eqset9} are
\be
\frac{4}{3} \Lambda = \mu + \frac{1}{3} B^2 + \frac{2}{3}E_1^2,	
\label{eqsetsimp1}
\ee
\be
\frac{1}{2}\frac{\partial^2 \beta}{\partial y^2} + \frac{1}{2}\frac{\partial^2 \beta}{\partial z^2} = 0,
\label{eqsetsimp2}
\ee
\be
- \frac{3}{4}\left(\frac{\partial \beta}{\partial y}\right)^2 - \frac{3}{4} \left(\frac{\partial \beta}{\partial z}\right)^2 - \Lambda =  B^2 - E_1^2,
\label{eqsetsimp3}
\ee
\be
\frac{1}{2}\left(\frac{\partial \beta}{\partial y}\right)^2 - \frac{1}{2} \left(\frac{\partial \beta}{\partial z}\right)^2 = B_3^2 - B_2^2,
\label{eqsetsimp4}
\ee
\be
- \frac{\partial}{\partial x} \left(\frac{\partial \beta}{\partial t} + \frac{1}{2} \frac{\partial}{\partial x} (\beta^2)\right) - 2 \Lambda = E_1^2,
\label{eqsetsimp5}
\ee
\be
- \frac{1}{2}\frac{\partial^2 \beta}{\partial x \partial y} = B_3 E_1,
\label{eqsetsimp6}
\ee
\be
- \frac{1}{2}\frac{\partial^2 \beta}{\partial x \partial z} = - B_3 E_1,
\label{eqsetsimp7}
\ee
\be
\frac{1}{2}\frac{\partial \beta}{\partial y}\frac{\partial \beta}{\partial z} = - B_2 B_3,
\label{eqsetsimp8}
\ee
\be
 - \frac{1}{4} \left(\frac{\partial\beta}{\partial y}\right)^2 
- \frac{1}{4} \left(\frac{\partial\beta}{\partial z}\right)^2 
+ \Lambda = \frac{1}{2} \left(2\mu + E_1^2 + B^2\right).
\label{eqsetsimp9}
\ee
The above equation sketches a possible new solution to Einstein's equations, considering a charged dust coupled to the warp drive space.

\subsection{Divergence of the specific electromagnetic field}

In this section, we will calculate the divergence of the electromagnetic energy-momentum tensor, considering that the electromagnetic field components are functions of the spacetime coordinate $x^\mu$, which are
\be
{T^{0 \alpha}}_{;\alpha}  = - \frac{\partial (\beta B^2)}{\partial x} 
- \frac{1}{2} \frac{\partial (B^2 + E_1^2)}{\partial t} 
- \frac{\partial (B_2 E_1)}{\partial z} 
+ \frac{\partial (B_3 E_1)}{\partial y} 
- \frac{\partial (\mu \beta)}{\partial x} 
- \frac{\partial\mu}{\partial t},
\label{divcomp0}
\ee
\be
{T^{1 \alpha}}_{;\alpha} =  \frac{1}{2} \frac{\partial (B^2 - E_1^2)}{\partial x}\,,
\label{divcomp1}
\ee
\be
{T^{2 \alpha}}_{;\alpha} = -  \frac{\partial (\beta B_3 E_1)}{\partial x} - \frac{\partial (B_2 B_3)}{\partial z} - \frac{\partial (B_3 E_1)}{\partial t} + \frac{1}{2}\frac{\partial}{\partial y}(B_3^2 - B_2^2 + E_1^2),
\label{divcomp2} 
\ee
\be
{T^{3 \alpha}}_{;\alpha} = \frac{\partial (\beta B_2 E_1)}{\partial x} + \frac{\partial (B_2 E_1)}{\partial t} - \frac{\partial (B_2 B_3)}{\partial y} + \frac{1}{2} \frac{\partial}{\partial z} (B_2^2 - B_3^2 + E_1^2).
\label{divcomp3}  
\ee
Considering $E_1$, $B_1$ and $B_2$ constants, Eqs.\,\eqref{divcomp0} to \eqref{divcomp3} can be written as
\be
{T^{0 \alpha}}_{;\alpha}  = - B^2 \frac{\partial \beta}{\partial x} 
- \frac{\partial (\mu \beta)}{\partial x} 
- \frac{\partial\mu}{\partial t} 
\label{divcompsimp0}
\ee
\be
{T^{1 \alpha}}_{;\alpha} =  0 
\label{divcompsimp1}
\ee
\be
{T^{2 \alpha}}_{;\alpha} = - B_3 E_1 \frac{\partial \beta}{\partial x} 
\label{divcompsimp2} 
\ee
\be
{T^{3 \alpha}}_{;\alpha} = B_2 E_1 \frac{\partial \beta}{\partial x} 
\label{divcompsimp3}  
\ee
A few important expressions are found by implying that the electromagnetic energy-momentum tensor must have null divergence. Eq.\,\eqref{divcompsimp0} is a continuity equation. Eq.\,\eqref{divcompsimp2} implies that 
\be 
B_3 = 0 \ \ , \ \ \text{or} \ \ E_1 = 0 \ \ , \ \ \text{or} \ \ \frac{\partial \beta} {\partial x} = 0 \,,
\ee 
and Eq.\,\eqref{divcompsimp3} implies that 
\be
B_2 = 0 \ \ , \ \ \text{or} \ \ E_1 = 0 \ \ , \ \ \text{or} \ \  \frac{\partial \beta}{\partial x} = 0 \,.
\ee 
Next, it analyzes how each of these cases can affect Eq.\,\eqref{eqsetsimp1} to Eq.\,\eqref{eqsetsimp9}.

\ni\textbf{Case $\frac{\partial \beta}{\partial x} = 0$ and $E_1 = 0$}

\ni These conditions are satisfied simultaneously, and it can be seen 
from Eq. \eqref{eqsetsimp5}.  Moreover, Eqs. \eqref{eqsetsimp6} 
and \eqref{eqsetsimp7} is identically zero, and the set of Einstein 
differential equations from Eq. \eqref{eqsetsimp1} to 
\eqref{eqsetsimp9} can be written as
\be
\Lambda = 0 = \frac{3}{4}\mu + \frac{1}{4} B^2,	
\label{eqsetbeta1}
\ee
\be
\frac{\partial^2 \beta}{\partial y^2} + 
\frac{\partial^2 \beta}{\partial z^2} = 0,
\label{eqsetbeta2}
\ee
\be
\left(\frac{\partial \beta}{\partial y}\right)^2 
+ \left(\frac{\partial \beta}{\partial z}\right)^2 
 = - \frac{4}{3}\Lambda  - \frac{4}{3} B^2,
\label{eqsetbeta3}
\ee
\be
\left(\frac{\partial \beta}{\partial y}\right)^2 
- \left(\frac{\partial \beta}{\partial z}\right)^2 
= 2 \left(B_3^2 - B_2^2\right),
\label{eqsetbeta4}
\ee
\be
\frac{\partial \beta}{\partial y}\frac{\partial \beta}{\partial z} 
= - 2 B_2 B_3,
\label{eqsetbeta5}
\ee
\be
\left(\frac{\partial\beta}{\partial y}\right)^2 
+ \left(\frac{\partial\beta}{\partial z}\right)^2 
= - 2 \Lambda - 2\left(2\mu + B^2\right).
\label{eqsetbeta6}
\ee
For this configuration, the cosmological constant is null, and the following relation between the electromagnetic field and the matter density is found
\be
\mu = - \frac{1}{3} B^2.
\label{matterdenneg}
\ee
So, the matter density will always be negative, but the energy conditions will still be satisfied. The shift vector will not depend on the $x$ spacetime coordinate, and there is no Burgers-type equation, i.e., no shock wave, but $\beta$ is a function of $(t,y,z)$, and it also satisfies the Laplace equation according to Eq.\,\eqref{eqsetbeta2}. Both Eqs.\,\eqref{eqsetbeta3} and \eqref{eqsetbeta6} are specific cases of the well-known Eikonal equation, and they imply that the solution for the shift vector is not unique and may have a complex component. It is clear that since the matter density is negative, the energy density for the energy-momentum tensor is positive $(T_{00} > 0)$. The energy-momentum tensor matrix form is 
\be
T_{\alpha \nu} =
\begin{pmatrix}
-\frac{1}{2}\mu(1 + \beta^2) & -\frac{1}{2}\beta B^2 & 0      & 0  \\
-\frac{1}{2}\beta B^2       & -\frac{1}{2}B^2       & 0       & 0 \\
0  & 0  & \frac{1}{2}(B_3^2 - B_2^2) & - B_2 B_3 \\ 
0  & 0  & - B_2 B_3  & - \frac{1}{2}(B_3^2 - B_2^2)
\end{pmatrix}.
\label{emmtsimp1}
\ee

\ni\textbf{Case $B_3 = 0$ and $B_2 = 0$}

\ni For this case, it is straightforward to see that $B_2 = B_3 = 0$ implies 
\be
\frac{\partial \beta}{\partial y} = 0 \ \  \text{and} \ \ \frac{\partial \beta}{\partial z} = 0,
\ee 
from Eqs.\,\eqref{eqsetsimp6} to \eqref{eqsetsimp8}. The set of equations for this case is
\be
\Lambda = 4 \pi E_1^2,	
\label{eqsetmag1}
\ee
\be
\mu = 0,	
\label{eqsetmag2}
\ee
\be
- \frac{\partial}{\partial x} \left(\frac{\partial \beta}{\partial t}
+ \frac{1}{2} \frac{\partial}{\partial x} (\beta^2)\right) = 
2 \Lambda + 8 \pi E_1^2.
\label{eqsetmag3}
\ee
These results violate the null and dominant energy conditions; the matter density $\mu$ and the magnetic field \textbf{B} are both zero, and the cosmological constant is positive and proportional to the electric field energy from Eq.\,\eqref{eqsetmag1}. This configuration's energy density is always negative $T_{00} < 0$. The new electromagnetic energy-momentum tensor in matrix form is
\be
T_{\mu \nu} =
\begin{pmatrix}
-\frac{1}{2}E^2(1 - \beta^2) & -\frac{1}{2}\beta E^2 & 0 & 0  \\
-\frac{1}{2}\beta E^2 & -\frac{1}{2}E^2 & 0 & 0 \\
0 & 0 & \frac{1}{2}E^2 & 0 \\
0 & 0 & 0 & \frac{1}{2}E^2
\end{pmatrix}.
\label{emmtsimp2}
\ee
Considering a specific configuration of charged dust as a source to the warp drive geometry and analyzing the Einstein equations, several expressions relating the components of the electromagnetic fields, the shift vector, and the matter density were found. Imposing the null divergence leads to two boundary conditions that may help solve the Einstein equations. The same Burgers-type equations found in previous chapters 4 and 5 can be found. 

\subsection{Discussion}

For the charged dust source, an energy-momentum tensor with charged matter density $\mu$ in an electromagnetic field and with a geometrical (cosmological) constant is coupled to the warp drive spacetime. This configuration is rather cumbersome for finding analytical solutions, so particular cases are analyzed where the electric and magnetic fields are orthogonal, like electromagnetic waves, transporting the charged dust and creating the warp bubble. An expression for the term $T_{00}$ of the energy-momentum tensor is analyzed, which is shown to be a fourth-order polynomial for powers of the shift vector $O(\beta^4)$. For $\beta = 0$, the warp drive metric becomes Minkowski, and the solution is the common vacuum with no warp drive. The coefficients $\omega_{(i)}$ of the polynomial $T_{00}(\beta)$ are dependent on the components of the electromagnetic field and the matter density. Several regimes are discussed: the non-relativistic $(\beta = 0)$, relativistic $O(\beta)$, and superluminal $O(\beta^n)$, where $n \geq 2$. An equation relating the cosmological constant, the electromagnetic field components, the matter density, and the shift vector is found.

The energy conditions may be satisfied depending on the relation between the electromagnetic field components. Superluminal regimes are possible even for small electric and magnetic field strength values. The shift vector has upper and lower bound limits, but specific configurations of the electromagnetic fields can exceed these bounds. Practical applications regarding this matter-energy source to the warp drive can be considered. Imposing the energy conditions on the particular case where the electric field is perpendicular to the magnetic field leads to expressions where those conditions can be satisfied. Still, superluminal regimes can be obtained for negative and positive matter density. The only energy condition that requires positive matter density is the null energy condition. An equation of state is obtained for this case, connecting the cosmological constant, the matter density, and the electric and magnetic field strength. The null divergence of the energy-momentum tensor is imposed, and limiting conditions are found for superluminal regimes of warp drive. Two specific forms of an energy-momentum source are given; however, no solutions to the Einstein equations are displayed in this work for the charged dust warp drive configuration. For the first form of the energy-momentum tensor, the term $T_{00}$ is always negative for real values of the shift vector, but it can be positive for negative energy density. For the second form of the energy-momentum tensor found, the term $T_{00}$ can only assume positive values for small values of the shift vector, only in relativistic cases with no superluminal regime.

Analyzing the right-hand side of the Einstein equations with the null divergence conditions showed that obtaining superluminal speed with negative and positive matter density may be possible, depending on the electric and magnetic field norms. Two possible energy-momentum tensors were found, considering null divergence imposition. In one of the configurations in Eq.\,\eqref{emmtsimp1} the matter density becomes negative and proportional to the magnetic field energy density, the shift vector $\beta$ being a function of only $(t,y,z)$, and the absence of the Burgers-type equation as found in Chap.\,4. Nevertheless, a specific case of the Eikonal equation in Eqs.\,\eqref{eqsetbeta3} and \eqref{eqsetbeta6} has to be solved, leading to a wave equation for $\beta$. In contrast, the second energy-momentum tensor in Eq.\,\eqref{emmtsimp2} allowed for null matter density and a Burgers-type equation but violated the null and dominant energy conditions since the matter density and the magnetic field are null. The cosmological constant\index[topics]{cosmological constant} is proportional to the electric energy density. For this case, the energy density is always negative $(T_{00} < 0)$. 

Considering the general electromagnetic energy-momentum tensor, the energy density $T_{00}$ is a polynomial of the fourth order in $\beta$, demonstrating that there are possible subluminal, luminal, and superluminal regimes for the charged dust. The coefficients for this polynomial are functions of the matter density and the electric and magnetic components. Neglecting $O(\beta^2)$ and higher for a weak warp drive implies that the shift vector has an upper bound, which can be very large even for weak electric and magnetic fields if $B_2 E_3 \approx B_3 E_2$.

This approach first tackles the right-hand side of Einstein's equations, analyzing the matter-energy sources instead of the left-hand (geometrical) side. This was an attempt to find stable warp bubble solutions for the warp drive, even if energy conditions were violated. One may propose a new metric that considers plasmas as a source and finds faster-than-light wave solutions connecting the warp drive to solitons via the Korteweg-de Vries-type equation. Another warp bubble formation mechanism to be considered is cavitation in plasma.

\section{Cosmological constant and perfect fluid}

This chapter presents the results for the perfect fluid coupled to the warp drive geometry and the addition of a cosmological constant\index[topics]{cosmological constant} \cite{nos4}. This attempt aims to improve results found for the perfect fluid and to show that even minor modifications in warp drive geometry may overcome energy condition violations. Four solutions are found for this spacetime; they are labeled Cases 1a, 1b, 2a, and 2b, and are shown in Table\,\ref{cosmotab1}. 

This chapter is organized as follows: the first section discusses the solution of the Einstein Equations considering the perfect fluid with the coupling of the Cosmological constant. The second section discusses the imposition of the null divergence of the perfect fluid energy-momentum tensor. The third section shows the calculations for the perfect fluid's classical energy conditions, discusses the results' physical significance, considers the imposition of a positive energy density, and explains the requirements. Finally, the results will be discussed in the last section.

\subsection{Solving Einstein equations}

The components for the Eulerian (normal) observers'\index[topics]{Eulerian observers} 4-velocities are given by,
\be
u^\alpha = \left(1, - \beta , 0, 0\right), \ \ 
u_\alpha = (- 1,0,0,0)\,\,.
\ee
Using the Einstein equations and contracting both sides of this equation with the Eulerian observers, 4-velocities result in Eq.\,\eqref{negeng2}, whose expression concerning the weak energy condition, as Alcubierre demonstrated it \cite{Alcubierre1994}, with the addition of the cosmological constant in the Einstein tensor yields
\be
T_{\alpha \beta} \, u^\alpha u^\beta = \frac{\Lambda}{8\pi} - \frac{1}{32\pi}
\left[\left(\frac{\partial \beta}{\partial y}\right)^2 + \left(\frac{\partial \beta}{\partial z}\right)^2 \right] \,.
\label{edwarp drive1}
\ee
This result is similar to the one found by Alcubierre. However, depending on the sign of $\Lambda$, the weak energy condition is no longer non-positive, even if considering only normal-valued forms for the shift vector and no vacuum solutions. The Einstein tensor components are given by the set of equations from Eq.\,\eqref{et00} to Eq.\,\eqref{et33}, and the important expressions used to solve the Einstein equations for different matter sources are given by equations \eqref{burgerseq}, \eqref{expet1}, \eqref{gradpress}, \eqref{eqstaterel}. Just like the case for the perfect fluid and the anisotropic fluid, this configuration also unfolds in two sets of solutions, which are discussed and separated as Cases 1 and 2. In the first case.

\ni\textbf{Case 1) $\displaystyle{\frac{\partial \beta}{\partial z} = 0}$}

\ni\textbf{Case 1a:} $\left[\displaystyle{\frac{\partial \beta}{\partial z} = 0} \ \text{and} \ \displaystyle{\frac{\partial\beta}{\partial x} = 0}\right]$ 

\ni Eqs.\,\eqref{et00} to \eqref{et33} from Einstein equations, using the conditions mentioned for this case, result in the following set of equations
\be
\Lambda =  \frac{3}{4} \left(\mu - \frac{p}{3}\right) \,,	
\label{eqset1case1}
\ee
\be
\left(\frac{\partial\beta}{\partial y}\right)^2 = 4 (\Lambda - \mu)\,,
\label{eqset2case1a}
\ee
\be
\left(\frac{\partial \beta}{\partial y}\right)^2 = \frac{4}{3}(\Lambda - p) \,.
\label{eqset3case1a}
\ee
These last equations imply that the shift vector $\beta$ is not uniquely defined. It is a function that depends only on $(t,y)$ spacetime coordinates, and in this case, it may be a real-valued function depending on the sign of $\Lambda$. For the next subcase.

\ni\textbf{Case 1b:} $\left[\displaystyle{\frac{\partial \beta}{\partial z} = 0} \ \text{and} \ \displaystyle{\frac{\partial\beta}{\partial y} = 0}\right]$ 
 
\ni For this subcase, one has to solve the following equations,
\be
\Lambda = \mu = p = 0 \,,	
\ee
\be
- \frac{\partial}{\partial x} \left(\frac{\partial \beta}{\partial t} + \frac{1}{2} \frac{\partial}{\partial x}(\beta^2)\right) = 0 \,.
\ee
This last equation is the same Burgers-type equation that connects the warp drive to shock waves, as discussed in Ref.\,\cite{nos1}. The cosmological constant, fluid pressure, and matter density all equal zero, and the warp drive metric is a vacuum solution for the Einstein equations. Now, for the second set of solutions. Table \ref{cosmotab1} below summarizes the results and solutions.

\ni\textbf{Case 2) $\displaystyle{\frac{\partial \beta}{\partial y} = 0}$}
	
\ni\textbf{Case 2a:} $\left[\displaystyle{\frac{\partial \beta}{\partial y} = 0} \ \text{and} \ \displaystyle{\frac{\partial\beta}{\partial x} = 0}\right]$ 
		
For this configuration, the set of relevant expressions from the Einstein equations, considering the two conditions for this subcase, is simplified to
\be
\Lambda = \frac{3}{4}\left(\mu - \frac{p}{3}\right) \,,	
\ee
\be
\left(\frac{\partial\beta}{\partial z}\right)^2 = 4 (\Lambda - \mu)\,,
\ee
\be
\left(\frac{\partial \beta}{\partial z}\right)^2 = \frac{4}{3}\ (\Lambda - p)\,.
\ee
The above set of equations is very similar to Case 1a, but is not uniquely defined; in this case, $\beta$ depends on the $(t,z)$ coordinates.

\ni\textbf{Case 2b:} $\left[\displaystyle{\frac{\partial \beta}{\partial y} = 0} \ \text{and} \ \displaystyle{\frac{\partial\beta}{\partial z} = 0}\right]$ 
 
In this case, one has to solve the following equations
\be
\Lambda = \mu = p = 0\,,	
\ee
\be
- \frac{\partial}{\partial x} \left(\frac{\partial \beta}{\partial t} + \frac{1}{2} \frac{\partial}{\partial x} (\beta^2)\right) = 0\,,
\ee
which is the same case as Case 1b, and the result found for dust in Ref.\,\cite{nos1}, a vacuum solution to the Einstein equations with a Burgers-type equation.

\begin{table}[ht]
\begin{tabular}{| m{3cm} | m{3cm} | m{8cm} |}
\hline 
Case & Condition & Results \\ 
\hline 
\multirow{2}{*}{$1) \
\displaystyle{\frac{\partial \beta}{\partial z} = 0}$}
& $1a) \ \displaystyle{\frac{\partial \beta}{\partial x} = 0}$ &
$\begin{array} {ll} 
\Lambda =  6\pi \left(\mu - \frac{p}{3}\right) \\ [6pt]
\beta = \beta(y,t)\\ [6pt]
\displaystyle{\frac{\partial \beta}{\partial y} 
= \pm\sqrt{4 (\Lambda - 8\pi \mu)}} \\ [8pt]
\displaystyle{\frac{\partial \beta}{\partial y} 
= \pm\sqrt{\frac{4}{3} (\Lambda - 8\pi\mu)}} \\ [8pt]
\displaystyle{\beta \frac{\partial \mu}{\partial x} 
+ \frac{\partial \mu}{\partial t} = 0 
\ \ \text{(null divergence)}} \\ [8pt]
\end{array}$ \\ [28pt]
\cline{2-3} &
$1b) \ \displaystyle{\frac{\partial \beta}{\partial y} = 0}$ &
$\begin{array} {ll} 
\Lambda = \mu = p = 0 \\ [6pt]
\beta = \beta(x,t)\\ [6pt]
\displaystyle{
\frac{\partial \beta}{\partial t}
+ \frac{1}{2} \frac{\partial}{\partial x} 
(\beta^2) = h(t)\, } \\ [6pt]
\text{null divergence is trivially satisfied.}\\ [2pt]
\text{This is the solution we found in Chap.\,4} \\ [8pt]
\end{array}$ \\ [28pt] 
\hline 
\multirow{2}{*}{$2) \
\displaystyle{\frac{\partial \beta}{\partial y} = 0}$}
& $2a) \ \displaystyle{\frac{\partial \beta}{\partial x} = 0}$ &
$\begin{array} {ll} 
\Lambda =  6 \pi \left(\mu - \frac{p}{3}\right) \\ [6pt]
\beta = \beta(z,t)\\ [6pt]
\displaystyle{\frac{\partial \beta}{\partial z} 
= \pm\sqrt{4 (\Lambda - 8\pi\mu)}} \\ [8pt]
\displaystyle{\frac{\partial \beta}{\partial z} 
= \pm\sqrt{\frac{4}{3} (\Lambda - 8\pi\mu)}} \\ [8pt]
\displaystyle{\beta \frac{\partial \mu}{\partial x} 
+ \frac{\partial \mu}{\partial t} = 0 
\ \ \text{(null divergence)}} \\ [8pt]
\end{array}$ \\ [28pt]
\cline{2-3}   
& $2b) \ \displaystyle{\frac{\partial \beta}{\partial z} = 0}$ &
$\begin{array} {ll} 
\Lambda = \mu = p = 0 \\ [6pt]
\beta = \beta(x,t)\\ [6pt]
\displaystyle{\frac{\partial \beta}{\partial t} 
+ \frac{1}{2} \frac{\partial}{\partial x}(\beta^2)
= h(t)} \\ [6pt]
\text{null divergence is trivially satisfied.}\\ [2pt]
\text{This is the solution we found in Chap.\,4} \\ [8pt]
\end{array}$ \\ [28pt] 
\hline 
\end{tabular}
\caption{Summary of all solutions of the Einstein equation with the Cosmological constant and the warp drive metric with the perfect fluid as the source.}
\label{cosmotab1}
\end{table}

Adding the cosmological constant\index[topics]{cosmological constant} with an undetermined sign is a simple and objective way to change the warp drive geometry and find non-negative energy densities. The solutions from Case 1b and Case 2b are identical to the ones found for the perfect fluid with no constant. However, Case 1a and Case 2a present the possibility of a real-valued function for the shift vector even with negative matter density and pressures.

Case 1a and Case 2a are analogous to each other, however, with the slight difference that the shift vector is a function of $(t,Y)$ coordinates, that is, $\beta = \beta(t,y)$, for Case 1a and Case 2a, the shit vector is a function of $(t,z)$ coordinates, $\beta = \beta(t,z)$. In those two cases, real-valued solutions for the shift vector rely on the sign of $\Lambda - 8\pi\mu$, where $\Lambda$ is the cosmological constant, and $p$ is the static pressure for the perfect fluid. If $\Lambda > 8\pi\mu$, then $\beta$ is real-valued. Otherwise, the shift vector would be a complex-valued function. With the addition of the cosmological constant, the solutions for these two cases are no longer unique but are divided into two subsets of solutions.

Case 1b and Case 2b are identical to the dust solution found in Chap.\,4, which connects the warp drive to shock waves via the Burgers-type equation. This result implies that there is no matter density and static pressure, hence there is a vacuum solution to the Einstein equations. 
It also shows that the dust might not be enough matter-energy source to create a warp drive with the exact physical requirements imposed by Alcubierre. Maybe only the matter density is not enough to make a warp bubble. This is clear from the energy density originally calculated by Alcubierre \cite{Alcubierre1994} considering his warp drive metric. 
\be
T_{\alpha \beta} \, u^\alpha u^\beta = - \frac{1}{32\pi}
\left[\left(\frac{\partial \beta}{\partial y}\right)^2 + \left(\frac{\partial \beta}{\partial z}\right)^2 \right].
\ee

The addition of the cosmological constant made the shift vector not unique, since now there are two sets of partial equations to be solved for cases 1a and 2a. The null divergence of the energy-momentum tensor still gives the same results as the perfect fluid from Chap.\,5. The energy conditions are only satisfied if the matter density is positive; however, the shift vector does not need to be complex-valued in this configuration with the Cosmological constant \index[topics]{cosmological constant}. Case 1b and Case 2b are identical to the vacuum solution found for dust in Chap.\,4, connecting the warp drive geometry to shock waves via a Burgers-type equation.

\subsection{Divergence for the perfect fluid} 

The null divergence for the energy-momentum tensor reveals important properties for sources in general relativity, especially when dealing with fluid solutions of the Einstein equations. In this section, the null divergence is imposed, and expressions relating to the Cosmological constant, matter density, static pressure, and shift vector are found. Calculating the divergence for the perfect fluid energy-momentum tensor, one arrives at the following equations,
\be
{T^{0 \nu}}_{;\nu} = - (\mu + p) \frac{\partial \beta}{\partial x}
- \beta \frac{\partial (p + \mu)}{\partial x} 
- \frac{\partial \mu}{\partial t},
\label{nulldiv0}
\ee
\be
{T^{1 \nu}}_{;\nu} = \frac{\partial p}{\partial x},
\label{nulldiv1}
\ee
\be
{T^{2 \nu}}_{;\nu} = \frac{\partial p}{\partial y},
\label{nulldiv2}
\ee
\be
{T^{3 \nu}}_{;\nu} = \frac{\partial p}{\partial z}.
\label{nulldiv3}
\ee
Besides imposing the null divergence condition, Eqs.\,\eqref{nulldiv0} to \eqref{nulldiv3} imply that the pressure $p$ does not depend on the spatial coordinates and is constant. This implies that the Eq.\,\eqref{nulldiv0} is simplify to
\be
\beta \frac{\partial \mu}{\partial x} + \frac{\partial \mu}{\partial t} = 0.
\label{nulldiv01a2a}
\ee
Considering cases 1a and 2a, the state equation of these two solutions is given by 
\be
\Lambda = \frac{3}{4}\left(\mu - \frac{p}{3} \right),
\ee
and since $\Lambda$ is a constant, and from the null divergence imposition, the pressure is also considered constant, it implies that the matter density is also constant. This means that Eq.\,\eqref{nulldiv01a2a} is immediately satisfied since the partial derivatives of the matter density are all equal to zero. This is the same result found for the perfect fluid with no Cosmological constant. For Cases 1b and 2b Eq.\,\eqref{nulldiv0} is trivially satisfied since $\mu = p = 0$.

It is important to notice that the shift vector is not uniquely defined in Case 1a and Case 2a since the following equations need to be solved simultaneously to define $\beta$
\be
\left(\frac{\partial \beta}{\partial y} \right)^2 
= 4 (\Lambda - \mu),
\label{betasol1}
\ee
\be
\left(\frac{\partial \beta}{\partial y} \right)^2
= \frac{4}{3} (\Lambda - p).
\label{betasol2}
\ee
Even though adding a Cosmological constant helped generate possible positive energy densities for the warp drive, it also included this caveat regarding the uniqueness of solutions to the Einstein equations. The shift vector is a consistent equation only if the right-hand side of both Eq.\,\eqref{betasol1} and Eq.\,\eqref{betasol2} are the same, and this happens only when the cosmological constant $\Lambda$ assumes the following particular value
\be
\Lambda = \frac{3\mu - p}{2}.
\ee
This condition ensures the internal consistency of the Einstein equations by eliminating the ambiguity in the definition of the shift vector.

\subsection{Energy conditions} 

This section presents the calculations and results for the energy conditions with the energy-momentum tensor for the perfect fluid with the cosmological constant. The energy conditions are immediately satisfied since Case 1b and Case 2b are vacuum solutions. For Case 1a and Case 2a, even though the state equation now relates the matter density and the pressure by adding a constant $\Lambda$, these two solutions still present the same results for the energy conditions as the one found in Ref.\,\cite{nos2} discussed for the perfect fluid. The weak and dominant energy conditions are only satisfied if the matter density is positive. Still, this does not imply that the shift vector needs to be complex-valued because of the addition of the cosmological constant $\Lambda$. The strong and null energy conditions require the combination of the pressure and the matter density to produce a positive value. However, this does not mean that necessarily $p > 0$ and $\mu > 0$.

\subsubsection{Weak energy conditions} 

For this case, the energy-momentum tensor at each point in the spacetime must obey the following inequality
\be
T_{\alpha \sigma} \, u^\alpha u^\sigma \geq 0, 
\label{wecaux}
\ee
for any timelike vector $\textbf{u} \, (u_\alpha u^\alpha < 0)$ and any null zero vector $\textbf{k} \, (k_\alpha k^\alpha = 0)$. For an observer with unit tangent vector $\textbf{v}$ at a certain point of the spacetime, the local energy density measured by any observer is non-negative \cite{HawkingEllis1973}. For the perfect fluid energy-momentum the expression $T_{\alpha \sigma} \, u^\alpha u^\sigma$ is
\be
T_{\alpha \sigma} \, u^\alpha u^\sigma = \mu,    
\ee
and the weak energy condition from Eq.\,\eqref{wecaux} is satisfied
if the matter density $\mu$ is positive. This is also the case for the dust.

\subsubsection{Dominant energy conditions}

For every timelike vector $u_\alpha$, the following inequality must 
be satisfied,
\be
T^{\alpha \beta} \, u_\alpha u_\beta \geq 0, \quad \text{and} \quad F^\alpha  F_\alpha  \leq 0, 
\ee
where $F^\alpha = T^{\alpha \beta} u_\beta$ is a non-spacelike vector, and the following condition must also be satisfied  
\be
T^{00} \geq |T^{\alpha \beta}|, \ \text{for each} \ \alpha, \beta.
\ee
Evaluating the first condition for the perfect fluid leads to
\be
T^{\alpha \beta} \, u_\alpha u_\beta = \mu.
\label{cosmodec1}
\ee
The other condition $F^\alpha  F_\alpha$ is given by the result
\be
F^\alpha  F_\alpha = - \mu^2 \leq 0.
\label{cosmodec2}
\ee
Hence, the dominant energy condition is satisfied for $\mu > 0$, as can be seen in Eq.\ \eqref{cosmodec1}. Besides, Eq.\,\eqref{cosmodec2} is always satisfied no matter the sign of the matter density. This condition also holds if one considers the dust to be the case for the perfect fluid with null pressure.

\subsubsection{Strong energy conditions}

For the strong energy condition, the expression
\be
\Bigg(T_{\alpha \beta} - \frac{1}{2}T \, g_{\alpha \beta} \Bigg) 
u^\alpha u^\beta \geq 0 
\label{seccond}
\ee
it is true for any timelike vector $u$. Computing the strong energy condition in Eq.\,\eqref{seccond}, 
\be
\Bigg(T_{\alpha \beta} - \frac{1}{2}T \, g_{\alpha \beta} \Bigg) 
u^\alpha u^\beta = \frac{1}{2}(3p + \mu),
\ee
and the strong energy condition stated in Eq.\,\eqref{seccond} is
satisfied if 
\be
3p + \mu \geq 0.
\ee
The same is valid for the dust energy-momentum tensor, considering $p=0$ for the perfect fluid, if $\mu \geq 0$.

\subsubsection{Null energy conditions}

The null energy conditions are satisfied in the limit of observers with null 4-velocity. For the null vector $\textbf{k}$ the following conditions
must be satisfied,
\be
T_{\alpha \sigma} \, k^\alpha k^\sigma \geq 0, \qquad \text{for any null 
vector} \ k^\alpha.
\ee
Assuming that the following null vector $k^\alpha$ is given by,
\be
k^\alpha = (a,b,0,0),    
\ee
we have that the relation between the components $a$ and $b$ is obtained by solving $k_\alpha k^\alpha = 0$. The two solutions are given by,
\be
\frac{a}{b} = \frac{1}{\beta \pm 1}.
\ee
Then, the null energy condition reads,
\be
T_{\alpha \sigma} \, k^\alpha k^\sigma = 
\left(\frac{b}{\beta \pm 1}\right)^2
\left(\mu + p\right),
\label{nec1}
\ee
And the null energy condition may be satisfied if the following conditions are met:
\be
\mu + p \geq 0.
\label{nec2}
\ee
Table \ref{tableenergycond} summarizes the results found for the energy conditions for the perfect fluid with the cosmological constant. 

\begin{table}[ht]
\caption{Summary results for the perfect fluid energy conditions} 
\centering 
\begin{tabular}[c]{l @{\hspace{50pt}} l} 
\hline\hline 
Energy condition & Results \\ [0.5ex] 
\hline 
Weak         &  $\mu \geq 0$  \\ 
Strong       &  $\mu \geq 0$  \\
Dominant     &  $\mu + 3p \geq 0$  \\
Null         &  $\mu + p \geq 0.$ \\ 
[1ex] 
\hline 
\end{tabular}
\label{tableenergycond} 
\end{table}

\subsection{Discussion}

The proposal to include the cosmological constant in Einstein's equations aims to demonstrate that, with minor modifications, it may be possible to obtain warp drives with positive energy densities respecting the energy conditions. 

Four solutions to the Einstein equations are found for the perfect fluid with the cosmological constant. Solutions for Case 1b and Case 2b are identical to the ones found in Chap.\,3-6, presenting vacuum solutions to the Einstein equations and connecting the warp drive geometry to shock waves via a Burgers-type equation. These two solutions immediately satisfy the energy conditions and the null divergence. Solutions for Case 1a and Case 2a present a new state equation, connecting the static pressure and the matter density by adding the cosmological constant. The shift vector no longer needs to be complex-valued, even if the pressure and the matter density are positive. This result can satisfy the weak and dominant energy conditions since the matter density can be positive, and $\beta$ can still be real-valued. However, the shift vector is not unique since two partial differential equations can determine it. The shift vector is uniquely defined only with specific values for $\Lambda$. The strong and null energy conditions now depend on the combination of the static pressure and the matter density, resulting in positive values. However, this does not mean that matter and pressure must be positive if their sum results in a positive value. These conditions can be seen in Table \ref{tableenergycond}.

The imposition of the null divergence for the energy-momentum tensor implies that the static pressure $p$ is constant. Because the state equation connects the pressure to the matter density and the cosmological constant for Case 1a and Case 2a, it also implies that the matter density is constant. If the warp drive geometry is modified by a dynamical factor instead of a constant, then it would be necessary to solve also Eq.\,\eqref{nulldiv01a2a}. 

The Burgers-type equation is a vacuum solution for the Einstein equations for different sources. This is an intriguing result. It seems that shock waves are present in Alcubierre warp drive classes. This could be tied to the presence of event horizons inside the bubble. There is also the question of the shift vector being dependent only on a few spacetime coordinates. Because of that, the expansion volume of the warp bubble does not follow the behavior proposed by Alcubierre, which is contracting spacetime in front of the bubble and expanding spacetime behind it, propelling the warp bubble at superluminal speed throughout spacetime.

\section*{Conclusion}	
\addcontentsline{toc}{section}{Conclusion}

This thesis presents the outcomes from a series of results obtained in chapters 3-8, coupling the warp drive spacetime to different sources of matter and energy. Solutions to the Einstein equations are presented. Considering the warp drive metric, the sources are dust (Chap.\,4), perfect fluid (Chap.\,5), and anisotropic fluid (Chap.\,6), charged dust (Chap.\,7), and the perfect fluid with a cosmological constant (Chap.\,8). 

Departing from the dust, the warp drive was discovered to be a vacuum solution connecting faster-than-light travel with shock waves via a Burgers-type equation. The matter density is zero, so the shock wave warp drive is a vacuum solution to the Einstein equations. A question arises from the shift vector spacetime dependencies since it depends only on time and the space coordinates in the direction of the movement of the warp bubble. There is an isomorphism between the final warp drive metric and Minkowski, suggesting that this may be an interior solution of a junction of two flat spacetimes enclosing a warped region with limited thickness. For a shift vector depending only on $(t,x)$, the Ricci tensor components are not flat, but they are all null in regions where the Burgers-type equation is satisfied. This indicates that this may be a junction condition between warp drive and Minkowski. The interior of the warp bubble and its exterior are flat, and the bubble's walls are a warped region with non-zero curvature.

Chaps.\,5-6 use two matter-energy sources: the perfect fluid and the anisotropic fluid. Four solutions for the perfect fluid are found with isotropic static pressure $p$ and matter density $\mu$. Two of those solutions are identical to the one found for dust, with the Burgers-type equation and shock waves connected to the warp bubble formation. The other two solutions are similar but differ regarding the coordinates' dependencies on the shift vector. One of these two solutions depends only on $(t,y)$ and the other only on $(t,z)$, with the possibility of obeying the energy conditions for negative matter density or a complex-valued form of the shift vector. In these two cases, the equations of state are the same: $p = \omega \mu$, where $p$ is the fluid's static pressure, $\mu$ is the matter density, and $\omega = 3$. This equation of state is not the usual one found for common baryonic mass, where $\omega = 0$ for non-relativistic particles, $\omega = 1/3$ for ultra-relativistic particles, and $\omega = - 1$ for the cosmological constant and accelerated cosmic inflation, and $\omega = 3$ would be tied to superluminal particles in this case.

For the anisotropic fluid, with anisotropic pressures $A, B, C$, the flow of momentum $D$, intrinsic static pressure $p$, and matter density $\mu$. Four solutions are found for this source and the warp drive. The Burgers-type equation also appears in this configuration, as well as the solutions of the Einstein equations connecting the warp drive and shock waves. These results show a fundamental connection between warp drive and shock wave solutions that can be tied to event horizons within the warp bubble. The other two solutions for the anisotropic fluid are similar in form but differing in spacetime coordinate dependencies for the shift vector, exactly as for the perfect fluid. One of these solutions depends on $(t,y)$, and the other depends on $(t,z)$. Expressions between the pressures $A$ and $D$ can be imposed so that the energy conditions are obeyed; however, in such situations, the shift vector will be of the complex-valued form, but depending on the values of the pressures $A, D, p$, the matter density can be positive. This demonstrates that with more parameters in the Einstein equations, it may be possible to form warp bubbles with common baryonic mass and apparent negative energy densities due to pressure gradients. The equation of state $p = \omega \mu$ for this anisotropic fluid depends on pressures $A, D, p$ and the square of the shift vector norm $\beta^2$. The two solutions with the Burgers-type equation are vacuum solutions to the Einstein equations, with null matter density, null pressures, and obeying the energy conditions immediately. 

For the charged dust source, an energy-momentum tensor with charged matter density $\mu$ in an electromagnetic field and with a cosmological constant is coupled to the warp drive spacetime. This configuration is rather cumbersome for finding analytical solutions, so particular cases are analyzed where the electric and magnetic fields are orthogonal, like electromagnetic waves, transporting the charged dust and creating the warp bubble. An expression for the term $T_{00}$ of the energy-momentum tensor is analyzed, which is shown to be a fourth-order polynomial for powers of the shift vector $O(\beta^4)$. For $\beta = 0$, the warp drive metric becomes Minkowski, and the solution is the common vacuum with no warp drive. The coefficients $\omega_{(i)}$ of the polynomial $T_{00}(\beta)$ are dependent on the components of the electromagnetic field and the matter density. Several regimes are discussed: the non-relativistic $(\beta = 0)$, relativistic $O(\beta)$, and superluminal $O(\beta^n)$, where $n \geq 2$. An equation relating the cosmological constant, the electromagnetic field components, the matter density, and the shift vector is found. The energy conditions may be satisfied depending on the relation between the electromagnetic field components. Superluminal regimes are possible even for small electric and magnetic field strength values.

In some cases, the shift vector has upper and lower bound limits, but specific configurations of the electromagnetic fields can grow to unlimited values. Practical applications regarding this matter-energy source to the warp drive can be considered. Imposing the energy conditions on the particular case where the electric field is perpendicular to the magnetic field leads to expressions where those conditions can be satisfied. Superluminal regimes can be obtained for negative and positive matter density. The only energy condition that requires positive matter density is the null energy condition. An equation of state is obtained for this case, connecting the cosmological constant, the matter density, and the electric and magnetic field strength. The null divergence of the energy-momentum tensor is imposed, and limiting conditions are found for superluminal regimes of warp drive. Two specific forms of an energy-momentum source are given; however, no solutions to the Einstein equations are displayed in this work for the charged dust warp drive configuration. For the first form of the energy-momentum tensor, the term $T_{00}$ is always negative for real values of the shift vector, but it can be positive for negative energy density. For the second form of the energy-momentum tensor found, the term $T_{00}$ can only assume positive values for small values of the shift vector, only in relativistic cases with no superluminal regime. 

In Chap.\,8, the cosmological constant is proposed when coupling the perfect fluid source to the warp drive. With this minor modification in warp drive, it is possible that, depending on this constant sign, the need for negative energy conditions may be contingent. So, even though Alcubierre's warp drive violates energy conditions, this does not necessarily mean that other classes of warp drives will also violate all energy conditions. The four solutions for this configuration are similar to the original ones for the perfect fluid. Still, adding a cosmological constant impacts the signs of matter density, and if the shift vectors are real-valued functions, the warp drive can be attained concerning the energy conditions.

There are still a lot of questions to be answered regarding warp drive spacetimes, regarding the negative energy densities, and the violation of energy conditions; however, the above-presented results raise questions about whether the negative energy densities imposition may be circumvented with matching conditions between spacetimes that allow for a mixture of anisotropic charged fluids with heat flux, creating regions of apparent negative energy due to non-null pressure gradients when the contact hypersurfaces are defined between these fluids.
\bibliographystyle{unsrturl}
\addcontentsline{toc}{section}{References}
\bibliography{references}
\begin{appendix}
\section{Sage manifold codes}

The computational scripts for this thesis were developed in  \href{https://sagemanifolds.obspm.fr/}{Sage Manifold}

\begin{lstlisting}[language=Python]
%display latex
# Defining a Lorentzian manifold
M = Manifold(4, 'M', structure='Lorentzian')
# Defining the chart
X.<t,x,y,z> = M.chart(r"t x:(-oo,+oo) y:(-oo,+oo) z:(-oo,+oo)");
# Defining shift vector as a function of spacetime coordinates
shift = function('shift', latex_name = "\\beta", nargs=4)(t,x,y,z)
# Defining the alpha lapse function as a function of spacetime coordinates
alpha = var('alpha',latex_name = "\\alpha")
cosmo = var('cosmo',latex_name="\\Lambda")
# Defining function with alias a, as function of spacetime coordinates
a = function('a',latex_name= "A")(t,x,y,z)
# Defining function with alias b, as function of spacetime coordinates
b = function('b',latex_name= "B")(t,x,y,z)
# Defining function with alias c, as function of spacetime coordinates
c = function('c',latex_name= "C")(t,x,y,z)
# Defining function with alias d, as function of spacetime coordinates
d = function('d',latex_name= "D")(t,x,y,z)
# Defining pressure as function of spacetime coordinates
p = function('p',latex_name= "p")(t,x,y,z)
# Defining matter density as function of spacetime coordinates
mu = function('mu',latex_name= "\\mu")(t,x,y,z)
# Defining density rho as function of spacetime coordinates
rho = function('rho',latex_name="\\rho")(t,x,y,z)
# Defining variable named sigma
sigma = var('sigma',latex_name="\\sigma")
R = var('R',latex_name="R")
#
xs = function('xs',latex_name = "x_s")(t)
rs = function('rs',latex_name = "r_s")(xs,x,y,z)
rs = sqrt((x-xs)^2+y^2+z^2)
vs = function('vs',latex_name="v_s")(t)
#
fs = function('fs',latex_name="f(r_s)")(rs)
fs = (tanh(sigma(rs+R)) - tanh(sigma(rs-R)))/(2*sigma)
# Defining the metric tensor components
g = M.metric()
g.set_name('g')
# g = M.tensor_field(2,0, name='T');
g[0,0] = shift**2 - 1
g[1,1] = 1
g[2,2] = 1
g[3,3] = 1
g[0,1] = - shift
g[1,0] = - shift
# Defining the inverse metric
gg = g.up(g)
gg.set_name('g')
\end{lstlisting}

\section{Tensor components of Warp Drive metric in reduced form}

This appendix presents the Riemann, Ricci, and Einstein tensor components calculated for the warp drive metric below,
\be
\dd s^2 = - (1-\beta^2) \, \dd t^2 - 2 \beta \, \dd t \, \dd x
+ \dd x^2 + \dd y^2 + \dd z^2 \,,
\label{wdmetricnos}
\ee
where $\beta = \beta(t,x,y,z)$ is the first component $\beta^x$ 
of the shift vector $\beta^a = (\beta^x, \beta^y,\beta^z)$, 
noting that the upper index in $\beta^x$ was dropped since 
there are no redundancies due to the presence of only one 
component.

\subsection{Riemann components} \label{riemann}

\begin{lstlisting}[language=Python]
Rie = g.riemann()
Rie.display_comp(only_nonredundant=True)
\end{lstlisting}

\be
\mathrm{R}_{\phantom{t}t t x}^{t \phantom{t} \phantom{t} 
\phantom{x}} = -\beta \left(\frac{\partial\beta}{\partial x}\right)^2
- \beta^{2} \frac{\partial^2\beta}{\partial x^2} + \frac{1}{4}  
\beta \left(\frac{\partial\beta}{\partial y}\right)^{2} + \frac{1}{4}  
\beta \left(\frac{\partial\beta}{\partial z}\right)^{2} - 
\beta \frac{\partial^2\beta}{\partial t\partial x} 
\label{rtttx}
\ee

\be
\mathrm{R}_{\phantom{t}t t y}^{t \phantom{t} \phantom{t} \phantom{y}} 
= -\beta^{2} \frac{\partial^2\,\beta}{\partial x\partial y} 
- \beta \frac{\partial\beta}{\partial x} 
\frac{\partial\beta}{\partial y} - \frac{1}{2} \, 
\beta \frac{\partial^2\beta}{\partial t\partial y} 
\label{rttty}
\ee

\be
\mathrm{R}_{ \phantom{ t}  t  t  z }^{  t \phantom{ t} \phantom{ t} 
\phantom{ z} } = -\beta^{2} \frac{\partial^2\beta}{\partial x\partial z} 
- \beta \frac{\partial\beta}{\partial x} \frac{\partial\beta}{\partial z} 
- \frac{1}{2}  \beta \frac{\partial^2\beta}{\partial t\partial z} 
\label{rtttz}
\ee

\be
\mathrm{R} _{ \phantom{ t}  t  x  y }^{  t \phantom{ t} \phantom{ x} 
\phantom{ y} } = \frac{1}{2}  \beta \frac{\partial^2\beta}
{\partial x\partial y} 
\label{rttxy}
\ee

\be
\mathrm{R}_{ \phantom{ t}  t  x  z }^{t \phantom{ t} \phantom{ x} 
\phantom{ z} } = \frac{1}{2}  \beta \frac{\partial^2\beta}
{\partial x\partial z} 
\label{rttxz}
\ee

\be
\mathrm{R}_{ \phantom{ t}  x  t  x }^{  t \phantom{ x} \phantom{ t} 
\phantom{ x} } = \left(\frac{\partial\beta}{\partial x}\right)^{2} 
+ \beta \frac{\partial^2\beta}{\partial x ^ 2} - \frac{1}{4}  
\left(\frac{\partial\beta}{\partial y}\right)^{2} - \frac{1}{4}  
\left(\frac{\partial\beta}{\partial z}\right)^{2} + 
\frac{\partial^2\beta}{\partial t\partial x} 
\label{rtxtx}
\ee

\be
\mathrm{R}_{ \phantom{ t}  x  t  y }^{  t \phantom{ x} \phantom{ t} 
\phantom{ y} } = \beta \frac{\partial^2\beta}{\partial x\partial y} 
+ \frac{\partial\beta}{\partial x} \frac{\partial\beta}{\partial y} 
+ \frac{1}{2}  \frac{\partial^2\beta}{\partial t\partial y} 
\ee

\be
\mathrm{R}_{ \phantom{ t}  x  t  z }^{  t \phantom{ x} \phantom{ t} 
\phantom{ z} } = \beta \frac{\partial^2\beta}{\partial x\partial z} 
+ \frac{\partial\beta}{\partial x} \frac{\partial\beta}{\partial z} 
+ \frac{1}{2}  \frac{\partial^2\beta}{\partial t\partial z} 
\ee

\be
\mathrm{R}_{\phantom{t}x x y}^{t \phantom{x} \phantom{x} 
\phantom{y}} = -\frac{1}{2} \frac{\partial^2\beta}{\partial x\partial y} 
\ee

\be
\mathrm{R}_{\phantom{t}x x z }^{t \phantom{x} \phantom{x} 
\phantom{z}} = -\frac{1}{2}\frac{\partial^2\beta}{\partial x\partial z} 
\ee

\be
\mathrm{R}_{\phantom{t} y t x}^{t \phantom{y} \phantom{t} \phantom{x}} 
= \frac{1}{2}  \beta \frac{\partial^2\beta}{\partial x\partial y} + 
\frac{\partial\beta}{\partial x} \frac{\partial\beta}{\partial y} + 
\frac{1}{2}\frac{\partial^2\beta}{\partial t\partial y} 
\ee

\be
\mathrm{R}_{\phantom{t}y t y}^{t \phantom{y} \phantom{t} \phantom{y}} 
= \frac{3}{4}\left(\frac{\partial\beta}{\partial y}\right)^{2} + 
\frac{1}{2}  \beta \frac{\partial^2\beta}{\partial y^2} 
\ee 

\be
\mathrm{R}_{\phantom{t}y t z }^{t \phantom{y} \phantom{t} \phantom{z}} 
= \frac{1}{2}  \beta \frac{\partial^2\beta}{\partial y\partial z} 
+ \frac{3}{4}\frac{\partial\beta}{\partial y} \frac{\partial\beta}{\partial z} 
\ee

\be
\mathrm{R}_{\phantom{t}y x y}^{t \phantom{y} \phantom{x} \phantom{y}} 
= -\frac{1}{2}  \frac{\partial^2\beta}{\partial y^2} 
\ee

\be
\mathrm{R}_{\phantom{t} y x z }^{t \phantom{y} \phantom{x} \phantom{z}} 
= -\frac{1}{2}\frac{\partial^2\beta}{\partial y\partial z} 
\ee

\be
\mathrm{R}_{\phantom{t}z t x}^{t \phantom{z} \phantom{t} \phantom{x}} 
= \frac{1}{2} \beta \frac{\partial^2\beta}{\partial x\partial z} 
+ \frac{\partial\beta}{\partial x} \frac{\partial\beta}{\partial z} 
+ \frac{1}{2}  \frac{\partial^2\beta}{\partial t\partial z} 
\ee

\be
\mathrm{R}_{\phantom{t}z t y}^{t \phantom{z}\phantom{t}\phantom{y}} 
= \frac{1}{2} \beta \frac{\partial^2\beta}{\partial y\partial z} +
\frac{3}{4}\frac{\partial\beta}{\partial y} 
\frac{\partial\beta}{\partial z} 
\ee

\be
\mathrm{R}_{\phantom{ t}z t z}^{t \phantom{z} \phantom{t}\phantom{z}} 
= \frac{3}{4} \left(\frac{\partial\beta}{\partial z}\right)^{2} 
+ \frac{1}{2}  \beta \frac{\partial^2\beta}{\partial z^2} 
\ee

\be
\mathrm{R}_{\phantom{t}z x y}^{t \phantom{z}\phantom{x}\phantom{y}} 
= -\frac{1}{2}\frac{\partial^2\beta}{\partial y\partial z} 
\ee

\be
\mathrm{R}_{\phantom{t}z x z}^{t \phantom{z} \phantom{x}\phantom{z}} 
= -\frac{1}{2}\frac{\partial^2\beta}{\partial z^2} 
\ee

\begin{align}
\nonumber
\mathrm{R}_{\phantom{x}t t x}^{x \phantom{t} \phantom{t} \phantom{x}} 
&= -\left(\beta^{2} - 1\right) 
\left(\frac{\partial\beta}{\partial x}\right)^2
+ \frac{1}{4}\left(\beta^2 - 1\right) 
\left(\frac{\partial\beta}{\partial y}\right)^2 
+ \frac{1}{4} \left(\beta^2 - 1\right) 
\left(\frac{\partial\beta}{\partial z}\right)^2 
\\
&- \left(\beta^{2} - 1\right) \frac{\partial^2\beta}{\partial t\partial x} 
- {\left(\beta^{3} - \beta\right)}\frac{\partial^2\beta}{\partial x^2} 
\end{align}

\be
\mathrm{R}_{\phantom{x}t t y}^{x \phantom{t} \phantom{t} \phantom{y}} = 
-\left(\beta^{2} - 1\right) \frac{\partial\beta}{\partial x} 
\frac{\partial\beta}{\partial y} 
- \frac{1}{2}  {\left(\beta^{2} - 1\right)} 
\frac{\partial^2\beta}{\partial t\partial y} - 
{\left(\beta^{3} - \beta\right)} \frac{\partial^2\beta}{\partial x\partial y} 
\ee

\be
\mathrm{R}_{\phantom{x}t t z}^{x \phantom{t} \phantom{t} \phantom{z}} = 
-{\left(\beta^{2} - 1\right)} \frac{\partial\beta}{\partial x} 
\frac{\partial\beta}{\partial z} - \frac{1}{2}  {\left(\beta^2 - 1\right)} 
\frac{\partial^2\beta}{\partial t\partial z} - {\left(\beta^3 - \beta\right)} 
\frac{\partial^2\beta}{\partial x\partial z} 
\ee

\be
\mathrm{R}_{\phantom{x}t x y}^{x \phantom{t} \phantom{x} \phantom{y}} = 
\frac{1}{2}\left(\beta^{2} - 1\right)\frac{\partial^2\beta}{\partial x\partial y} 
\ee

\be
\mathrm{R}_{\phantom{x}t x z}^{x \phantom{t} \phantom{x} \phantom{z}} = 
\frac{1}{2}\left(\beta^{2} - 1\right) \frac{\partial^2\beta}{\partial x\partial z} 
\ee

\be
\mathrm{R}_{\phantom{x}x t x}^{x \phantom{x} \phantom{t} \phantom{x}} = 
\beta \left(\frac{\partial\beta}{\partial x}\right)^{2} + 
\beta^{2} \frac{\partial^2\beta}{\partial x ^ 2} - 
\frac{1}{4} \beta \left(\frac{\partial\beta}{\partial y}\right)^2 - 
\frac{1}{4} \beta \left(\frac{\partial\beta}{\partial z}\right)^2 + 
\beta \frac{\partial^2\beta}{\partial t\partial x} 
\ee

\be
\mathrm{R}_{\phantom{x}x t y}^{x \phantom{x} \phantom{t} \phantom{y}} = 
\beta^{2} \frac{\partial^2\beta}{\partial x\partial y} + 
\beta \frac{\partial\beta}{\partial x} \frac{\partial\beta}{\partial y} 
+ \frac{1}{2}  \beta \frac{\partial^2\beta}{\partial t\partial y} 
\ee

\be
\mathrm{R}_{\phantom{x}x t z}^{x \phantom{x} \phantom{t} \phantom{z}} 
= \beta^{2} \frac{\partial^2\beta}{\partial x\partial z} + 
\beta \frac{\partial\beta}{\partial x} \frac{\partial\beta}{\partial z} 
+ \frac{1}{2}  \beta \frac{\partial^2\beta}{\partial t\partial z} 
\ee

\be
\mathrm{R}_{\phantom{x}x x y}^{x \phantom{x} \phantom{x} \phantom{y}} 
= -\frac{1}{2}  \beta \frac{\partial^2\beta}{\partial x\partial y} 
\ee

\be
\mathrm{R}_{\phantom{x}x x z}^{x \phantom{x} \phantom{x} \phantom{z}} 
= -\frac{1}{2}  \beta \frac{\partial^2\beta}{\partial x\partial z} 
\ee

\be
\mathrm{R}_{\phantom{x}y t x}^{x \phantom{y} \phantom{t} \phantom{x}} 
= \beta \frac{\partial\beta}{\partial x} \frac{\partial\beta}{\partial y} 
+ \frac{1}{2} \beta \frac{\partial^2\beta}{\partial t\partial y} + 
\frac{1}{2}\left(\beta^2 + 1\right) \frac{\partial^2\beta}{\partial x\partial y} 
\ee

\be
\mathrm{R}_{\phantom{x}y t y}^{x \phantom{y} \phantom{t} \phantom{y}} = 
\beta \left(\frac{\partial\beta}{\partial y}\right)^{2} + 
\frac{1}{2} \left(\beta^2 + 1\right) \frac{\partial^2\beta}{\partial y^2} 
\ee

\be
\mathrm{R}_{\phantom{x}y t z}^{x \phantom{y} \phantom{t} \phantom{z}} = 
\beta \frac{\partial\beta}{\partial y} \frac{\partial\beta}{\partial z} 
+ \frac{1}{2}\left(\beta^2 + 1\right) \frac{\partial^2\beta}{\partial y\partial z} 
\ee

\be
\mathrm{R}_{\phantom{x}y x y}^{x \phantom{y} \phantom{x} \phantom{y}} = 
-\frac{1}{4} \left(\frac{\partial\beta}{\partial y}\right)^{2} - 
\frac{1}{2}\beta \frac{\partial^2\beta}{\partial y ^ 2} 
\ee

\be
\mathrm{R}_{\phantom{x}y x z}^{x \phantom{y} \phantom{x} \phantom{z}} = 
-\frac{1}{2}\beta \frac{\partial^2\beta}{\partial y\partial z} - 
\frac{1}{4}\frac{\partial\beta}{\partial y} \frac{\partial\beta}{\partial z} 
\ee

\be
\mathrm{R}_{\phantom{x}z t x}^{x \phantom{z} \phantom{t} \phantom{x}} = 
\beta \frac{\partial\beta}{\partial x} \frac{\partial\beta}{\partial z} + 
\frac{1}{2}\beta \frac{\partial^2\beta}{\partial t\partial z} + 
\frac{1}{2} \left(\beta^{2} + 1\right) \frac{\partial^2\beta}{\partial x\partial z} 
\ee

\be
\mathrm{R}_{\phantom{x}z t y}^{x \phantom{z} \phantom{t} \phantom{y}} = 
\beta \frac{\partial\beta}{\partial y} \frac{\partial\beta}{\partial z} + 
\frac{1}{2} \left(\beta^{2} + 1\right) \frac{\partial^2\beta}{\partial y\partial z} 
\ee

\be
\mathrm{R}_{\phantom{x}z t z}^{x \phantom{z} \phantom{t} \phantom{z}} = 
\beta \left(\frac{\partial\beta}{\partial z}\right)^{2} + 
\frac{1}{2} \left(\beta^{2} + 1\right) \frac{\partial^2\beta}{\partial z^2} 
\ee

\be
\mathrm{R}_{\phantom{x}z x y}^{x \phantom{z} \phantom{x} \phantom{y}} = 
-\frac{1}{2} \beta \frac{\partial^2\beta}{\partial y\partial z} - 
\frac{1}{4} \frac{\partial\beta}{\partial y} \frac{\partial\beta}{\partial z} 
\ee

\be
\mathrm{R}_{\phantom{x}z x z}^{x \phantom{z} \phantom{x} \phantom{z}} = 
-\frac{1}{4} \left(\frac{\partial\beta}{\partial z}\right)^{2} - 
\frac{1}{2} \beta \frac{\partial^2\beta}{\partial z^2} 
\ee

\be
\mathrm{R}_{\phantom{y}t t x}^{y \phantom{t} \phantom{t} \phantom{x}} = 
\beta \frac{\partial^2\beta}{\partial x\partial y} + 
\frac{\partial\beta}{\partial x} \frac{\partial\beta}{\partial y} + 
\frac{1}{2}  \frac{\partial^2\beta}{\partial t\partial y} 
\ee

\be
\mathrm{R}_{\phantom{y}t t y}^{y \phantom{t} \phantom{t} \phantom{y}} = 
\frac{1}{4}\left(\beta^{2} + 3\right) \left(\frac{\partial\beta}{\partial y}\right)^2 
+ \beta \frac{\partial^2\beta}{\partial y^2} 
\ee

\be
\mathrm{R}_{\phantom{y}t t z}^{y \phantom{t} \phantom{t} \phantom{z}} = 
\frac{1}{4}\left(\beta^{2} + 3\right)\frac{\partial\beta}{\partial y}
\frac{\partial\beta}{\partial z} + 
\beta \frac{\partial^2\beta}{\partial y\partial z} 
\ee

\be
\mathrm{R}_{\phantom{y}t x y}^{y \phantom{t} \phantom{x} \phantom{y}} = 
-\frac{1}{4} \beta \left(\frac{\partial\beta}{\partial y}\right)^{2} 
- \frac{1}{2} \frac{\partial^2\beta}{\partial y^2} 
\ee

\be
\mathrm{R}_{\phantom{y}t x z}^{y \phantom{t} \phantom{x} \phantom{z}} = 
-\frac{1}{4}\beta \frac{\partial\beta}{\partial y} \frac{\partial\beta}{\partial z} 
- \frac{1}{2} \frac{\partial^2\beta}{\partial y\partial z} 
\ee

\be
\mathrm{R}_{\phantom{y}x t x}^{y \phantom{x} \phantom{t} \phantom{x}} = 
-\frac{1}{2} \frac{\partial^2\beta}{\partial x\partial y} 
\ee

\be
\mathrm{R}_{\phantom{y}x t y}^{y \phantom{x} \phantom{t} \phantom{y}} = 
-\frac{1}{4}\beta \left(\frac{\partial\beta}{\partial y}\right)^{2} - 
\frac{1}{2}  \frac{\partial^2\beta}{\partial y ^ 2} 
\ee

\be
\mathrm{R}_{\phantom{y}x t z}^{y \phantom{x} \phantom{t} \phantom{z}} = 
-\frac{1}{4}\beta \frac{\partial\beta}{\partial y} \frac{\partial\beta}{\partial z} 
- \frac{1}{2}  \frac{\partial^2\beta}{\partial y\partial z} 
\ee

\be
\mathrm{R}_{\phantom{y}x x y}^{y \phantom{x} \phantom{x} \phantom{y}} = 
\frac{1}{4}  \left(\frac{\partial\beta}{\partial y}\right)^{2} 
\ee

\be
\mathrm{R}_{\phantom{y}x x z}^{y \phantom{x} \phantom{x} \phantom{z} } = 
\frac{1}{4}  \frac{\partial\beta}{\partial y} \frac{\partial\beta}{\partial z} 
\ee

\be
\mathrm{R}_{\phantom{z} t t x}^{z \phantom{t} \phantom{t} \phantom{x}} =
\beta \frac{\partial^2\beta}{\partial x\partial z} + 
\frac{\partial\beta}{\partial x} \frac{\partial\beta}{\partial z} + 
\frac{1}{2}  \frac{\partial^2\beta}{\partial t\partial z} 
\ee

\be
\mathrm{R}_{\phantom{z}t t y}^{z \phantom{t} \phantom{t} \phantom{y}} = 
\frac{1}{4} \left(\beta^{2} + 3\right) \frac{\partial\beta}{\partial y} 
\frac{\partial\beta}{\partial z} + \beta \frac{\partial^2\beta}{\partial y\partial z} 
\ee

\be
\mathrm{R}_{\phantom{z}t t z}^{z \phantom{t} \phantom{t} \phantom{z}} = 
\frac{1}{4}\left(\beta^{2} + 3\right)\left(\frac{\partial\beta}{\partial z}\right)^2 
+ \beta \frac{\partial^2\beta}{\partial z^2} 
\ee

\be
\mathrm{R}_{\phantom{z}t x y}^{z \phantom{t} \phantom{x} \phantom{y}} = 
-\frac{1}{4} \beta \frac{\partial\beta}{\partial y} \frac{\partial\beta}{\partial z} 
- \frac{1}{2}  \frac{\partial^2\beta}{\partial y\partial z} 
\ee

\be
\mathrm{R}_{\phantom{z}t x z}^{z \phantom{t} \phantom{x} \phantom{z} } = 
-\frac{1}{4} \beta \left(\frac{\partial\beta}{\partial z}\right)^{2} 
- \frac{1}{2}  \frac{\partial^2\beta}{\partial z^2} 
\ee

\be
\mathrm{R}_{\phantom{z}x t x}^{z \phantom{x}\phantom{t} \phantom{x}} = 
-\frac{1}{2}  \frac{\partial^2\beta}{\partial x\partial z} 
\ee

\be
\mathrm{R}_{\phantom{z}x t y}^{z \phantom{x} \phantom{t} \phantom{y}} =
-\frac{1}{4} \beta \frac{\partial\beta}{\partial y} \frac{\partial\beta}{\partial z} 
- \frac{1}{2}  \frac{\partial^2\beta}{\partial y\partial z} 
\ee

\be
\mathrm{R}_{\phantom{z}x t z}^{z \phantom{x} \phantom{t} \phantom{z} } = 
-\frac{1}{4} \beta \left(\frac{\partial\beta}{\partial z}\right)^2 
- \frac{1}{2}  \frac{\partial^2\beta}{\partial z^2} 
\ee

\be
\mathrm{R}_{\phantom{z}x x y}^{z \phantom{x} \phantom{x} \phantom{y}} = 
\frac{1}{4} \frac{\partial\beta}{\partial y} \frac{\partial\beta}{\partial z} 
\ee

\be
\mathrm{R}_{\phantom{z}x x z}^{z \phantom{x} \phantom{x} \phantom{z}} = 
\frac{1}{4} \left(\frac{\partial\beta}{\partial z}\right)^{2}
\ee
\subsection{Ricci components}

\begin{lstlisting}[language=Python]
Ric = g.ricci()
Ric.display_comp(only_nonredundant=True)
\end{lstlisting}

\begin{align}
R_{tt} &= {\left(\beta^{2} - 1\right)} \left(\frac{\partial\beta}
{\partial x}\right)^{2} - \frac{1}{2}  {\left(\beta^{2} + 1\right)} 
\left(\frac{\partial\beta}{\partial y}\right)^{2} - \frac{1}{2}  
{\left(\beta^{2} + 1\right)} \left(\frac{\partial\beta}
{\partial z}\right)^{2} 
\\ 
&+ {\left(\beta^{2} - 1\right)} \frac{\partial^2\beta}
{\partial t\partial x} + {\left(\beta^{3} - \beta\right)} 
\frac{\partial^2\beta}{\partial x ^ 2} - \beta\frac{\partial^2\beta}
{\partial y ^ 2} - \beta\frac{\partial^2\beta}{\partial z ^ 2}\,,
\end{align}

\be
R_{tx} = -\beta \left(\frac{\partial\beta}{\partial x}\right)^2 
- \beta^{2} \frac{\partial^2 \beta}{\partial x^2} 
+ \frac{1}{2}  \beta \left(\frac{\partial\beta}{\partial y}\right)^{2} 
+ \frac{1}{2}  \beta \left(\frac{\partial\beta}{\partial z}\right)^{2} 
- \beta \frac{\partial^2\beta}{\partial t\partial x} 
+ \frac{1}{2} \frac{\partial^2\beta}{\partial y^2} 
+ \frac{1}{2} \frac{\partial^2\beta}{\partial z^2}\,,
\ee

\be
R_{ty} = - \beta \frac{\partial\beta}{\partial x} 
\frac{\partial\beta}{\partial y} 
-\frac{1}{2}\beta \frac{\partial^2\beta}{\partial t\partial y} 
-\frac{1}{2}\left(\beta^2 + 1\right) 
\frac{\partial^2\beta}{\partial x\partial y}\,,
\ee

\be
R_{tz} = - \beta \frac{\partial\beta}{\partial x} 
\frac{\partial\beta}{\partial z} 
- \frac{1}{2}\beta \frac{\partial^2\beta}{\partial t\partial z} 
- \frac{1}{2}\left(\beta^2 + 1\right) 
\frac{\partial^2\beta}{\partial x\partial z}\,,
\ee

\be
R_{xx} = \left(\frac{\partial\beta}{\partial x}\right)^2 
+ \beta \frac{\partial^2\beta}{\partial x ^ 2} 
- \frac{1}{2}  \left(\frac{\partial\beta}{\partial y}\right)^2
- \frac{1}{2}  \left(\frac{\partial\beta}{\partial z}\right)^2 
+ \frac{\partial^2\beta}{\partial t\partial x}\,,
\ee

\be
R_{xy} = \frac{1}{2}\beta\frac{\partial^2\beta}{\partial x\partial y} 
+ \frac{\partial\beta}{\partial x} \frac{\partial\beta}{\partial y} 
+ \frac{1}{2}\frac{\partial^2\beta}{\partial t\partial y}\,,
\ee

\be
R_{xz} = \frac{1}{2}\beta \frac{\partial^2\beta}{\partial x\partial z} 
+ \frac{\partial\beta}{\partial x} \frac{\partial\beta}{\partial z} 
+ \frac{1}{2} \frac{\partial^2\beta}{\partial t\partial z}\,,
\ee

\be
R_{yy} = \frac{1}{2}\left(\frac{\partial\beta}{\partial y}\right)^2\,,
\ee

\be
R_{yz} = \frac{1}{2}\frac{\partial\beta}{\partial y} 
\frac{\partial\beta}{\partial z}\,,
\ee

\be
R_{zz} = \frac{1}{2}\left(\frac{\partial\beta}{\partial z}\right)^2
\ee

\subsection{Einstein tensor components}

the Einstein tensor components with added geometrical constants are
\be
G_{00} =  \Lambda(1-\beta^2) - \frac{1}{4} 
(1 + 3\beta^2)
\left[
\left(\frac{\partial \beta}{\partial y} \right)^2 +  
\left(\frac{\partial \beta}{\partial z} \right)^2 
\right] 
- \beta \left(\frac{\partial^2 \beta}{\partial y^2} + 
\frac{\partial^2 \beta}{\partial z^2}\right),
\label{et00-app}
\ee
\be
G_{01} =  \Lambda \beta + \frac{3}{4} 
\beta \left[
\left(\frac{\partial \beta}{\partial y}\right)^2 
+ \left(\frac{\partial \beta}{\partial z}\right)^2 
\right] 
+ \frac{1}{2}\left(
\frac{\partial^2 \beta}{\partial y^2} 
+ \frac{\partial^2 \beta}{\partial z^2}
\right),
\label{et01-app}
\ee
\be
G_{02} = - \frac{1}{2}
\frac{\partial^2 \beta}{\partial x \partial y} 
- \frac{\beta}{2} 
\left(2\frac{\partial \beta}{\partial y}
\, \frac{\partial \beta}{\partial x} +
\beta \frac{\partial^2 \beta}{\partial x \partial y} +
\frac{\partial^2 \beta}{\partial t \partial y}\right),
\label{et02-app}
\ee
\be
G_{03} = - \frac{1}{2}
\frac{\partial^2 \beta}{\partial x \partial z} 
- \frac{\beta}{2} 
\left(2\frac{\partial \beta}{\partial z}
\, \frac{\partial \beta}{\partial x} +
\beta \frac{\partial^2 \beta}{\partial x \partial z} +
\frac{\partial^2 \beta}{\partial t \partial z}\right),
\label{et03-app}
\ee
\be
G_{11} = \Lambda - \frac{3}{4} \left[
\left(\frac{\partial \beta}{\partial y}\right)^2 
+ \left(\frac{\partial \beta}{\partial z}\right)^2
\right], 
\label{et11-app}
\ee
\be
G_{12} = \frac{1}{2}\left(
2 \frac{\partial \beta}{\partial y} \, 
\frac{\partial \beta}{\partial x} 
+ \beta \frac{\partial^2 \beta}{\partial x \partial y} 
+ \frac{\partial^2 \beta}{\partial t \partial y}\right),
\label{et12-app}
\ee
\be
G_{13} = \frac{1}{2}\left(
2 \frac{\partial \beta}{\partial z} \, 
\frac{\partial \beta}{\partial x} 
+ \beta \frac{\partial^2 \beta}{\partial x \partial z} 
+ \frac{\partial^2 \beta}{\partial t \partial z}\right),
\label{et13-app}
\ee
\be
G_{23} = \frac{1}{2} \frac{\partial \beta}{\partial z} 
\, \frac{\partial \beta}{\partial y},
\label{et23-app}
\ee
\be
G_{22} = - \Lambda - \frac{\partial}{\partial x}\left[
\frac{\partial \beta}{\partial t}
+ \frac{1}{2} \frac{\partial}{\partial x} (\beta^2)
\right]
- \frac{1}{4}\left[
\left(\frac{\partial \beta}{\partial y}\right)^2
- \left(\frac{\partial \beta}{\partial z}\right)^2
\right],
\label{et22-app}
\ee
\be
G_{33} = - \Lambda - \frac{\partial}{\partial x}\left[
\frac{\partial \beta}{\partial t}
+ \frac{1}{2} \frac{\partial}{\partial x} (\beta^2)
\right]
+ \frac{1}{4}\left[
\left(\frac{\partial \beta}{\partial y}\right)^2
- \left(\frac{\partial \beta}{\partial z}\right)^2
\right]\,\,.
\label{et33-app}
\ee
\subsection{Christoffel symbols}

\begin{lstlisting}[language=Python]
print(gg)
gg.display_comp()
\end{lstlisting}

\be
\Gamma_{ \phantom{\, t} \, t \, t }^{ \, t \phantom{\, t} \phantom{\, t} } = \beta^2 \frac{\partial \beta}{\partial x}
\ee

\be
\Gamma_{ \phantom{\, t} \, t \, x }^{ \, t \phantom{\, t} \phantom{\, x} } = -\beta \frac{\partial \beta}{\partial x}
\ee

\be
\Gamma_{ \phantom{\, t} \, t \, y }^{ \, t \phantom{\, t} \phantom{\, y} } = -\frac{1}{2} \beta \frac{\partial \beta}{\partial y}
\ee

\be
\Gamma_{ \phantom{\, t} \, t \, z }^{ \, t \phantom{\, t} \phantom{\, z} } = -\frac{1}{2} \beta \frac{\partial \beta}{\partial z}
\ee

\be
\Gamma_{ \phantom{\, t} \, x \, x }^{ \, t \phantom{\, x} \phantom{\, x} } = \frac{\partial \beta}{\partial x}
\ee

\be
\Gamma_{ \phantom{\, t} \, x \, y }^{ \, t \phantom{\, x} \phantom{\, y} } = \frac{1}{2} \frac{\partial \beta}{\partial y}
\ee

\be
\Gamma_{ \phantom{\, t} \, x \, z }^{ \, t \phantom{\, x} \phantom{\, z} } = \frac{1}{2} \frac{\partial \beta}{\partial z}
\ee

\be
\Gamma_{ \phantom{\, x} \, t \, t }^{ \, x \phantom{\, t} \phantom{\, t} } = \left(\beta^3 - \beta\right) \frac{\partial \beta}{\partial x} - \frac{\partial \beta}{\partial t}
\ee

\be
\Gamma_{ \phantom{\, x} \, t \, x }^{ \, x \phantom{\, t} \phantom{\, x} } = -\beta^2 \frac{\partial \beta}{\partial x}
\ee

\be
\Gamma_{ \phantom{\, x} \, t \, y }^{ \, x \phantom{\, t} \phantom{\, y} } = -\frac{1}{2} \left(\beta^2 + 1\right) \frac{\partial \beta}{\partial y}
\ee

\be
\Gamma_{ \phantom{\, x} \, t \, z }^{ \, x \phantom{\, t} \phantom{\, z} } = -\frac{1}{2} \left(\beta^2 + 1\right) \frac{\partial \beta}{\partial z}
\ee

\be
\Gamma_{ \phantom{\, x} \, x \, x }^{ \, x \phantom{\, x} \phantom{\, x} } = \beta \frac{\partial \beta}{\partial x}
\ee

\be
\Gamma_{ \phantom{\, x} \, x \, y }^{ \, x \phantom{\, x} \phantom{\, y} } = \frac{1}{2} \beta \frac{\partial \beta}{\partial y}
\ee

\be
\Gamma_{ \phantom{\, x} \, x \, z }^{ \, x \phantom{\, x} \phantom{\, z} } = \frac{1}{2} \beta \frac{\partial \beta}{\partial z}
\ee

\be
\Gamma_{ \phantom{\, y} \, t \, t }^{ \, y \phantom{\, t} \phantom{\, t} } = -\beta \frac{\partial \beta}{\partial y}
\ee

\be
\Gamma_{ \phantom{\, y} \, t \, x }^{ \, y \phantom{\, t} \phantom{\, x} } = \frac{1}{2} \frac{\partial \beta}{\partial y}
\ee

\be
\Gamma_{ \phantom{\, z} \, t \, t }^{ \, z \phantom{\, t} \phantom{\, t} } = -\beta \frac{\partial \beta}{\partial z}
\ee

\be
\Gamma_{ \phantom{\, z} \, t \, x }^{ \, z \phantom{\, t} \phantom{\, x} } = \frac{1}{2} \frac{\partial \beta}{\partial z}
\ee

\section{Warp drive thought experiment}

Considering two points in space, one labeled $A$ and the other labeled $B$, separated by a spatial proper distance $D$ that is large enough to neglect any gravitational interaction between those two points. Two observers remain in points $A$ and $B$ in flat spacetime. The mass particle departures from rest of point $A$ with smooth acceleration and no spacetime disturbance until it reaches a distance $d \ll D$ from point $A$, and then it begins to move in a rapidly changing coordinate acceleration $a$ propelled by the warp bubble until it reaches the middle point between $A$ and $B$ and starts to decelerate with $- a$ up to a point in space separated by a distance $d$ from point $B$ and the warp bubble would cease to disturb spacetime and the particle could complete covering the distance $d$ with a smooth travel. The time for the particle to cover this distance is less than a ray of light would take. 

\begin{figure}[!h]
 \centering
 \includegraphics[scale=0.55]{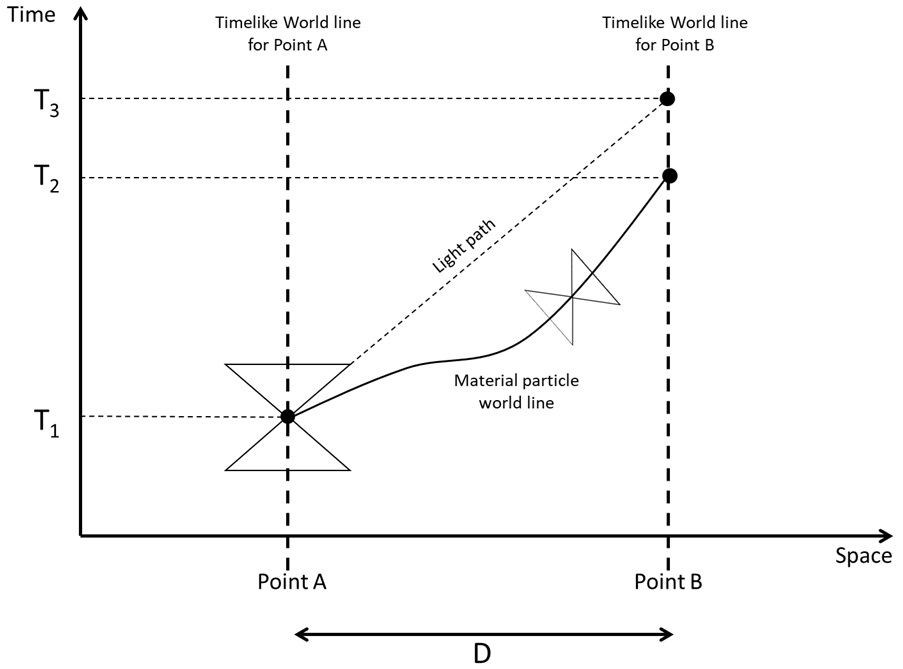}
 \caption[Spacetime diagram for bubble worldline.]{For an observer $A$, a ray of light takes a time $2D/c$ to make a round trip between $A$ and $B$. A particle with mass can make the run in a time less than $2D/c$ according 
 to the Warp Drive Theory.}
 \label{gedanken_1}
\end{figure}

Fig.\,\ref{gedanken_1} is schematic for this effective superluminal method of propulsion. A particle with mass is initially at rest with synchronized watches with another observer at the same spatial point. If at $t = 0$, the observer at rest emitted a photon, and the spaceship departed from the same spatial point. If the particle could accelerate instantaneously from rest and follows Alcubierre's proposition, it would appear to the observer left behind at rest to be in a spacelike geodesic, like the theoretical particle tachyon. It would make the run faster than the emitted photon. Locally, the particle would always be inside a lightcone in a timelike geodesic because the warp bubble would distort spacetime and tilt the lightcone, preserving causality for the particle in the interior of this bubble.

\begin{figure}[!h]
 \centering
 \includegraphics[scale=0.70]{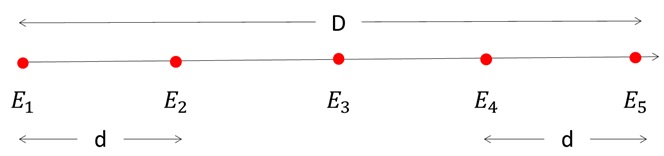}
 \caption[Alcubierre thought experiment schematics.]{Sequence of events on Alcubierre superluminal travel thought experiment. Points $A$ and $B$ are separated by a distance $D$ and coincide with $E_1$ and $E_2$, respectively.}
 \label{gedanken_2}
\end{figure}

A ray of light sent from $A$ to $B$ would travel the distance of those two points with a time $D/c$, and if the ray of light is reflected from $B$ to $A$, the total time of travel would be $2D/c$. It is well known that within the framework of Special Relativity, a particle with mass can't cover the distance $D$ between $A$ and $B$ with a coordinate time less than $2D/c$ as perceived by observers $A$ and $B$, when it comes to General Relativity spacetime is no longer exclusively flat since it can be distorted by any source allowing a very intricate and nonlinear dynamic between spacetime coordinates, which means that spacetime can vary over time and space depending on the source of matter, momentum, energy distribution and also by geometry itself. With all that in mind, Alcubierre described how a particle with mass would make the round trip between $A$ and $B$ with time less than $2D/c$. The warp drive thought experiment can be broken down into the following sequence of events $(E_1, E_2, E_3, E_4, E_5)$, see Fig.\,\ref{gedanken_2}. 

All the events in the warp drive thought experiment are described below.

\begin{itemize}
\item Event 1 ($E_1$): The particle leaves an inertial reference frame in point $A$ with constant velocity $v < c$ toward another inertial reference frame represented in spacetime by a point $B$. The points $A$ and $B$ are separated by a proper distance $D$. In this representation, the event $E_1$ coincides with point $A$.
\item Event 2 ($E_2$): The particle stops at point $E_2$, at a distance $d \ll D$ from point $A$, and the warp bubble begins to distort spacetime, propelling the particle inside the bubble to instantly acquire an acceleration $a$ heading to point $B$.
\item Event 3 ($E_3$): This event takes place at the midpoint between $A$ and $B$. The warp bubble starts to decelerate with equal but negative acceleration $-a$ until it reaches a point separated by a distance $d$ from point $B$.
\item Event 4 ($E_4$): The warp bubble with the particle in its interior stops decelerating until it stops at a distance $d$ from point $B$ and starts to move with the same constant speed $v < c$ from the beginning in event $(E_1)$.
\item Event 5 ($E_5)$: The particle arrives at point $E_5$ that coincides with point $B$ in space, then repeats the same steps as it returns to point $A$.
\end{itemize}
\vspace{3pt}

For events $E_1$ and $E_5$ in flat spacetime where the spaceship covers a distance $d$ in each event with a constant velocity $v < c$, the time it takes to cover the distance is 
\be
T_1 = \frac{2d}{v} \,.   
\ee
In the region where the spacetime is disturbed by the warp bubble between events $E_2$ and $E_4$ for distant observers, it takes
\be
T_2 = 2\sqrt{\frac{D - 2d}{a}} 
\label{T2wp}
\ee
for the particle to move along its trajectory, so the total coordinate time $(T = T_1 + T_2)$ elapsed and measured by far away observers is given by the following relation,
\be
T = 2 \left( \frac{d}{v} + \sqrt{\frac{D-2d}{a}} \, \right)\,.
\label{Twp}
\ee
The spaceship proper time for the first part of the 
trip is 
\be
\tau_1 = 2\frac{d}{\gamma v} \,,
\label{gamma}
\ee
where $\gamma$ is the Lorentz factor, as it is well-known from special relativity given by 
\be
\gamma = \frac{1}{\sqrt{1 - \frac{v^2}{c^2}}} \,.
\ee

The proper time for the second part of the trajectory is the same as $T_2$ in equation \eqref{T2wp}. The total proper time $(\tau = \tau_1 + \tau_2)$ measured by the spaceship is given by the following
\be
\tau = 2 \left(\frac{d}{\gamma v} + \sqrt{\frac{D-2d}{a}} \,\right).
\label{tau1wp}
\ee
This relation shows that the time dilation on the first term of the right-hand side of the equation \eqref{tau1wp} comes only from the trip stages where the particle moves through flat spacetime. The second term on the right-hand side of both equations \eqref{Twp} and \eqref{tau1wp} comes from when spacetime is dynamic and distorted by the bubble. If the condition $(d \ll D)$ holds, then as a first-order Taylor expansion in $d/D$ has the following result for both the time measured by the observer in point $A$ and the particle's proper time
\be
\tau \approx T \approx 2 \sqrt{\frac{D}{a}}\,.
\label{approx}
\ee
Eq.\,\eqref{approx} implies that the time $T$ measured by an observer located at point $A$ or $B$ and the proper time measured by the particle inside the warp bubble can be made arbitrarily small by increasing the value of the acceleration $a$. The particle will then be able to travel interstellar distances with superluminal speed propelled by spacetime distortion. At the same time, it will always remain on a timelike trajectory in its local lightcone. 

The thought experiment proposed by Alcubierre \cite{Alcubierre1994} considered a spaceship inside a warped bubble with a determined radius and thickness that was able to distort spacetime, expanding space behind and contracting it in front of it, being propelled forward by spacetime itself in a superluminal fashion. this warp bubble would be embedded in flat spacetime, being spacetime itself, since the spaceship inside of it would also be in a flat spacetime in an inertial reference frame, thus respecting the laws of special relativity. A distant observer at rest in flat spacetime would measure the time that the spaceship would take to make a trip between points in space that are less than would take light to do it.

\begin{figure}[!h]
 \centering
 \includegraphics[scale=0.55]{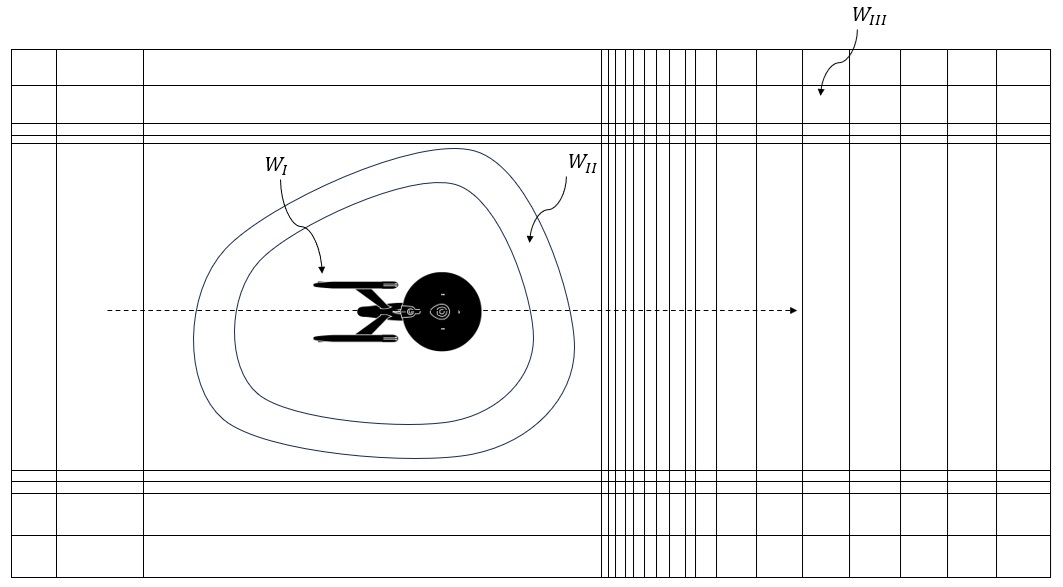}
 \caption[Spaceship embedded in WD spacetime.]{Schematic representation of the thought experiment proposed by Alcubierre. The image shows a spaceship inside the warp bubble moving across spacetime, expanding spacetime behind it, and contracting	it in front of it.}
 \label{warp_drive_3}
\end{figure}

Fig.\,\eqref{warp_drive_3} is a schematic representation of Alcubierre's thought experiment, and it must be noted that the image contains artistic liberty since it shows a classical misinterpretation of what spacetime is and how it behaves, using the analogy of the \textit{fabric of spacetime}. Fig.\,\eqref{warp_drive_3} shows the spaceship inside the bubble with fixed thickness and radius. There are three regions represented: $W_I$ is the interior of the warp bubble, a flat spacetime where the spaceship is transported, $W_{II}$ is the warped region forming a thin shell that separates $W_I$ from  $W_{III}$, which which is the exterior and more general flat spacetime where the bubble resides. The vertical and horizontal lines are representations of the \textit{fabrics of spacetime} being distorted by the warped region $W_{II}$. The lines approach as they approach the warp bubble from the front but displace from each other as they move far from the backside of the bubble. 

One of the main caveats of warp drive theory lies in the neighborhood of region $W_{III}$ since it connects the other two spacetimes. It encodes many controversial aspects of General Relativity, i.e., event horizons, black and white hole behaviors, effective negative pressure, apparent negative energy density, complex-valued solutions, possibly vacuum solutions, etc. 

\textbf{Disclaimer}:

\textit{It must be pointed out that this image is an aid for the reader to capture the nature of the thought experiment proposed by Alcubierre \cite{Alcubierre1994}. In no way is this a formal representation of how General Relativity works. Even though this \textit{fabric of spacetime} analogy is widely adopted and accepted by some scientific promoters, layman public, or even beginning undergraduate students at physics courses, being an incredible aid to get acquainted with the intricacies of the theory of general relativity, it is imprecise.}
\section{Shock Wave Theory}

\index[authors]{Rankine, W.\,J.\,M.}
\index[authors]{Hugoniot, H.}
The foundations of shock wave theory and thermodynamic implications emerged through the works of Rankine and Hugoniot. Rankine \cite{rankine1870} was the first to demonstrate that shocks involve a non-adiabatic process characterized by the irreversible generation of entropy within the shock. This critical insight provided a framework for understanding the discontinuous changes in state variables across a shock front, forming the basis of the Rankine-Hugoniot conditions.

Hugoniot \cite{hugoniot1887, hugoniot1889} showed that, in the absence of viscosity and heat conduction, the conservation of energy implies two key principles: conservation of entropy in smooth regions of flow and a distinct jump in entropy across a shock. These results provided a rigorous mathematical and thermodynamic basis for analyzing shock wave behavior, offering insights into the nature of discontinuities in fluid motion. Hugoniot's equations, which describe the relationship between pre- and post-shock states, have become a cornerstone of compressible fluid dynamics.

\index[authors]{Courant, R.}
\index[authors]{Friedrichs, K.\,O.}
\index[authors]{ZelDovich, Y.\,B.}
\index[authors]{Raizer, Y.\,P.}
\index[authors]{Salas, M.\,D.}
\index[authors]{Forbes, J.\,W.}
Several texts have expanded on these foundational works, applying the principles of shock wave theory to broader contexts. Courant and Friedrichs \cite{courant1999} provided a detailed mathematical treatment of supersonic flow and shock waves, emphasizing the stability and structure of shocks in various flow regimes. Similarly, ZelDovich and Raizer \cite{zeldovich2012} extended the theory to high-temperature hydrodynamic phenomena, exploring the interplay between shock waves, chemical kinetics, and plasma dynamics. Salas provided a review of the events and contributions that led to the modern understanding of shock phenomena \cite{salas2006}. Forbes \cite{forbes2012} provided an entry point for studying shock compression in condensed matter, offering insights into phase transitions, material properties, and the Hugoniot curve under high-pressure conditions. 

\index[authors]{Liepmann, H.\,W.}
\index[authors]{Roshko, A.}
\index[authors]{Shapiro, A.\,H.}
\index[authors]{Landau, L.\,D.}
\index[authors]{Lifschitz, E.\,M.}
Liepmann and Roshko \cite{liepmann1957} introduced gas dynamics, focusing on isentropic flow, shock waves, and unsteady wave motion. Shapiro \cite{shapiro1953} advanced these ideas of compressible fluid dynamics, incorporating practical applications. Landau and Lifshitz \cite{landau1959} addressed the broader fluid mechanics framework, integrating compressible flows' behavior within a comprehensive theoretical context. In their later work \cite{landau2013}, they revisited the Rankine-Hugoniot jump conditions, approaching them from the perspective of conservation laws for physical quantities such as energy, momentum, and heat flow. Providing a unifying theoretical framework for understanding shock waves as a manifestation of the fundamental principles of conservation, highlighting their role in both classical fluid dynamics and high-energy phenomena. 

\index[authors]{Smoller, E.}
\index[authors]{Temple, B.}
Smoller and Temple \cite{smoller1993} extended the study of shock wave theory to the relativistic Euler equations, addressing the global existence of solutions under specific initial conditions. Their work analyzes the interplay between relativistic effects, nonlinear dynamics, and the formation of shock waves, studying the stability and behavior of relativistic fluids. 
\end{appendix}



\newglossaryentry{local light cone}
{
name = Local light cone,
text = local light cone,
description={In general relativity, the local light cone at a given spacetime point represents the boundary of causal influence for that point. It is defined by the set of all possible light rays (null geodesics) that pass through the point, separating events that can be causally connected from those that cannot. The cone's interior consists of timelike trajectories corresponding to the physically realizable motion of massive particles. In contrast, the cone's surface corresponds to light-like (null) paths followed by photons and other massless particles. The light cone structure governs causality, ensuring that no information or influence can propagate faster than the speed of light.}
}

\newglossaryentry{Alcubierre metric}
{
    name = Alcubierre Metric,
    description = {A theoretical solution to the Einstein field equations, proposed by Miguel Alcubierre, describing a spacetime geometry allowing superluminal travel within a "warp bubble" that contracts space in front and expands it behind.}
}

\newglossaryentry{ADM formalism}
{
    name = ADM (Arnowitt-Deser-Misner) Formalism,
    description = {A mathematical framework in General Relativity that decomposes spacetime into a three-dimensional hypersurface evolving, facilitating the study of dynamic spacetimes like warp drives.}
}

\newglossaryentry{anisotropic fluid}
{
    name = Anisotropic fluid,
    text = anisotropic fluid,
    description = {A type of fluid used in modeling matter-energy sources for spacetime geometries, where pressures are not the same in all space directions as in the perfect fluid with isotropic pressure.}
}

\newglossaryentry{Burgers equation}
{
    name = Burgers equation,
    description = {is a fundamental partial differential equation (PDE) used in fluid mechanics, nonlinear acoustics, and traffic flow modeling. It is one of the simplest nonlinear PDEs that exhibits shock wave formation and turbulent behavior.}
}

\newglossaryentry{Cosmological Constant}
{
    name = Cosmological constant,
    text = {cosmological constant},
    description = {is a fundamental term in Einstein’s field equations of general relativity that represents a uniform energy density filling space. It was originally introduced by Albert Einstein in 1917 to allow for a static universe but later became crucial in explaining its accelerated expansion.}
}

\newglossaryentry{Einstein equations}
{
    name = Einstein equations,
    description = {The fundamental equations of General Relativity, relating the curvature of spacetime (via the Einstein tensor) to the energy and momentum of matter and radiation.}
}

\newglossaryentry{energy conditions}
{
    name = Energy conditions,
    text = energy conditions,
    description = {Criteria imposed on the energy-momentum tensor in General Relativity to ensure physically reasonable solutions, including the Weak Energy Condition (WEC), Strong Energy Condition (SEC), Null Energy Condition (NEC), and Dominant Energy Condition (DEC).}
}

\newglossaryentry{extrinsic curvature}
{
    name = Extrinsic curvature,
    text = extrinsic curvature,
    description = {In differential geometry and general relativity, extrinsic curvature describes how a surface or hypersurface is embedded in a higher-dimensional space. It measures the degree to which the surface bends relative to the surrounding space and is formally defined in terms of the change in the unit normal vector along the surface. The second fundamental form characterizes extrinsic curvature and is crucial in formulating the junction conditions in general relativity, such as studying black hole horizons and cosmological boundaries. It contrasts with intrinsic curvature, which describes curvature inherent to the surface, independent of its embedding.}
}

\newglossaryentry{geodesic}
{
    name = Geodesic,
    text = geodesic,
    description = {The path of least distance in curved spacetime, representing the trajectory of free-falling objects under gravity.}
}

\newglossaryentry{junction conditions}
{
    name = Junction conditions,
    text = junction conditions,
    description = {Boundary conditions connecting different regions of spacetime, such as between a warp drive spacetime and Minkowski spacetime, ensuring physical continuity.}
}

\newglossaryentry{negative energy density}
{
    name = Negative energy density,
    text = negative energy density,
    description = {A controversial requirement for warp drive spacetimes, where the energy density becomes less than zero, typically associated with exotic matter.}
}

\newglossaryentry{perfect fluid}
{
    name = Perfect fluid,
    text = perfect fluid,
    description = {A hypothetical fluid with uniform properties used to model the matter-energy content of spacetime, often employed in warp drive solutions.}
}

\newglossaryentry{ricci tensor}
{
    name = Ricci tensor ($R_{\mu\nu}$),
    description = {A contraction of the Riemann curvature tensor that measures spacetime curvature due to matter-energy content, central to Einstein’s equations.}
}

\newglossaryentry{shock wave}
{
    name = Shock wave,
    text = shock wave,
    description = {is a propagating disturbance that moves faster than the local speed of sound in the medium. Like an ordinary wave, a shock wave carries energy and can propagate through a medium but is characterized by an abrupt, nearly discontinuous change in pressure, temperature, and density.}
}

\newglossaryentry{spacetime foliation}
{
    name = Spacetime foliation,
    text = spacetime foliation,
    description = {The division of spacetime into a series of spatial slices, each representing a "snapshot" of the universe at a given time, integral to ADM formalism.}
}

\newglossaryentry{warp bubble}
{
    name = Warp bubble,
    text = warp bubble,
    description = {A localized distortion of spacetime proposed in warp drive theories, enabling superluminal travel while maintaining subluminal speeds within the bubble.}
}

\newglossaryentry{warp drive}
{
    name = Warp drive,
    text = warp drive,
    description = {A theoretical propulsion system based on the Alcubierre metric, enabling faster-than-light travel through spacetime distortions.}
}

\newglossaryentry{riemanntensor}
{
    name = Riemann tensor ($R^{\alpha}_{\beta\mu\nu}$),
    description = {A four-dimensional tensor describing the curvature of spacetime, capturing how a vector changes when parallel transported around a closed loop in spacetime.}
}

\newglossaryentry{minkowskispacetime}
{
    name = Minkowski spacetime,
    description = {A flat spacetime geometry used in Special Relativity, serving as the simplest solution to the Einstein field equations in the absence of matter or energy.}
}

\newglossaryentry{exoticmatter}
{
    name = Exotic matter,
    text = exotic matter,
    description = {Hypothetical matter with negative energy density, required to stabilize spacetime structures such as warp bubbles and wormholes, though its physical existence is speculative.}
}

\newglossaryentry{shiftvector}
{
    name = Shift vector,
    text = shift vector,
    description = {A vector in ADM formalism that describes how spatial coordinates shift between successive hypersurfaces, critical for defining the warp bubble dynamics.}
}

\newglossaryentry{lapsefunction}
{
    name = Lapse function ($\alpha$),
    text = lapse function,
    description = {A scalar function in ADM formalism that relates proper time to coordinate time, influencing the rate of evolution of hypersurfaces in spacetime.}
}

\newglossaryentry{hyperbolic}
{
    name = Hyperbolic,
    text = hyperbolic,
    description = {A mathematical property of spacetime geometries ensuring causality and predictability, implying that the Einstein field equations are well-posed for initial value problems.}
}

\newglossaryentry{causality violation}
{
    name = Causality violation,
    text = causality violation,
    description = {The occurrence of closed timelike curves or other phenomena that allow information or matter to travel backward in time, a potential issue in warp drive geometries.}
}

\newglossaryentry{vacuum solution}
{
    name = Vacuum solution,
    text = vacuum solution,
    description = {A solution to the Einstein field equations in regions where the energy-momentum tensor vanishes, such as in empty spacetime or idealized warp bubble edges.}
}

\newglossaryentry{timelike vector}
{
    name = Timelike vector,
    text = timelike vector,
    description = {A vector representing an observer's trajectory through spacetime, with a magnitude corresponding to proper time along the path.}
}

\newglossaryentry{null vector}
{
    name = Null vector,
    text = null vector,
    description = {A vector with zero magnitude under the spacetime metric, corresponding to the trajectory of light or other massless particles.}
}

\newglossaryentry{gravitational wave}
{
    name = Gravitational waves,
    text = Gravitational waves,
    description = {Ripples in spacetime caused by accelerating masses, which can propagate through spacetime and interact with warp bubble dynamics.}
}

\newglossaryentry{Null Energy Condition}
{
    name = Null energy condition,
    text = null energy condition,
    description = {An energy condition stating that the energy-momentum tensor contracted with any null vector must be non-negative, often violated in exotic spacetimes.}
}

\newglossaryentry{Dominant Energy Condition}
{
    name = Dominant energy condition,
    text = dominant energy condition,
    description = {is a condition in general relativity that imposes restrictions on the types of physically reasonable matter and energy distributions in spacetime. It ensures that energy density, as measured by any observer, is non-negative and that energy-momentum flows causally and with subliminal speed.}
}

\newglossaryentry{Weak Energy Condition}
{
    name = Weak energy condition,
    text = weak energy condition,
    description = {is a fundamental requirement in general relativity that ensures that energy density, as measured by any observer, is always non-negative. It is one of several energy conditions that describe physically reasonable distributions of matter and energy.}
}

\newglossaryentry{Strong Energy Condition}
{
    name = Strong energy condition,
    text = strong energy condition,
    description = {is a criterion that restricts the types of energy and matter distributions allowed in spacetime. It ensures that gravity remains attractive by requiring that the combined energy density and pressure of a system contribute positively to spacetime curvature. It implies that the total energy density plus the sum of the pressures in all spatial directions remains non-negative, which is crucial in classical cosmology. However, it is known to be violated by dark energy and certain exotic fields, leading to accelerated cosmic expansion.}
}

\newglossaryentry{tachyon}
{
    name = Tachyon,
    text = tachyon,
    description = {A hypothetical particle traveling faster than light, often considered in theoretical discussions of superluminal travel and causality.}
}

\newglossaryentry{metrictensor}
{
    name = Metric tensor ($g_{\mu\nu}$),
    text = metric tensor,
    description = {A mathematical object describing the geometry of spacetime, determining distances, angles, and causal structure.}
}

\newglossaryentry{closedtimelikecurve}
{
    name = Closed timelike curve (CTC),
    text = Closed timelike curve (CTC),
    description = {A loop in spacetime allowing a return to the same point in time, potentially leading to paradoxes in causality.}
}

\newglossaryentry{warpfield}
{
    name = Warp field,
    text = Warp field,
    description = {The region of spacetime distorted to create a warp bubble, enabling apparent superluminal travel.}
}

\newglossaryentry{eventhorizon}
{
    name = Event horizon,
    text = event horizon,
    description = {A boundary in spacetime beyond which events cannot influence an outside observer, potentially arising in warp bubble geometries.}
}

\newglossaryentry{spacetimesymmetry}
{
    name = Spacetime symmetry,
    text = spacetime symmetry,
    description = {Invariants of the spacetime geometry under certain transformations, important in the analysis of warp drive metrics and energy conditions.}
}

\newglossaryentry{Christoffel Symbols}
{
    name = Christoffell symbols,
    description = {In differential geometry and general relativity, Christoffell symbols are mathematical objects that describe how coordinate bases change in curved spacetime. They are used to define the covariant derivative and express how vectors and tensors are transported in a curved manifold. Although they are not tensors themselves, they play a crucial role in formulating the geodesic equation, which describes the motion of free-falling particles. Christoffell symbols are computed from the metric tensor and encode the effects of spacetime curvature on the motion of objects.}
}

\newglossaryentry{regulating form function}
{
    name = Regulating form function,
    text = regulating form function,
    description = {In Miguel Alcubierre’s warp drive model is a smooth mathematical function that defines the shape and distribution of the warp bubble, determining how spacetime is contracted in front of and expanded behind the bubble. This function ensures a localized and finite energy distribution, preventing singularities or discontinuities in spacetime. The standard form function used in the Alcubierre metric is:  
\[
f(r_s) = \frac{\tanh \big[\sigma (r_s + R)\big] - \tanh\big([\sigma (r_s - R)\big]}{2\tanh(\sigma R)}
\]
where $r_s = \sqrt{(x - x_s(t))^2 + y^2 + z^2}$ is the radial distance from the bubble’s center at position $x_s(t)$, $R$ represents the characteristic radius of the warp bubble, $\sigma$ is a parameter controlling the steepness of the transition between the inside and outside of the bubble.}
}

\printglossaries
\addcontentsline{toc}{section}{Subject Index}
\printindex[topics]

\addcontentsline{toc}{section}{Author Index}
\printindex[authors]

\end{document}